\documentclass[prb,url]{revtex4}
\usepackage[toc,page]{appendix}
\usepackage{graphicx,pdflscape,epsf,epsfig,amsmath}% Include figure files
\usepackage{alltt,dsfont,amsmath,amssymb,bm}

\usepackage{hyperref}
\usepackage{color}

\newcommand{\llabel}[1]{\label{#1} }

\newcommand{\iden}{ \mathds{ 1}}

\renewcommand{\lll}{\langle \langle}

\newcommand{\rrr}{\rangle \rangle}
\newcommand{\A}{{\cal A}}
\newcommand{\up}{\Upsilon}

\newcommand{\G}{{\cal{G}}}
\newcommand{\GI}{{\cal{G}}^{-1}}

\newcommand{\GLI}{{\hat{G}}^{-1}}

\newcommand{\GH}{{\bf g}}
\newcommand{\GHI}{\GH^{-1}}

\newcommand{\LL}{{\bf L}}
\newcommand{\ttau}{{ (\cal{T}) }}

\newcommand{\del}[1]{\delta_{#1}}

\newcommand{\X}[2]{X_{{#1}}^{#2}}
\newcommand{\si}{\sigma}
\newcommand{\sib}{\bar{\sigma}}

\newcommand{\tJ}{\ $t$-$J$ \ }

\newcommand{\U}{{\cal U}}

\newcommand{\V}{{\cal V}}

\newcommand{\bb}[1]{{{\mathbf #1}}}
\newcommand{\nn}{\nonumber}
\newcommand{\chem}{{\bm \mu}}

\newcommand{\beq}{\begin{equation}}
\newcommand{\eeq}{\end{equation}}
\newcommand{\barray}{\begin{eqnarray}}
\newcommand{\earray}{\end{eqnarray}}
\newcommand{\disp}[1]{Eq.~(\ref{#1})}
\newcommand{\refdisp}[1]{Ref.~(\onlinecite{#1})}
\newcommand{\figdisp}[1]{Fig.~(\ref{#1})}

\begin{document}
\title{Extremely Correlated Fermi Liquids: The Formalism }
\author{ B. Sriram Shastry}
\affiliation{ Physics Department, University of California, Santa Cruz, CA 95064, USA}
\date{ February 27, 2013}
\begin{abstract}
We present the detailed formalism of the extremely correlated Fermi liquid theory, developed  for  treating  the physics of the \tJ model. We start from the exact Schwinger equation of motion for the  Greens function for projected electrons, and develop a   systematic expansion  in  a parameter $\lambda$, relating to the double occupancy. The resulting Greens function  has a canonical part  arising from an effective Hamiltonian of the auxiliary electrons, and a caparison part, playing the role of a frequency dependent  adaptive spectral weight.  This adaptive weight balances the requirement at low $\omega$, of the  invariance of the Fermi volume, and at high $\omega$ of decaying as $\frac{c_0}{i \omega}$, with a correlation depleted  $c_0 <1$. The effective Hamiltonian $H_{eff}$ describing the auxiliary Fermions is given a natural interpretation
with an effective interaction $V_{eff}$  containing both the exchange $J_{ij}$, and the hopping parameters $t_{ij}$. {\color{black} It is made Hermitian by adding suitable terms that ultimately vanish, in the {\em symmetrized theory} developed in this paper.}
Simple but important {\em shift invariances}  of the \tJ model are noted with respect to translating its  parameters  uniformly.  These play a crucial role in constraining the form of $V_{eff}$ and also provide checks for  further approximations.  The  auxiliary and physical Greens function satisfy
two sum rules, and the Lagrange multipliers for these are identified.  A complete   set of expressions for the Greens functions to second order in $\lambda$     is  given, satisfying various invariances.   A systematic iterative procedure for higher order approximations is detailed. A superconducting instability of the theory is noted at the simplest level with a high transition temperature.

\end{abstract}
\maketitle
%\tableofcontents
%%%%%%%%%%%%%%%%Removed duplicates%%%%%%%%%%%%%

\section{Introduction}
This work presents the detailed formalism of a newly developed framework for  systematic calculation of the {\color{black} dynamical}  properties of the \tJ model, starting from the basic parameters $t$ and $J$ of the model.   A subsequent paper  \refdisp{hansen-shastry}    presents   self consistent numerical results from  the  initial  application of this theory, for the case of two dimensional square lattice relevant to  cuprate superconductors. We will refer to {\em extreme correlations} as the limit $U\gg t$, so that   the single occupancy constraint is enforced.  The \tJ model \disp{ham} is the prime example of such a situation. In practice this theory applies already when $U \gtrapprox Zt$,  where $Z$ is the coordination number of the lattice.
The theory and calculations presented are in the {\em extremely correlated Fermi liquid} (ECFL) phase discussed in \refdisp{ecfl}. This phase is liquid like,  and connects  continuously to  the Fermi liquid  phase of  weak coupling models such as the Hubbard model, while accommodating the extreme local interaction  $U\to \infty$.

  The \tJ  model  described by \disp{ham}, is one of the standard models of condensed matter physics. It  has been the focus of intense effort for the last few decades, as reviewed in \refdisp{gros}. Interest in the model   grew particularly  after its identification by  Anderson in \refdisp{pwa}, as governing many of  the rich and complex set of phenomena in high $T_c$ cuprates superconductors.  The origin of the  exchange part of the \tJ model in an inverse expansion in the interaction $U$  is familiar from  superexchange theory.   The relation $J = \frac{ 4 t^2}{U}$ is  found  starting from the Hubbard model  as in \refdisp{harris}, so that large $U$ leads to a small $J$. An early  account of the model and the various sum rules can be found in the  \refdisp{harris}.
  More recently Zhang and Rice\cite{zhang-rice} gave an argument for reducing the three band copper oxygen model to an effective single band model, with a \tJ form.  Their method, apart from being more realistic, gives  independent  magnitudes  for $t$ and $J$ unconstrained by relations of the type inherent in superexchange within a single band model.

 Controlled  calculations within this model are beset by two fundamental difficulties: (a) the non canonical nature of the single occupancy (Gutzwiller\cite{gutzwiller}) projection of the  electrons that changes the canonical anticommutation relations to a more nontrivial Hubbard (Lie) algebra  and (b) the absence of any obvious small parameter for generating a systematic theory.
The present author has recently formulated a method in \refdisp{ecfl} and \refdisp{anatomy} that overcomes  these difficulties to a large extent. The basic idea is to approach the system starting  from the limit of low particle density $n= N_e/N_s$ (ratio of electron number to the number of sites), i.e. a generalized virial expansion. The density can be increased towards half filling systematically,  as described below.  Early applications to Angle Resolved Photo Emission (ARPES)  experiments  in \refdisp{gweon} are promising,
and the general  structure of the solution already leads to non trivial and experimentally  testable predictions in \refdisp{asymmetry}.
The present work  gives the details of the method introduced in \refdisp{ecfl}, and carries out a calculation to the lowest nontrivial order in a parameter $\lambda$  described below. The   main elements involved in this framework can be summarized as follows:
\begin{itemize}
\item {\bf The Schwinger method:}
 \refdisp{ecfl}  utilizes the key observation  that the Schwinger method dispenses with  Wicks theorem, and replaces that step of canonical theory by a formal matrix (operator) inversion.  The Schwinger equation for the Greens function typically involves a time derivative and a functional derivative with respect to a source potential $\V$ (defined more fully  below). It has the great advantage over standard equations of motion in that the functional derivative  generates all required  higher order Greens functions. This  is  unlike say the BBGKY hierarchy of quantum statistical mechanics,  where one needs to import higher order correlations from elsewhere. For the \tJ model, \refdisp{ecfl} obtains an exact  Schwinger equation described below in  \disp{eom_2} and \disp{xydef}. For our purpose,  that equation   may be illustrated schematically by the symbolic  equation:
\beq
 (\GLI_0 (\chem) -   \ Y_1 -   \  X  )\cdot \G =  \ (\iden -  \ \gamma), \label{symbolic-1}
\eeq 
where $\GLI_0$ (\disp{g0})  is a non interacting Greens function and $\gamma$ (\disp{gamma-1}) is essentially the spatially localized but time dependent Greens function itself $  \sim \G_{local}$. Further  $Y_1$ is a Hartree type energy and $X \sim \mbox{(something)} \times  \frac{\delta}{\delta \V}$, contains  the all important functional derivative with respect to $\V$ (both $X,Y$  are defined in \disp{xydef}). {\color{black}  The undefined ``something'' lumps together constants and  the interaction potential,   but is independent of $\G$.}  This is a convenient launching pad provided by Schwinger's method, since it is exact. However it is also intractable as it stands!  There is no  obvious small  parameter, and the presence of the time dependent $\gamma$ on the right hand side represents the removal of states (and double occupancy)  from the canonical theory and   creates a new set of problems.  We must understand and overcome these, in order to   create a practical and controlled scheme for calculations. We therefore push forward to the next set of steps.

\item {\bf Non canonical nature of the  problem and its consequences:}

The non canonical nature of the problem is reflected in the $\gamma$ term on the right of \disp{symbolic-1}, it is a time dependent Greens function obtainable from $\G$ itself (\disp{gamma}).  This $\gamma$ term contains an  essential difficulty of the problem;
it has a technical origin that we first discuss, and also an important physical aspect that we describe below.
\subitem{($\star$)}
Consider first the canonical theories, such as the Hubbard model  (see \disp{symbolic-hubbard} below),  where one only has the $\iden$ term in the right hand side of  \disp{symbolic-1}. 
In order to get rid of the functional derivative operator $X$ in favour of a  (multiplicative) self energy, one uses  $X \sim \mbox{(something)} \times  \frac{\delta}{\delta \V}$ to write:
\barray
X \cdot \G \to \Sigma \ \G & \equiv & \mbox{(something)} \times\G \  \Gamma \ \G, \;\;\; \mbox{using} \ \frac{\delta }{\delta \V} \G = \G \ \Gamma \ \G, \nn \\
 \mbox{following from} \  \Gamma & \equiv & - \frac{\delta }{\delta \V} \ \GI, 
\earray
wherein the vertex $\Gamma$ is introduced.  This gives the Schwinger Dyson  relationship between  the self energy $\Sigma$ and  vertex:
\beq
\Sigma = \mbox{(something)} \times \G \ \Gamma, \;\;\;\mbox{so that}\;\;\; (\GLI_0 (\chem) -   \ Y_1 -   \Sigma  )\  \G =  \iden. 
\eeq
This Schwinger Dyson construction  necessarily requires that the vertex $\Gamma$  reduce to unity at high frequencies, i.e. should be   ``asymptotically free''. In case of the non canonical theory \disp{symbolic-1}, a similar procedure fails. It  is easily verified  that  the required good behaviour  is lost because
 of the time dependent term $\gamma$
   on the right hand side of \disp{symbolic-1}, as shown in \refdisp{ecql}. The  so defined vertex   grows linearly with  frequency, and invalidates the Dysonian self energy scheme.

\subitem{($\star$)} The physical problem that is related to the non canonical $\gamma$ term has to do with the spectral weight of the projected electrons in  a \tJ model. Here basic sum rules give us insight into the origin, as well as a resolution of this fundamental  problem. For non canonical electrons, the high frequency behavior  of the Greens function is $\G \sim \frac{c_0}{i \omega}$ with $c_0 = 1- \frac{n}{2}$, rather than the familiar result for canonical electrons $c_0=1$. The depletion of $c_0$ from unity arises from the physics of single occupancy projection of the (non canonical) electrons $\hat{c}_{i \si}$ (denoted by the Hubbard operator $\X{i}{\si 0}$ below).  Consider  the relation  $ c_0 = \langle \hat{c}_{i \si} \hat{c}_{i \si}^\dagger+ \hat{c}_{i \si}^\dagger \hat{c}_{i \si} \rangle $ , the process $ \langle \hat{c}_{i \si} \hat{c}_{i \si}^\dagger \rangle$ suffers from the inhibiting requirement that in order to create an electron with spin $\si$, the spin state  $\sib$ at site $i$ must also  be unoccupied (so that a double occupancy is not created by this process),  resulting in $c_0 < 1$.
 On the other hand, if the numerator of $\G(i \omega)$ remains as $c_0$ {\em at all frequencies}, then the Fermi surface must enlarge in volume, and thereby violate the Luttinger Ward theorem of invariance of this volume \cite{simple-remark}.  {\em We thus arrive at an appreciation of the fundamental  tension between the  conflicting requirements;  at   high frequency  of fixing  a known coefficient  $c_0 < 1$, and at  low frequency  of a  numerator almost unshifted from unity,  for preserving the Fermi surface volume}.  A   resolution  is provided  by the possibility of an adaptive (or smart) spectral weight, i.e the  numerator of the Greens function. If a frequency dependent spectral weight can be found, so as to interpolate smoothly between the high and low frequency requirements, then both could be satisfied. 
\subsubitem{($\star$)} {\bf The product ansatz}: The above  points  suggest that the Greens function of the \tJ model  is usefully thought of as a product of two terms in frequency space i.e. $\G \sim \GH \times \mu$ (\disp{product}), where $\GH$ is a canonical Greens function and $\mu$ the caparison factor playing the role of an adaptive (or  smart) spectral weight factor.  The $\GH$ term (i.e. the denominator) is required to be  a canonical object with its poles and cuts as usual in a Fermi liquid, and defines the auxiliary Fermi liquid in this theory.
 The frequency dependent  $\mu$ term (in the  numerator) plays the role of the smart spectral weight; it  reduces to the correct coefficient $c_0$ at high frequencies while recovering weight at lower frequencies.
 Thus a convolution in time domain into two suitable time dependent pieces could resolve this conundrum, and motivates the   {\em product ansatz} in \disp{product}. This  {\em product ansatz} is  at the heart    of the procedure described here and is seen to lead to a pair of {\em exact} equations for the two parts $\GH$ and $\mu$ below in \disp{ginvdef} and  \disp{fundapair}.

\subsubitem ($\star$)The $\mu$ term   is also termed the caparison factor in \refdisp{ecfl}, keeping in mind that it provides a second layer  of dressing, over and above the dressing  provided by the usual Fermi liquid type processes in $\GH$ itself. 
\item {\bf Small parameter in theory:}

 The \tJ model is  the sum of two highly non trivial terms, the  kinetic energy projected to the space of single occupancy,  and the exchange energy. It has no obvious small parameters making it especially difficult to deal with.
 Some  inspiration is gained by examining  the form of the analogous Schwinger  equation for canonical theories, such as the Hubbard model. Again omitting details, the relevant equation can be written  symbolically as:
\beq
 (\GLI_0 (\chem) -   U \ G -   \  U \frac{\delta}{\delta \V}  )\cdot G =  \ \iden , \label{symbolic-hubbard}
\eeq 
where $U$ is the Coulomb repulsion in the Hubbard model.   Comparing with  \disp{symbolic-1} suggests a simple approach 
to introduce  a new parameter $\lambda$. In its simplest form,  we propose  to study the modified problem  symbolically expressed as:
\beq
 (\GLI_0 (\chem) -  \lambda \ Y_1 -  \lambda \  X  )\cdot \G. =  \ (\iden -  \lambda \ \gamma), \label{symbolic-2}
\eeq 
with $0 \leq \lambda \leq 1$, so that this equation \disp{symbolic-2} interpolates smoothly between the  Fermi gas and the \tJ model. This appearance of the parameter   parallels the way  the Hubbard parameter U enters \disp{symbolic-hubbard}. The  complication of the non canonical $\gamma$ term on the right is handled  analogously to  the Hartree term $Y_1$.  Unlike the repulsive Hubbard case, with an infinite interval  $[0, \infty] $ for $U$, the parameter $\lambda$ lives in a small and finite interval $[0,1]$. The expectation is  that low order perturbation expansion in $\lambda$ has a reasonable chance of capturing the physics of extreme correlations at $\lambda =1$.
We  show in the Appendix \ref{app-1} that in the atomic limit, the role of $\lambda$ can be explicitly related to that of the fraction of double occupancy (and thus also density), so that  tuning $\lambda$  smoothly adjusts this fraction between its two limits. {\color{black} Further in \disp{eqreplambda} below, a suggestive expression  for the Fermionic operators  is noted that relates  $\lambda < 1$  to a soft version of   Gutzwiller projection.}

\item { \bf Effective Hamiltonian for the auxiliary Fermions with a pseudo potential:} 

Setting aside the caparison factor $\mu$  for a moment, 
 we  examine further   the  equations of motion  (\disp{a} and \disp{eom_1})  for the auxiliary Fermion $\GH$ following from  \disp{symbolic-2} together with the product ansatz $\G = \GH \times \mu$. We would  like  to interpret these as  the actual (canonical) equations of a suitable Fermi liquid, obtainable from a  Hermitian Hamiltonian. However, we find that the equations (\disp{a} and \disp{eom_1}) as they stand,  do not  immediately cooperate with this task. {\color{black}  They  require a process of { symmetrization} described next, where one adds  extra  terms that vanish when treated exactly,  and after this  lead to a Hermitean theory for $\GH$. We term the resulting equations as  the {\em symmetrized theory}, as outlined  in this paper.

  The  theory based on  \disp{a} and \disp{eom_1} without symmetrization,  is of course also exact, and is potentially useful in its own right.  We develop such a {\em minimal theory} elsewhere, with the expectation that  this  {\em minimal theory} would   not admit a Hermitean Hamiltonian to describe the auxiliary $\GH$. Also in an approximate treatment, e.g. through an expansion in the parameter $\lambda$ to any fixed but finite  order, we would expect the symmetrized and minimal versions of the theory to be different, converging only when {\em all orders} are taken into account.

  Returning to the symmetrization procedure,}  we construct an effective Hamiltonian $H_{eff}$ for canonical electrons ($f_{i \si}, f^\dagger_{i \si}$),  with the property that the (imaginary time)  Heisenberg equation of motion for canonical electrons  $\dot{f}_{i \si} = - [f_{i \si},H_{eff}]$,  match exactly the Heisenberg equation of motion  for projected electrons  $ \dot{\hat{c}}_{i \si}= - [ \hat{c}_{i \si}, H_{t-  J}]$, except for terms that vanish on enforcing the single occupancy constraint on the auxiliary $f_{i \si}$ electrons. Thus we require
 \beq
 [f_{i \si},H_{eff}] = \left( [ \hat{c}_{i \si}, H_{t-  J}]\right)_{(\hat{c}, \hat{c}^\dagger) \to (f, f^\dagger) } + \mbox{(expressions involving $f, f^\dagger$ that vanish at single occupancy)}. \label{strategy}
 \eeq
We can then add these missing terms with $(f, f^\dagger) \to (\hat{c}, \hat{c}^\dagger)$ to the Heisenberg  equation of motion (EOM)  for $\hat{c}$ and thereby obtain an auxiliary Fermi liquid that would be also  ``natural'', i.e. have all the standard properties of a Fermi liquid\cite{agd,mahan}. One should therefore be able to use standard Feynman diagrams (\refdisp{agd}) to compute the properties of this auxiliary theory in powers of $\lambda$, if one were so inclined. 

  We find it straightforward to find such an effective Hamiltonian $H_{eff}$ (\disp{heff-1}) as described below in Sec.~(\ref{heff}).  {\color{black} {\bf The physical meaning of   $H_{eff}$} becomes clearer with the following remarks. The kinetic energy of the projected electrons could also be written differently.  An   alternate  representation, occasionally used in literature, relates:
\beq\hat{c}^\dagger_{i \si} = \X{ \si 0}{i} \to f^\dagger_{i \si} \ ( 1- n_{i \sib}  ),\;\;\hat{c}_{i \si} = \X{ 0 \si }{i} \to f_{i \si} \ ( 1-  n_{i \sib} ), \label{eqrep}
\eeq
 with $\sib = - \si$ and  $n_{i \si}= f^\dagger_{i \si} f_{i \si}$.}
  Within this  representation,   the Hilbert space continues to allow for double occupancy, i.e. is canonical,  but the various operators representing the physical processes act only upon the singly occupied subspace, {\em and produce states that are likewise singly occupied}. Thus we may write
the kinetic energy part as:
\beq
KE= - \sum_{ij} t_{ij} \ (1- n_{i \sib}) \ f^\dagger_{i \si} f_{j \si}  \ (1- n_{j \sib}).
\eeq
  Since
  the exchange energy $\sum_{ij} J_{ij} \ \vec{S}_i \cdot \vec{S}_j$ automatically 
 conserves single occupancy, we will not write it out. The kinetic energy is thus a multi Fermi operator and represents both the propagation and interaction between particles. To separate these functionalities, we introduce a parameter $\lambda$ here, it will turn out to be  the same parameter as in \disp{symbolic-2}, and write
 \barray
KE(\lambda)& = & - \sum_{ij} t_{ij} \ (1- \lambda \ n_{i \sib}) \ f^\dagger_{i \si} f_{j \si}  \ (1- \lambda \ n_{j \sib}) \nn \\
&=&- \sum_{ij} t_{ij}  \ f^\dagger_{i \si} f_{j \si}  + \lambda \  \sum_{i j} t_{ij} \ \ f^\dagger_{i \si} f_{j \si}  \left(  n_{i \sib} +  n_{j \sib}\right) + \lambda^2 \  H_d  \label{heff-0} \\
H_d&= & -  \sum_{i j} t_{ij} \ \ f^\dagger_{i \si} f_{j \si} \ \left(  n_{i \sib}  n_{j \sib}\right), \;\;\;H_d \to \mbox{dropped}.
\earray
The term $H_d$ acts on the doubly occupied subspace and is null in the singly occupied space, and hence  it may  be dropped altogether.  The remaining part of the  kinetic energy term $KE(\lambda)$ has the structure of a four Fermi interaction between the canonical Fermions, and turns out to be  a large part   of $H_{eff}$ in \disp{heff-1}.  {\color{black} The introduction of the parameter $\lambda$, can thus be viewed as replacing \disp{eqrep} by a ``softer'' representation of the Gutzwiller projection:
\beq\hat{c}^\dagger_{i \si}  \to f^\dagger_{i \si} \ ( 1- \lambda \  n_{i \sib}  ),\;\;\hat{c}_{i \si}  \to f_{i \si} \ ( 1- \lambda \ n_{i \sib} ). \label{eqreplambda}
\eeq
This  $\lambda$ representation  discourages but does not completely eliminate double occupancy. However    as $\lambda \to 1$, it  does  become the exact projected operators \disp{eqrep}, and further provides a simple interpolation between standard (canonical) Fermions and the projected electrons by varying $\lambda$ in the range  $0 \leq \lambda \leq 1$. Thus \disp{eqreplambda} suggests the interpretation of the  parameter $\lambda$, as the   controller of  the (partial) Gutzwiller  projection.}

In this representation (with $\lambda =1$),  the physical electron Greens function $\G_{ij}$  corresponds to
the correlator $- \lll  (1- n_{i \sib_i}) f_{i \si_i}, f^\dagger_{j \si_j} (1- n_{i \sib_j}) \rrr$, while $- \lll  f_{i \si_i}, f^\dagger_{j \si_j}  \rrr$ would  represent  the auxiliary Greens function $\GH[i,j]$.  The caparison factor $\mu$ seems hard to interpret in this language though.
The ECFL formalism developed here presents a procedure to splice together $\GH$ and $\mu$ precisely,  to yield the physical $\G$.  Its otherwise  formal structure becomes clearer upon making  the above  connection;  in particular  \disp{heff-0}   helps  in  developing  some  intuition for $\GH$. For instance   a physical interpretation of the auxiliary Fermions is provided by the $f_{i \si}$ themselves, and thereby requiring the same number of auxiliary Fermions as the physical ones, as done below, is perfectly natural.

 \item {\bf Invariances of the  effective Hamiltonian  $H_{eff}$ and the emergence of the second chemical potential  $u_0$:}
 
 In  $H_{eff}$ (\disp{heff-1}),  the hopping parameter $t_{ij}$ is elevated to the role of an  interaction coupling, in addition to  its role a band hopping parameter.  This feature needs attention, since we know that a constant (k independent)  shift of the band energies $\varepsilon_k \to \varepsilon_k + u_t $, or adding an onsite interaction through $J_{ij} \to J_{ij} + \delta_{ij} \ u_J$,  is inconsequential for  the \tJ model, but makes a difference in  \disp{heff-0}, {\color{black} and in various approximations for the \tJ model}.  This  ``pure''  gauge invariance  is of primary importance in this kind of a theory,  and must be addressed at the very outset to obtain a consistent and meaningful description of the \tJ model.  Such shifts could potentially  lead to a change of the interaction strengths in $H_{eff}$, unless they can be explicitly  eliminated in the theory. This issue  is  addressed   by first  listing these {\em shift symmetries} of the model in Sec.~(\ref{tjmodel}), and then requiring the approximation scheme to be shift  invariant,  at each order of $\lambda$.

Imposing the shift symmetries on $H_{eff}$  \disp{heff-1} causes it  to have a term with a Hubbard  Coulomb like interaction with strength $u_0$, such  that arbitrary  shifts of $t$ and $J$  can be absorbed into the  parameter $u_0$. {\color{black}  Analogous  to the standard  chemical potential $\chem$, this $u_0$ is a Lagrange multiplier of a term in the Hamiltonian $H_{eff}$. However it multiplies an interaction term that is  {\em quartic} in the canonical Fermions, unlike $\chem$ that multiplies the usual (quadratic) number operator.}
  The chemical potential $\chem$ and the second chemical potential   $u_0$  are jointly determined   by  {\em two  sum rules} \disp{sumrule-1} and \disp{sumrule-2},  one for the number of physical electrons and the other for the (identical) number of auxiliary canonical electrons.   
\end{itemize}

In this work, we  obtain a set of equations for the Greens function. These are  essentially  of the same form as in our recent earlier Letter  \refdisp{ecfl}, but differ in a few details due to the   usage of  the idea of the effective Hamiltonian and its shift invariances.  An iterative framework is  carefully established, and calculations of the Greens function to second order in $\lambda$ are carried out explicitly.

The outline of the paper is as follows. In Sec.~(II),  we list the {\em shift symmetries} of the \tJ model and obtain the  exact equation satisfied by the Greens function. We also determine  the form of the effective Hamiltonian $H_{eff}$ for the auxiliary Fermions,  such that the Heisenberg equations for the field operators are satisfied in  a Hermitian framework. In Sec.~(III-IV), we use the  {\em product ansatz} for the Greens function to introduce and find the exact  equations for the auxiliary Fermions and the caparison factor $\mu$. In Sec.~(V) we turn off the time dependent sources  and write the exact momentum space relations between the self energy, the caparison factor and the physical Greens functions- these are the analogs of the Schwinger-Dyson equations for this problem.
Section.~(VI) summarizes in tabular form the necessary equations needed for the next step in the iterative process that is    analogous to the skeleton graph expansion.
 Sec.~(VII)  describes   the $\lambda$ expansion of various objects and the precise nature of the iterative expansion. Several detailed calculations are needed to obtain the second order equations, and are detailed in Appendix.~(B).  Sec.~(VIII) details the Ward identities of this theory, which splits into two parts following the splitting of the  Greens functions. Sec.~(IX) gives the set of vertices defining the random phase approximation  for this theory and Sec.~(X) gives the formal results for the charge and  spin susceptibilities within RPA and its low order expansion. Sec.~(XII) concludes with some comments including a calculation of the superconducting transition temperature in this theory.

Appendix.~(A) gives a detailed calculation in the atomic limit. The simple calculation here may  be useful in providing the reader some insight into the interpretation  of the $\lambda$ expansion in terms of the number of doubly occupied sites.  Appendix.~(B) contains the detailed calculations of the various objects need to compile the second order Greens function.

\section{  The \tJ  Model and its shift invariance \label{tjmodel}}
We write the projected Fermi operators in terms of the Hubbard $X$ operators as usual
$\hat{c}_{i \si} \to \X{i}{0 \si}$, $\hat{c}^\dagger_{i \si} \to \X{i}{ \si 0 }$ and 
$\hat{c}^\dagger_{i \si'}\hat{c}_{i \si} \to \X{i}{\si' \si}$.  
We study the \tJ model given by
\barray
H & =& - \sum_{i,j,\si} t_{ij} \X{i}{\sigma 0}\X{j}{ 0 \sigma} -\chem \sum_{i,\si}  \X{i}{\sigma \sigma} + \frac{1}{2} \sum_{i,j} J_{ij} \{ \vec{S}_i . \vec{S}_j - \frac{1}{4} n_i n_j \} , \nn \\
&=&  - \sum_{i,j,\si} t_{ij} \X{i}{\sigma 0}\X{j}{ 0 \sigma} -\chem \sum_{i,\si}  \X{i}{\sigma \sigma}  
+ \frac{1}{4} \sum_{ij, \si} J_{ij}  \left( \X{i}{\si \sib} \X{j}{\sib \si} - \X{i}{\si \si} \X{j}{\sib \sib} \right) 
 \llabel{ham}
\earray
We will treat the two terms on an equal footing as far as possible, and allow terms with  $i=j$. The statement of the  model
is invariant under  a particular  ``pure gauge''  transformation that we next discuss. 
Let us first  note the  {\em    shift invariance} of the two parameters in  $H$.  Consider the   uniform (i.e. space independent) shifts of the basic  parameters:
\beq t_{ij} \to t_{ij} - u_t \ \delta_{i j}, \;\; J_{ij} \to J_{ij} + u_{J} \ \delta_{ij}, \label{shift-1} \eeq
with independent parameters $u_t, u_J$.  Under this transformation the Hamiltonian shifts as
\beq
H\to H + \left( u_{t} + \frac{1}{4} u_{J} \right) \  \hat{N} \label{shift-2}
\eeq
where $\hat{N}= \sum_{i \si} \X{i}{\si \si}$ is the number operator for the electrons. Let us note two simple theorems encoding this invariance:
\begin{itemize}
\item  {\em   Shift theorem-(I):}  A   shift of either $t$ or $J$  can be absorbed into suitable parameters,  leaving the physics unchanged. 

 \item  {\em  Shift  theorem-(II):} The two shifts  of $t$ and $J$ cancel each other when $u_J = - 4 \times u_t $. 
\end{itemize}

The first theorem is illustrated in the initial Hamiltonian \disp{ham}, where  the shift in \disp{shift-2}  can be absorbed in the chemical potential $\chem \to \chem+ u_t + \frac{1}{4} u_J$. Later  it serves to identify a second  generalized chemical potential $u_0$ encountered in the following.
 The second theorem is  subtle as  it leaves the chemical potential $\chem$ unchanged (see \refdisp{fn-shift}). It provides  a measure of the equal handed treatment of $t$ and $J$.  We will find these  almost trivial theorems  of  great use in devising and validating  various approximation schemes later.  

In further work we need to add a source term via the operator $\A$  
\beq 
 \A = \int_0^\beta \A(\tau) \ d \tau =  \sum_{j,\si_1,\si_2} \int_0^\beta \; d \tau \; \V_j^{\si_1 \si_2}( \tau ) \X{j}{\si_1 \si_2}(\tau)
 +\sum_{ij, \si_1 \si_2} \int_0^\beta \; d \tau \; \V_{i j}^{\si_1 \si_2}( \tau ) \X{i}{\si_1 0}(\tau) \X{j}{0 \si_2}(\tau)
 , \llabel{action}
\eeq
with the usual  imaginary time Heisenberg picture  $\tau $ dependence of the operators  $Q(\tau)= e^{ \tau H }Q e^{- \tau H}$, and the Bosonic sources,  $\V_j^{\si_1 \si_2}( \tau )  $ at every site and also $\V_{i j}^{\si_1 \si_2}( \tau )  $  for every pair of sites, as 
 arbitrary functions of time.   {\color{black} We will denote these sources  in a compact notation where the site index also carries the time argument as $\V^{\si_1 \si_2}_i \equiv \V^{\si_1 \si_2}_i(\tau_i)$ and $\V^{\si_1 \si_2}_{ij} \equiv \V^{\si_1 \si_2}_{ij}(\tau_i) \ \delta(\tau_i-\tau_j)$.   } For any variable we define a modified expectation 
\beq
\langle \langle Q(\tau_1,\tau_2,..) \rangle \rangle = \frac{ Tr\left[  e^{- \beta H } T_\tau  e^{-\A}\; Q(\tau_1,\tau_2,..) \right] }{Tr\left[  e^{- \beta H } T_\tau ( e^{-\A} )\right] }, \label{double-bracket}
\eeq
with a compact notation that includes  the (imaginary) time ordering symbol $T_\tau$ and the exponential factor automatically. With the abbreviation $i\equiv (R_i, \tau_i)$ for spatial $\vec{R}_i$ and imaginary time ($\tau$) coordinates,   the physical electron  is described by  a Greens function:
\beq
\G_{\si_i \si_f}[i,f]= - \langle \langle X_i^{0 \si_i} \; X_f^{\si_f 0} \rangle \rangle.
\eeq
From this,   the variation  can be found from functional differentiation as
\beq
\frac{\delta}{\delta \V_j^{\si_1 \si_2}(\tau_1) } \lll Q(\tau_2) \rrr =  \lll Q(\tau_2) \rrr \; \lll \X{j}{\si_1 \si_2}(\tau_1)\rrr - \lll  \X{j}{\si_1 \si_2}(\tau_1) Q(\tau_2)  \rrr .
\eeq
We note the fundamental anticommutator between the destruction and creation operators:
\beq\left\{ \X{i}{0 \si_1}, \X{j}{ \si_2 0}  \right\} = \delta_{ij} \left( \delta_{\si_1 \si_2} - ({\si_1 \si_2}) \  \X{i}{\sib_1 \sib_2}  \right). 
\eeq

\subsection{The Heisenberg Equation of Motion \label{heisenberg-eom}}

Let us now  study the time evolution of the destruction operator through its important commutator: 
\beq
[\X{i}{0 \si_{i}}, H] = - \sum_j t_{ij} \left[ \del{  \si_i \si_j }- { (\si_{i} \si_{j}}) \  \X{i}{\sib_{i} \sib_{j}}  \right] \X{j}{0 \si_{j}} \underbrace{ + \frac{1}{4}  J_0 \ \X{i}{0 \si_i}}
 - \chem \X{i}{0 \si_{i}} -
 \frac{1}{2} \sum_{j \neq i} J_{ij}     (\si_i \si_j) \  \X{j}{\sib_i \sib_j}  \ \X{i}{0 \si_j}. \label{eq18}
\eeq
Here $J_0 $ is the zero wave vector, (i.e. $J_{ii}$ the  onsite) exchange constant.
The  term in underbraces here and in the next equation ensures that the commutator reproduces the
term with $J_{ij} \to J_{ij} + u_J \ \delta_{ij} \ $ correctly.  We note that under the transformation \disp{shift-2}, the last term in \disp{eq18} adds nothing, in view of the ordering of the operators as written,   while the  term with underbraces provides the correct transformation factor.
Let us call this commutator as:
\barray
[\X{i}{0 \si_{i}}, H] & = & - \sum_j t_{ij} \X{j}{0 \si_i}  \underbrace{+  \frac{1}{4}  J_0 \ \X{i}{0 \si_i}}
 - \chem \X{i}{0 \si_i} + A_{i \si_i} \label{adef} \\
A_{i,\si_i}&=& \sum_{j \si_j} t_{ij} (\si_i \si_j) \  \X{i}{\sib_i \sib_j} \ \X{j}{0 \si_j} - \frac{1}{2} \sum_{j \neq i} J_{ij} \ (\si_i \si_j) \  \X{j}{\sib_i \sib_j}   \X{i}{0 \si_j} \label{a}
\earray
We next express the EOM for the Greens function in terms of $A$.
\subsection{Equation of motion for $\G$}
Let us compute the time derivative of $ \G$. For this we need the derivative
\barray
\partial_{ \tau_i} T_\tau \left( e^{- \A} \X{i}{0 \si_i}(\tau_i)\right) & = & - T_\tau \left( e^{ - \A} [  \X{i}{0 \si_i}(\tau_i), H ] \right) + T_\tau \left( e^{ - \A}[\A(\tau_i), \X{i}{0 \si_i}(\tau_i)] \right) \nn \\
~ [\A(\tau_i), \X{i}{0 \si_i}(\tau_i)]  &=&    \V_i^{\si_1 \si_2}( \tau_i ) 
   [ \X{i}{\si_1 \si_2}(\tau_i),\X{i}{0 \si_i }(\tau_i)] - \sum_{j } \V_{i j}^{\si_1 \si_2}(\tau_i)  \left\{ \X{i}{\si_1 0}(\tau_i),\X{i}{0 \si_i}(\tau_i) \right\} \ \X{j}{0 \si_2}(\tau_i)\nn \\
   &=& -\V_i^{\si_i \si_2} \ \X{i}{0 \si_2} - \sum_j \ \V_{ij}^{\si_i \si_2} \X{j}{0 \si_2} + \sum_j \ \V_{ij}^{\si_1 \si_2} \  (\si_1 \si_i) \ \X{i}{ \sib_i \sib_1}  \X{j}{0 \si_2}. \llabel{eom_anyop}
\earray
This follows from the definition of the time ordering and \disp{action} for $\A $. Using this we find: 
\barray
\partial_{\tau_i} \G_{\si_{i} \si_{f}}[i, f] & = & - \delta(\tau_i- \tau_f) \delta_{i,f} 
  \lll \left( \delta_{ \si_{i} \si_{f}}- { \si_{i} \si_{f}} \X{i}{\sib_{i} \sib_{f}} \right)   \rrr+
   \lll [  \X{i}{0 \si_{i}} (\tau_i),H] \; \X{f}{\si_{f} 0 }(\tau_f)   \rrr  \nn \\
&& -  \V_i^{\si_i \si_2}(\tau_i) \G_{\si_2 \si_{f}}[i, f] - \sum_j \V_{ij}^{\si_i \si_2} \G_{\si_2 \si_f}[j,f]-  \sum_j \V_{ij}^{\si_1 \si_2} \ (\si_1 \si_i) \ \lll  \X{i}{ \sib_i \sib_1}(\tau_i) \X{j}{0 \si_2}(\tau_j)  \X{f}{\si_f 0}(\tau_f) \rrr. \nn \label{eom_0}\\
\earray
{\color{black} To simplify notation, in such  expressions for the Greens functions (or \disp{eom_1} below), the sum over an index  implies a sum over the corresponding site and also  an integration over the corresponding time, e.g. $\sum_j \V^{\si_1 \si_2}_{ij} f(\ldots,\tau_j,\ldots) \to \sum_{R_j} \int_0^\beta d \tau_j \  \V^{\si_1 \si_2}_{ij}(\tau_j) \delta(\tau_i- \tau_j) \  f(\ldots,\tau_j,\ldots) $. A further bold letter summation convention is used after \disp{eom_11}.
However, note that  in expressions for operators such as \disp{adef} or \disp{a}, the sum  only refers to the site index summation. }
We further  use the abbreviations, 
\begin{align}
\delta[i,j] & =  \delta_{i,j} \; \delta(\tau_i - \tau_j),
& t[i,j]  \  & =   t_{ij} \; \delta(\tau_i - \tau_j), \nn \\
J[i,j] &= J_{ij} \; \delta(\tau_i - \tau_j), &\V^{\si_{a} \si_{b}}_r &=  \V^{\si_{a}\si_{b}}_r[\tau_r]. \llabel{t-def}
\end{align}

In terms of these, and using \disp{adef}  we find the equation of motion in terms of $A$:
\barray
(\partial_{\tau_i} - \chem) \G_{\si_{i} \si_{f}}[i,f]  & = & - \delta[i,f]   \lll \delta_{\si_{i} ,\si_{f}} - { \si_{i} \si_{f}} \X{i}{\sib_{i} \sib_{f}}   \rrr  \nn \\
 && +   t[i,j]  \ \G_{\si_i \si_f}[j,f]   - \frac{1}{4} J_0 \  \G_{\si_i \si_f}[i,f] + \lll A_{i \si_i}(\tau_i) \X{f}{\si_f 0}(\tau_f)  \rrr  
-   \V_{i}^{\si_{i} \si_{j}}(\tau_i)\;\; \G_{\si_{j} \si_{f}}[i,f] \nn \\
&& - \sum_j \V_{ij}^{\si_i \si_2} \G_{\si_2 \si_f}[j,f]-  \sum_j \V_{ij}^{\si_1 \si_2} \ (\si_1 \si_i) \ \lll  \X{i}{\sib_i \sib_1 }(\tau_i) \X{j}{0 \si_2}(\tau_j)  \X{f}{\si_f 0}(\tau_f) \rrr
. \llabel{eom_1}
\earray
We recall from the introduction, the discussion regarding suitably generalizing $A$  of \disp{a}, in order to make connection with a Hermitian $H_{eff}$, and therefore turn to  this task next.

\subsection{Effective Hamiltonian \label{heff}}
We now construct an effective  Hamiltonian of canonical Fermions
that will turn out to govern  the auxiliary Fermi liquid theory.
The motivation for this construction is to  cast the auxiliary Fermionic part of the ECFL theory 
into a natural and canonical framework, so that the equation for  the $\GH$, i.e.  the auxiliary piece of the full $\G$ is obtainable from a Hamiltonian that is  Hermitian and respects the usual Fermi symmetry of interactions under exchange.  

After some inspections we find that  a suitable Hamiltonian is provided  by the expression:
\barray
H_{eff}& =& - \sum_{ij} t_{ij} f^\dagger_{i \si} f_{j \si}  + \sum_i ( \frac{1}{4} J_0 - \mu)   f_{i \si}^\dagger f_{i \si} + \lambda \  V_{eff}, \nn \\
V_{eff}&=& \frac{1}{4} \sum_{ij} t_{ij} (\si_1 \si_2) \  \left[ \left( f^\dagger_{i \si_1}  f^\dagger_{i \sib_1} + f^\dagger_{j \si_1}  f^\dagger_{j \sib_1}   \right) f_{i \sib_2} f_{j \si_2} + (h.c.) \right] - \frac{1}{4} \sum_{ij} J_{ij} (\si_1 \si_2 ) \  f_{i \si_1}^\dagger f_{j \sib_1}^\dagger f_{j \sib_2} f_{i \si_2} \nn \\
&& + \frac{1}{4} \sum_{i} u_0 (\si_1 \si_2 ) \  f_{i \si_1}^\dagger f_{i \sib_1}^\dagger f_{i \sib_2} f_{i \si_2} . \label{heff-1}
\earray 
with a Hermitian  effective potential   $V_{eff}^\dagger=V_{eff}$ (Fig.~(1)) and assume no   constraint on double occupancy for these auxiliary (canonical) Fermions $f_{i \si}$.   The $t$ and $J$ parts reproduce the exact equations of motion as shown below with certain additional terms that vanish under the constraint of single occupancy. The   parameter $\lambda$  is  set to unity at the end, and provides an interpolation to the Fermi gas. 
The parameter $u_0$ represent an effective Hubbard type interaction for these Fermions, giving a contribution $u_0 \sum_i f^\dagger_{i \uparrow} f_{i \uparrow} f^\dagger_{i \downarrow} f_{i \downarrow}. $ Its magnitude is arbitrary at the moment, since it disappears under exclusion of double occupancy. Here it enables us to enforce the invariance in  {\em Shift-theorem-(I)}, where the shift of $t$ and $J$ can be absorbed in $u_0$.  It  will turn out to play the role of a second  chemical potential or Lagrange multiplier, in fixing the second sum rule \disp{sumrule-2}.  To illustrate this remark, note that adding a constant to $t$ or $J$ 
as in \disp{shift-2}, adds an onsite four Fermi interaction term. In order to satisfy the {\em Shift theorem -(I)}, we must compensate for this suitably, leading to the extra onsite  term with coefficient $u_0$, which can absorb this shift. It is also verified that the {\em Shift theorem-(II)} is satisfied without the $u_0$ term.  { We emphasize that the $u_0$ term is  both natural and essential   for the purpose of satisfying the Shift theorem (I).}
Since the  structure of the $u_0$ term  is almost  identical to that of $J_{ij}$ we will most often ``hide it'' inside $J_{ij}$, and  explicitly display it at the end. Thus unless explicitly displayed, we should read $J_{ij} \to J_{ij} - u_0 \delta_{ij}$ below.  For analogous terms  involving the $\X{i}{\si \si'}$ operators as in \disp{a},  we can include $u_0$ in $J_{ij}$  without any errors,  since the $u_0$ term always vanishes due to the properties of these operators.  
 \begin{figure}[h]
\includegraphics[width=2in]{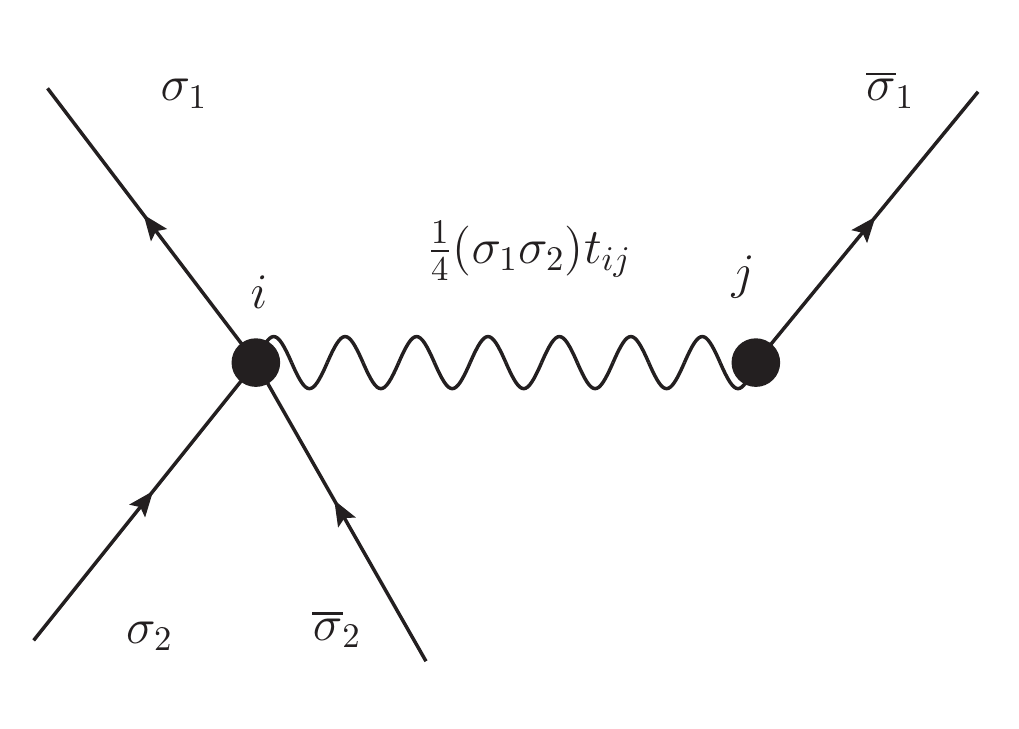}
\includegraphics[width=2in]{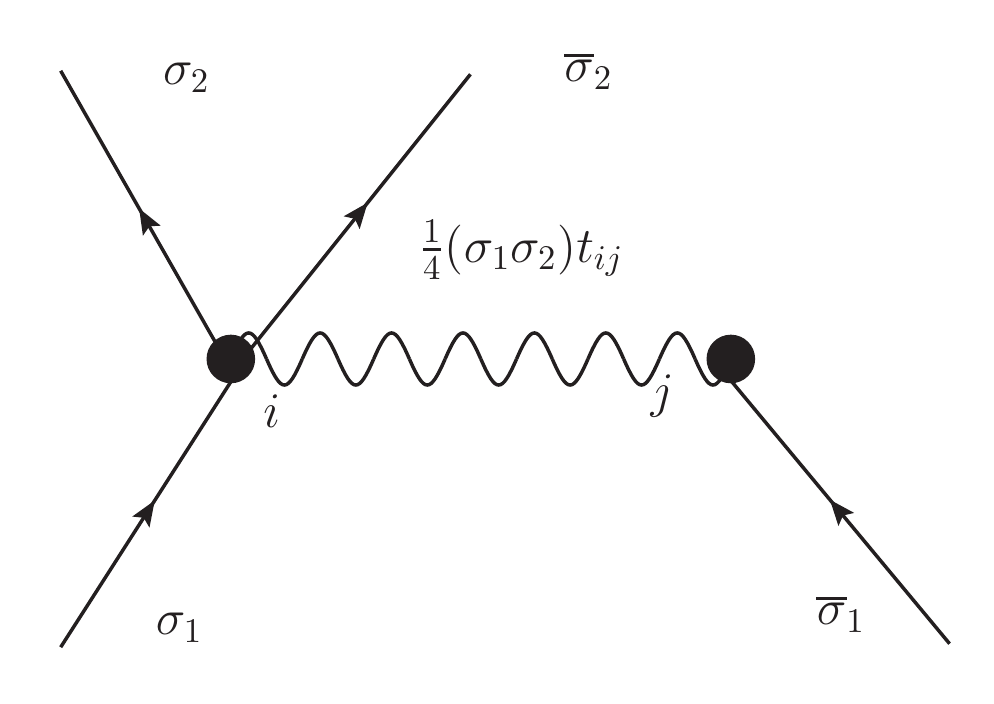}
\includegraphics[width=2in]{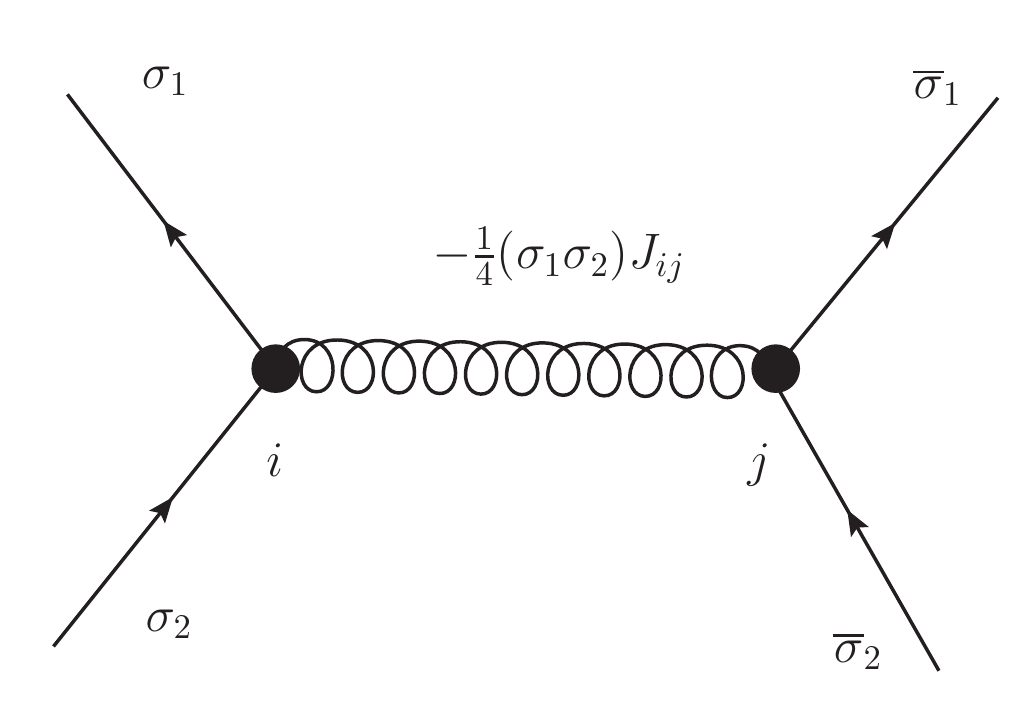}
\caption{The pseudopotential $V_{eff}$ in the real space representation, where the wavy line represents $t_{ij}$ and the coiled line represents $J_{ij}$. The first two interaction vertices have two undisplayed  symmetric partners with the exchange $i \leftrightarrow j$.  }
\label{Fig_1}
\end{figure}

Defining symmetric Cooper pair singlet operators
\barray
{\cal P}^\dagger(i,j)& =&  \sum \si f^\dagger_{i \si} f^\dagger_{j \sib} =  \left(  f^\dagger_{i \uparrow} f^\dagger_{j \downarrow}- f^\dagger_{i \downarrow} f^\dagger_{j \uparrow}  \right)  \nn \\
{\cal P}^\dagger(i,i)&=& \sum \si f^\dagger_{i \si} f^\dagger_{i \sib} = 2 f^\dagger_{i \uparrow} f^\dagger_{i \downarrow} 
\earray
with ${\cal P}^\dagger(j,i)={\cal P}^\dagger(i,j)$  we write
\beq
V_{eff}= \frac{1}{4} \sum_{ij} t_{ij} \left[ ({\cal P}^\dagger(i,i) + {\cal P}^\dagger(j,j)) \ {\cal P}(i,j) + (h.c.) \right]  - \frac{1}{4} \sum_{ij} J_{ij} \ {\cal P}^\dagger_{ij} \ {\cal P}_{ij}.
\eeq

In momentum representation the effective Hamiltonian  \disp {heff-1}
 reads
\barray
H_{eff}&=& \sum_k (\varepsilon_k + \frac{1}{4}  J_0 - \chem) f_{k \si}^\dagger f_{k \si} + \frac{\lambda}{ 4 \ N_s} \sum_{p}  \  (\si_1 \si_2)  \ W_{eff}(p_1,p_2;p_3,p_4)
\ f^\dagger_{p_1 \si_1}f^\dagger_{p_2 \sib_1} f_{p_3 \sib_2} f_{p_4 \si_2}, \nn \\
 W_{eff}(p_1,p_2; p_3,p_4)&=& - \delta_{p_1+p_2, p_3+p_4} \  \ \left\{\varepsilon_{p_1}+\varepsilon_{p_2}+\varepsilon_{p_3}+\varepsilon_{p_4}+J_{p_2-p_3} -u_0 \right\} \label{heff-2} 
\earray
(see Fig.~2), where the momentum independent term $u_0$ has been explicitly written out.
In this effective Hamiltonian, the band energies $\varepsilon_{p_j}$ of the original model are present, both in the  band energy of the $f$'s and  the interaction term. Therefore the shift \disp{shift-1} cannot be absorbed in the $\chem$ alone, and $u_0$ must also transform suitably to ensure that   the effective Hamiltonian  satisfies the  {\em Shift theorem-(I)}.
Thus in using the effective Hamiltonian
we   refine of this theorem to
\begin{itemize}
\item{\em Shift theorem-(I.1)}:  An arbitrary shift  \disp{shift-2} of $t$ and $J$, can be absorbed by  shifting the chemical potential $\chem \to \chem + u_t + \frac{1}{4} u_J$ and    $u_0$, as 
\beq
u_0 \to u_0 + 4 \ u_t + u_J \label{u-shift}.
\eeq
\end{itemize}
Note that the {\em Shift theorem-(II)} is manifestly satisfied: the combination of the band energies  $\varepsilon_{p_j}$ and the exchange term $J_p$ in \disp{heff-2} guarantees that their shift adds up to $u_J + 4 u_t \to 0$, which vanishes under the conditions of this theorem.

\begin{figure}[h]
\includegraphics[width=3in]{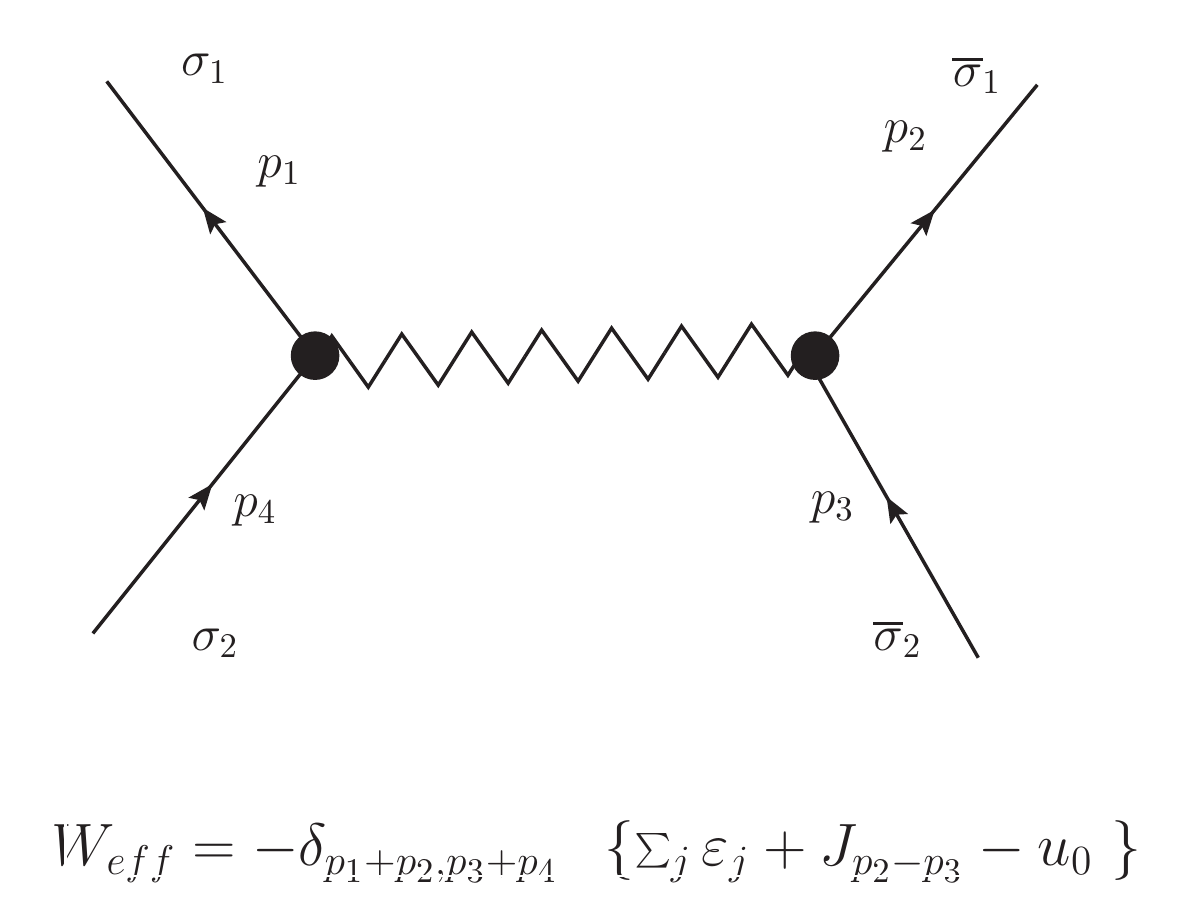}
\caption{The pseudopotential $W_{eff}$ in the momentum space representation. The zigzag line represents $W_{eff}$. Note that the momentum transfer in  the argument of $J$ is also expressible as $J_{p_1-p_4}$.}
\label{Fig_2}
\end{figure}
Since   the standard notation for interaction reads
$\sum \langle a b | V | a' b' \rangle f^\dagger_a f^\dagger_b f_{b'} f_{a'}$ for a conventional two body interaction, our notation corresponds to writing $W_{eff}(p_1,p_2;p_3,p_4) = \langle p_1 p_2 | W | p_4 p_3 \rangle$. Fermi symmetry  implies the  invariance $W_{eff}(p_1,p_2;p_3,p_4) = W_{eff}(p_2,p_1;p_4,p_3) $,  and Hermiticity implies the invariance $W_{eff}(p_1,p_2;p_3,p_4) = W_{eff}(p_3,p_4;p_1,p_2) $.
For this canonical theory,  we calculate the commutator:
\barray
[f_{i \si_i}, H_{eff}] & =& - \sum_j t_{ij} f_{j \si_i} +  ( \frac{1}{4} J_0  - \chem)  \  f_{i \si_i} + \hat{A}_{i \si_i} \nn \\
\hat{A}_{i \si_i} &=& [f_{i \si_i}, V_{eff}]. 
\earray
with 
 \beq
\hat{A}_{i \si_i} = \sum_{j \si_j} t_{ij} (\si_i \si_j) \ \left[ f_{i \sib_i}^\dagger f_{i \sib_j} f_{j \si_j}\underbrace{+  \frac{1}{2} f_{j \sib_i}^\dagger f_{j \sib_j} f_{j \si_j}+ \frac{1}{2} f_{j \sib_i}^\dagger f_{i \sib_j} f_{i \si_j}} \right] -  \frac{1}{2} \sum_{j \neq i} J_{ij} \ (\si_i \si_j) \   f_{j \sib_i}^\dagger f_{j \sib_j}    f_{i \si_j}  \label{aeff}
\eeq
Let us note that $\hat{A}_{i \si}$ \disp{aeff} differs from ${A}_{i \si}$ in  \disp{a}, through terms (in underbraces) that vanish identically  if we impose the single occupancy constraint on the auxiliary electrons.

\subsection{ Equation of Motion for $\G$ continued.}
We now return to the study of   the equation of motion for $\G$ in \disp{eom_1}, expressed  in terms of $A_{i \si}$ of \disp{a}, the commutator of the destruction operator with H.  This object  yields  the crucial Heisenberg equation of motion, therefore as discussed in \disp{strategy}, we next  look for terms that can be added to it to make it identical  to \disp{aeff}.
Comparing \disp{a} and \disp{aeff} we see that  these differ by terms (the second and third terms of the square bracket in \disp{aeff}) that are automatically  vanishing  for the $\X{i}{ab}$ operators on using their standard rules. Thus we can add such vanishing terms to \disp{a} that remain exact and also importantly preserve the Hermitian nature  of the auxiliary  Fermionic theory in approximate schemes. We thus rewrite also an exact, but  more useful result:
\barray
A_{i,\si_i}&=& \sum_{i j \si_j} t_{ij} (\si_i \si_j) \  \left[ \X{i}{\sib_i \sib_j} \ \X{j}{0 \si_j}  + \frac{1}{2}  \X{j}{\sib_i \sib_j} \X{j}{0 \si_j}+ \frac{1}{2}  \X{j}{\sib_i 0} \X{i}{0 \sib_j} \X{i}{0 \si_j} \right]
- \frac{1}{2} \sum_{j \neq i} J_{ij} \ (\si_i \si_j) \  \X{j}{\sib_i \sib_j}   \X{i}{0 \si_j}, \label{a2}
\earray
so that $A_{i,\si_i}$ and ${\hat A_{i,\si_i}}$ contain terms that are in one to one correspondence. We will use \disp{a2} in in place of \disp{a} in  \disp{eom_1} next. 

 The notation simplifies if we use the matrix notation for the spin indices introduced in \refdisp{ecql} and \refdisp{ecfl} e.g. $\G_{\si_i \si_f}[i,f] \to [\G[i,f]]_{\si_i \si_f}$, so that we may regard $\G$ as a $2 \times 2$ matrix. In short, the space-time indices are displayed but the spin indices are  hidden in the above  matrix structure. We next define $\gamma$ through:
\beq
 \gamma_{\si_a \si_b}[ i ] = \si_a \si_b \G_{\sib_b \sib_a}[i^-,i], \;\;\; \mbox{ or} \;\;\; \gamma[i] = \G^{(k)}[i^-,i], \label{gamma-1}
\eeq
where we denote  the $k$ conjugation  of any matrix $M$ by  $(M^{(k)})_{\si_1 \si_2}=M_{\sib_2 \sib_1} \si_1 \si_2 $. This conjugation corresponds to time reversal in the spin space. 
Let  $\iden$ be the identity matrix in the $2\times 2$ dimensional spin space.

We employ a  useful relation with an arbitrary operator  ${\cal Q}$ that follows from \disp{double-bracket}:
  We write
\barray
\lll \si_a \si_b \ \X{i}{\sib_a \sib_b}(\tau_i) \ {\cal Q} \rrr & =& \left( \gamma_{\si_a \si_b}[i] - D_{\si_a \si_b}[i] \right) \lll {\cal Q} \rrr \nn \\
\lll \si_a \si_b \ \X{i}{\sib_a 0}(\tau_i^+) \X{j}{0 \sib_b}(\tau_i) \ {\cal Q} \rrr &=& \left( \gamma_{\si_a \si_b}[i,j] - D_{\si_a \si_b}[i,j] \right) \lll {\cal Q} \rrr,
\earray
where  we denote {\color{black} With $\tau_j \equiv \tau_i^- $  and define }
\barray
\gamma_{\si_a \si_b}[i,j] & = & (\si_a \si_b) \ \G_{\sib_b \sib_a}[j \tau_i^-, i \tau_i] =\lll \si_a \si_b \ \X{i}{\sib_a 0} \X{j}{0 \sib_b}  \rrr,  \\
\gamma[i,i] & =& \gamma[i]
\earray
and
 \barray
 {D}_{\si_i \si_j}[i]& = &  \si_i \si_j { \frac{\delta}{\delta \V_i^{\sib_i \sib_j}(\tau_i)}} \nn \\
 {D}_{\si_i \si_j}[i,j] & = & \si_i \si_j { \frac{\delta}{\delta \V_{i,j}^{\sib_i \sib_j}(\tau_i)}}\nn \\
 \mbox{and} \;  D[i,i]& =& D[i]. \label{useful}
 \earray
In  $\gamma[i,i]$ and $  \gamma[i]$ we have  are equal time objects with creation operators to the left of destruction operators. Let us note the rewriting of the last term in \disp{eom_1}:
\barray
-  \sum_j \V_{ij}^{\si_1 \si_2} \ ( \si_i \si_1) \ \lll  \X{i}{\sib_1 \sib_i}(\tau_i) \X{j}{0 \si_2}(\tau_j)  \X{f}{\si_f 0}(\tau_f) \rrr &=& + \sum_j \V_{ij}^{\si_1 \si_2} \left(\gamma_{\si_i \si_1 }[i] - D_{ \si_i \si_1}[i] \right)  \G_{\si_2 \si_f}[j,f]. \label{four-three-operators}
\earray
With this preparation,  using \disp{a2} and  we  rewrite  \disp{eom_1} as
\barray
&& (\partial_{\tau_i} - \chem  +\frac{1}{4} J_0  ) \G_{\si_{i} \si_{f}}[i,f]   =  - \delta[i,f]  (\delta_{\si_i \si_f} -  \gamma_{\si_i \si_f}[i] ) -   \V_{i}^{\si_{i} \si_{\bb{j}}}\;\; \G_{\si_{\bb{j}} \si_{f}}[i,f] - \V_{i, \bb{j}}^{\si_{i} \si_{\bb{j}}}(\tau_i)\;\; \G_{\si_{\bb{j}} \si_{f}}[\bb{j},f]\nn \\
&& +  \V_{i\bb{j}}^{\si_1 \si_2} \left(\gamma_{\si_i \si_1 }(i) - D_{\si_i \si_1 }(i) \right)  \G_{\si_2 \si_f}(\bb{j},f)   \nn \\
&& + \ t[i,\bb{j}]    \left\{  \ \left(\iden - \gamma[i]+ D[i]- \frac{1}{2} \gamma[\bb{j}]+ \frac{1}{2} D[\bb{j}]  \right) 
\cdot \G[\bb{j},f] \ \right\}_{\si_i \si_f}  + \ t[i,\bb{j}]    \left\{ \left( - \frac{1}{2} \gamma[\bb{j}, i]+ \frac{1}{2} D[\bb{j}, i]  \right) 
\cdot \G[i,f]  \ \right\}_{\si_i \si_f}  \nn \\
&&  + \frac{1}{2}  J[i,\bb{j}] \ \  \left\{  \ \left( \gamma[\bb{j}]- D[\bb{j}]  \right) \cdot \G[i,f] \ \right\}_{\si_i \si_f},  \nn \\ \llabel{eom_11}
\earray
where the fixed variables are in normal letters and the repeated variables in bold letters are summed in space and  integrated in time. This may   be written    compactly in matrix form as
\barray
(\partial_{\tau_i} - \chem) \G[i,f]  & = & - \delta[i,f]  (\iden-  \gamma[i]) - \V_i \cdot \G[i,f] 
- \V_{i,\bb{j}} \cdot \G[\bb{j},f] + ( \gamma(i)-D[i]) \cdot \V_{i,\bb{j}} \cdot \G[\bb{j},f] \nn \\
&&   - X[i,\bb{j}] \cdot \G[\bb{j},f] -Y[i,\bb{j}] \cdot \G[\bb{j},f] \llabel{eom_2},
\earray  
where we used the definitions (with fixed $j$ and summed $\bb{k}$)
\barray
X[i,j]&=& - t[i,j]  \  (D[i]+\frac{1}{2} D[j]) + \delta[i,j] \ \frac{1}{2} \left( J[i,\bb{k}] \  D[\bb{k}] - t[i\bb{k}] \ D[\bb{k},i] \right)    \nn \\
Y[i,j]&=& - t[i,j]  \ ( \iden- \gamma[i] - \frac{1}{2} \gamma[j]  ) +\frac{1}{4} J_0 \  \iden -  \delta[i,j] \frac{1}{2} \left( J[i,\bb{k}] \    \gamma[\bb{k}] - t[i,\bb{k}] \ \gamma[\bb{k},i] \right) . \nn \\ \label{xydef}
\earray
These exact equations \disp{eom_2} and \disp{xydef} form the basis for the remaining discussion. The coefficients in $X$ and $Y$ differ slightly from the ones in \refdisp{ecfl}, in view of the usage of the effective Hamiltonian idea in this paper.
The extra  terms arise from   the form of \disp{aeff}, and actually vanish if we could treat either of these exactly. We will show that this formulation leads to approximations  obeying the {\em Shift Theorems (I-II)} discussed earlier;   note
however that \disp{eom_2} and  the forms of $X,Y$  in \disp{xydef} are manifestly invariant under these theorems.

\section{ Decomposition of  $\G$ into the auxiliary Fermion Greens function $\GH$ and the  caparison factor $\mu $ \label{decompose} }
 As discussed in the Introduction, we  next write the   {\em  product ansatz} for $\G$
 \barray
\G[a,b]&=& \GH[a,\bb{r}] \cdot \mu[\bb{r},b],  \label{product}
\earray
where $\GH$ is the canonical auxiliary Greens function and $\mu$ is the 
caparison factor,  or the  adaptive spectral weight. Since $\G$ satisfies antiperiodic boundary conditions under $\tau_a \to \tau_a+\beta$ and $\tau_b \to \tau_b+\beta$ separately, we must Fourier transform both  factors $\GH$ and $\mu$ with Fermionic frequencies $\omega_n = (2 n+1) \pi k_B T$.  
At this point  $\mu$  and $\GH$ are  undetermined. 
Let us first note in matrix notation the equal time  objects:
\barray
\gamma[i]&=& \G[i^-,i] \to (\GH[i,\bb{a}]\cdot \mu[\bb{a},i])^{(k)} = (\mu[\bb{a},i])^{(k)} \cdot (\GH[i,\bb{a}])^{(k)},\nn \\
\gamma[i,j]&=& \G[j^-,i] \to (\GH[j,\bb{a}]\cdot \mu[\bb{a},i])^{(k)} = (\mu[\bb{a},i])^{(k)} \cdot (\GH[j,\bb{a}])^{(k)}  . \label{gamma}
\earray

We define a three point  vertex functions
\barray
\Lambda^{\si_1 \si_2}_{\si_3 \si_4}(p,q;r) & \equiv&  - \frac{\delta}{\delta \V^{{\si_3 \si_4}}_r(\tau_r)} \ \{ \GHI_{\si_1 \si_2}[p,q] \ \}, \nn \\
\U^{\si_1 \si_2}_{\si_3 \si_4}[a,b;c] & \equiv &  \frac{\delta \mu_{\si_1 \si_2}[a,b]}{\delta \V_c^{\si_3 \si_4}(\tau_c)}, \label{vertexdef-3}
\earray
or as an implicit  matrix in the upper indices (but explicit in the lower ones):
\beq
 \Lambda_{\si_3 \si_4}(p,q;r) =  - \frac{\delta}{\delta \V^{{\si_3 \si_4}}_r} \ \{ \GHI[p,q] \ \}, \;\;\;\;\U_{{\si_3 \si_4}}[a,b;c] \equiv  \frac{\delta \mu[a,b]}{\delta \V_c^{{\si_3 \si_4}}}.
\eeq

{\color{black} In a similar vein, to obtain the four point vertex functions corresponding to the  source $\V_{rs}$ with a pair of points $r,s$ with $\tau_r = \tau_s$, we define:
 \beq
 \Lambda_*(p,q;r,s)= - \frac{\delta}{\delta \V^{*}_{r,s}(\tau_r) } \ \{ \GHI[p,q] \ \}, \;\;\; \U_{*}[a,b;c,d] \equiv  \frac{\delta \mu[a,b]}{\delta \V_{c,d}^{*}(\tau_c)} . \label{vertexdef-4}
 \eeq
 In some expressions involving  summations, it is  convenient to think of the vertices $\Lambda_*(p,q;r,s),    \U_{*}[p,q;r,s]$ with independent times $\tau_r,\tau_s$, with the constraint of equal times  imposed  by multiplying by  a delta function $\delta(\tau_r-\tau_s)$, as illustrated in \disp{fourpoint}.  }
 
\begin{figure}[h]
\includegraphics[width=3.in]{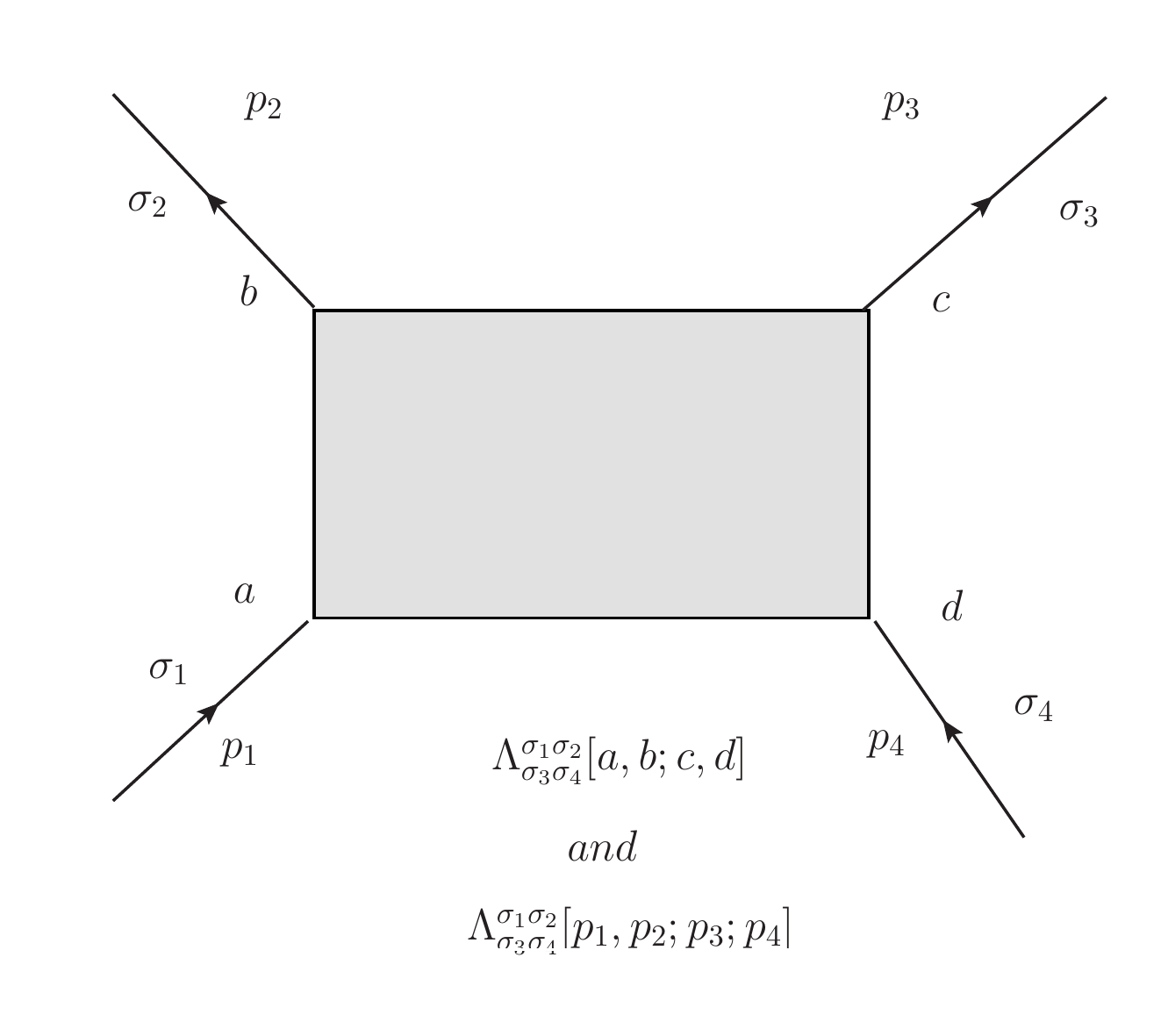}
\caption{The vertex  The four site vertex can be visualized from its definition for canonical theory:
 $- \lll f_{a \ \si_1} f^\dagger_{ b \ \si_2} f^\dagger_{c \  \si_3} f_{d \ \si_4} \rrr + \lll f_{a \ \si_1} f^\dagger_{ b \ \si_2} \rrr \lll f^\dagger_{c \  \si_3} f_{d \ \si_4} \rrr = \frac{\delta}{\delta \V_{c d}^{\si_3 \si_4}} \GH_{\si_1 \si_2} (a b) = \GH_{\si_1 \si'_1}[a, \bb{a}] \  \Lambda^{\si'_1 \si'_2}_{\si_3 \si_4}[\bb{a},\bb{b};c,d] \ \GH_{\si'_2 \si_2}[\bb{b}, b]. $
Therefore we may visualize that apart from the external legs,  $\Lambda^{\si_1 \si_2}_{\si_3 \si_4}[a,b;c,d] \sim \langle f_{a \ \si_1} f^\dagger_{ b \ \si_2} f^\dagger_{c \  \si_3} f_{d \ \si_4}\rangle$.  Note that   in this convention, the labels differ by  a cyclic permutation from those in Fig.~(\ref{Fig_2}). }
\label{Fig_3}
\end{figure}

 This set of vertices $\Lambda$ and $\U$ replace the single vertex $\Gamma$ of a canonical many body system, and we will also find equations determining these below.  Clearly in any exact treatment, the four point vertex contains the three point vertex by collapsing the points:
 \beq
 \Lambda^{\si_1 \si_2}_{\si_3 \si_4}(p,q;r)=\Lambda^{\si_1 \si_2}_{\si_3 \si_4}(p,q;r,s\to r), \label{four-three}
 \eeq
and similarly for $\U$.  However in any approximation scheme, this identity would  follow  only if the single occupancy  constraint at a given site $i$ namely: $\lll \ \X{i}{\si_1 \si_2} \ \X{i}{0 \si_3} \ldots \rrr=0$ is satisfied exactly, for all choices of the spin indices. Since typical approximations relax this constraint, if only slightly, it is therefore  useful to keep both the sets of vertices  in the theory as separate entities. Another attractive possibility  is to  require the identity \disp{four-three}, by  making a different  set of (controlled) approximations,   and is also discussed below. Fig.~(\ref{Fig_3}) illustrates the conventions used for the four point vertex, the three point vertex is obtained by the indicated contraction.

We now use a notation where  $^*$ is used as a place holder,  as  illustrated  in component form by 
\beq
\cdots \xi^*_{\si_a \si_b}\cdots  \frac{\delta}{\delta \V_\bb{j}^*}= \cdots \si_a \si_b \cdots \frac{\delta}{\delta \V_\bb{j}^{\sib_a, \sib_b }},
\eeq
with $\xi_{\si_a \si_b}= \si_a \si_b$, and an implicit spin flip in the indices of the attached derivative operator  ${\delta}/ {\delta \V_\bb{j}^{\sib_a, \sib_b }}$.

We would like to rewrite \disp{eom_2} in terms of the vertex functions. We  need to express 
\barray
X[i,\bb{j}] \cdot \G[\bb{j},f] & = & - t[i,\bb{j}]  \  (D[i]+\frac{1}{2} D[\bb{j}]) \cdot \G[\bb{j},f] + \ \frac{1}{2} \left( J[i,\bb{k}] \  D[\bb{k}] - t[i \bb{k}] \ D[\bb{k},i] \right) \cdot \G[i,f]
 \llabel{xgdef}
\earray
in terms of the vertex functions.
 
  Differentiating \disp{product} we find
\beq
\frac{\delta}{\delta \V^{{\si, \si'}}_r}  \G[a,b]  =  
\GH[a,\bb{c}] \cdot   \Lambda_{\si, \si'}(\bb{c},\bb{d};r) \ \cdot \G[\bb{d},b] +  \GH[a,\bb{c}]  \cdot \  \U_{\si, \si'}[\bb{c},{b};r ]
\eeq
Consulting \disp{useful} for the definition of $D_{\si_1 \si_2}[i]= \xi_{\si_1 \si_2} \frac{\delta}{\delta \V_i^{\sib_1 \sib_2}}$, where $\xi_{\si_1 \si_2}=\si_1 \si_2$,  we rewrite this as
\beq
D[r] \cdot \G[a,b]= \xi^* \cdot \GH[a,\bb{c}] \cdot \  \Lambda_*(\bb{c},\bb{d};r) \ \cdot \G[\bb{d},b] + \xi^* \cdot  \GH[a,\bb{c}]   \cdot \   \U_*[\bb{c},{b};r], \llabel{dgdef}
\eeq
where the spin flip in the derivatives is implied as stressed above.

Combining \disp{xgdef} and \disp{dgdef} we define the useful linear operator
\beq
\LL[i,j]  =  t[i,\bb{k}] \ \xi^* \cdot \GH[\bb{k},j]\cdot \left( \frac{\delta}{\delta \V_i^*} + \frac{1}{2}  \frac{\delta}{\delta \V_{\bb{k}}^*} \right) + \frac{1}{2} t[i,\bb{k}] \  \xi^* \cdot \GH[i,j] \cdot  \frac{\delta}{\delta \V_{\bb{k}, i }^*} - \frac{1}{2} J[i,\bb{k}] \  \xi^* \cdot \GH[i,j] \cdot  \frac{\delta}{\delta \V_{\bb{k}}^*} . \label{defL}
\eeq

Hence we may write  \disp{xgdef}  compactly as 
\barray
X[i,\bb{j}] \cdot \G[\bb{j},f] 
&\equiv& \Phi[i,\bb{b}]\cdot \G[\bb{b},f] + \Psi[i,f] \label{XG}
\earray
where the two central objects of this theory arise from the action of a common operator \disp{defL} on two seed objects $\GHI$ and $\mu$  as follows:
\barray
 &&\Phi[i,m] \equiv \LL[i,\bb{c}] \cdot \GHI[\bb{c},m] \nn \\
 &=& - t[i,\bb{j}] \ \xi^* \cdot  \GH[\bb{j},\bb{c}]
 \cdot \left( \Lambda_{*}[\bb{c}, m;i]+ \frac{1}{2}  \Lambda_{*}[\bb{c}, m;\bb{j}] \right) -  \  \frac{1}{2} t[i,\bb{k}] \  \   \xi^* \cdot  \GH[i,\bb{c}] \cdot  \Lambda_{*}[\bb{c}, m;\bb{k},i ]  +   \  \frac{1}{2} J[i,\bb{k}] \  \   \xi^* \cdot  \GH[i,\bb{c}] \cdot  \Lambda_{*}[\bb{c}, m;\bb{k}],  \nn \\ \label{phidef}
 \earray
 and
 \barray
&& \Psi[i,m] \equiv - \ \LL[i,\bb{c}] \cdot \mu[\bb{c},m]   \nn \\
 &=& - t[i,\bb{j}] \ \xi^* \cdot  \GH[\bb{j},\bb{c}]
 \cdot \left( \U_{*}[\bb{c}, m;i]+ \frac{1}{2}  \U_{*}[\bb{c}, m;\bb{j}] \right) -  \  \frac{1}{2} t[i,\bb{k}] \  \   \xi^* \cdot  \GH[i,\bb{c}] \cdot  \U_{*}[\bb{c}, m;\bb{k},i ]  +   \  \frac{1}{2} J[i,\bb{k}] \  \   \xi^* \cdot  \GH[i,\bb{c}] \cdot  \U_{*}[\bb{c}, m;\bb{k}].  \nn \\
 \label{psidef}
\earray

Writing \disp{xydef} as
\barray
Y[i,j]&=& - t[i,j]+Y_1[i,j] \nn \\
Y_1[i,j]&=&  t[i,j]  \ (   \gamma[i] + \frac{1}{2} \gamma[j] ) - \delta[i,j] \ \frac{1}{2} \left( J[i,\bb{k}] \ \gamma[\bb{k}] - t[i,\bb{k}] \gamma[\bb{k},i] \right). \label{ydef}
\earray

We also need to process the object:
\barray
( \gamma(i)-D[i]) \cdot \V_{i,\bb{j}} \cdot \G[\bb{j},f]& =&  \gamma(i)  \cdot \V_{i,\bb{j}} \cdot \G[\bb{j},f] - \xi^*\cdot \V_{i, \bb{j}}\cdot  \frac{\delta}{\delta \V_i^*} \ \G[\bb{j},f] \nn \\
& =&  \gamma(i)  \cdot \V_{i,\bb{j}} \cdot \G[\bb{j},f] - \xi^*\cdot \V_{i, \bb{j}}\cdot   \ \GH[\bb{j},\bb{c}] \cdot \Lambda_*[\bb{c}, \bb{r};i] \cdot \G[\bb{r}, f] -  \xi^*\cdot \V_{i, \bb{j}}\cdot   \ \GH[\bb{j},\bb{c}] \cdot \U_*[\bb{c}, f;i] \nn \\ \label{sourcery}
\earray

\section{Assembling the equations}
Let us rewrite the three relevant  equations symbolically:
\begin{enumerate}
\item  \disp{eom_2} for $\G$:
\barray
(\partial_{\tau_i} - \chem) \G  & = & - \delta  (\iden-  \gamma) - \V_i \cdot \G 
- \V_{i,\bb{j}} \cdot \G + ( \gamma-D_i) \cdot \V_{i,\bb{j}} \cdot \G 
   - X \cdot \G -Y \cdot \G \llabel{eom_21},
\earray  

\item
 \disp{sourcery} for the two site source $\V_{ij}$:

\beq
( \gamma-D_i) \cdot \V_{i,\bb{j}} \cdot \G =  \gamma(i)  \cdot \V_{i,\bb{j}} \cdot \G[\bb{j},f] - \xi^*\cdot \V_{i, \bb{j}}\cdot   \ \GH \cdot \Lambda_* \cdot \G -  \xi^*\cdot \V_{i, \bb{j}}\cdot    \GH \cdot \U_*
\eeq

\item
 \disp{XG}  the product rule:
\beq
X. \G = \Phi. \G + \Psi
\eeq
\end{enumerate}

Combining these  we rewrite \disp{eom_21} symbolically as
\beq
(\partial_{\tau_i} - \chem + Y+ \V_i + (\iden - \gamma) \cdot \V_{i, \bb{j}}+  \xi^*\cdot \V_{i, \bb{j}}\cdot   \ \GH \cdot \Lambda_* +  \Phi) \G   =  - \delta  (\iden-  \gamma)
 - \Psi -  \xi^*\cdot \V_{i, \bb{j}}\cdot    \GH \cdot \U_*
\eeq

Defining
\barray
\GHI_0[i,f]&=&
 \{ [( \chem - \partial_{\tau_i} - \frac{1}{4} J_0 )\iden - \V_i] \delta[i,f] + t[i,f] - \V_{i,f}(\tau_i) \ \delta(\tau_i-\tau_f)  \}, \label{g0}
 \earray 
 the exact  EOM  \disp{eom_2} can be written in matrix form:
\barray
&&\{ \GHI_0[i,\bb{j}] +      \gamma_i \cdot \V_{i, \bb{j}} - \xi^*\cdot \V_{i, \bb{a}}\cdot   \ \GH[\bb{a},\bb{b}] \cdot \Lambda_*(\bb{b},\bb{j};i)   -\ Y_1[i,\bb{j}] -  \Phi[i,\bb{j}] \} \ \cdot \GH[\bb{j},\bb{f}]\cdot \mu[\bb{f},f]   = \nn \\
&&  \delta[i,f]  \left( \iden -  \gamma[i] \right)   +  \ \Psi[i,f] + \xi^*\cdot \V_{i, \bb{a}}\cdot    \GH[\bb{a},\bb{b}] \cdot \U_*(\bb{b},f; i). \label{eom_3} 
\earray 
  At this point, a   convenient parameter $\lambda$ (finally set $\lambda \to 1$) is now inserted into this equation as follows: 
\barray
&&\{ \GHI_0[i,\bb{j}] +   \underbrace{  \lambda \  \gamma_i \cdot \V_{i, \bb{j}} - \lambda \  \xi^*\cdot \V_{i, \bb{a}}\cdot   \ \GH[\bb{a},\bb{b}] \cdot \Lambda_*(\bb{b},\bb{j};i) }   - \lambda \ \ Y_1[i,\bb{j}] - \lambda \   \Phi[i,\bb{j}] \} \ \cdot \GH[\bb{j},\bb{f}]\cdot \mu[\bb{f},f]   = \nn \\ 
&&  \delta[i,f]  \left( \iden - \lambda \  \gamma[i] \right)   + \lambda \  \ \Psi[i,f] + \underbrace{ \lambda \  \xi^*\cdot \V_{i, \bb{a}}\cdot    \GH[\bb{a},\bb{b}] \cdot \U_*(\bb{b},f; i )}. \nn \\ \label{EOM-1}
\earray 
Clearly this becomes the exact equation \disp{eom_3} at $\lambda=1$, and reduces to the Fermi gas Greens function \disp{g0} at $\lambda=0$. We may now split \disp{eom_3} exactly into a pair of   equations that are fundamental to the theory:
\barray
&&\{ \GHI_0[i,\bb{j}] +     \lambda \  \gamma_i \cdot \V_{i, \bb{j}} - \lambda \ \xi^*\cdot \V_{i, \bb{a}}\cdot   \ \GH[\bb{a},\bb{b}] \cdot \Lambda_*(\bb{b},\bb{j};i)   - \lambda \ \ Y_1[i,\bb{j}] - \lambda \   \Phi[i,\bb{j}] \} \ \cdot \GH[\bb{j},{f}]    = \delta[i,{f}] \label{ginvdef} \\ \nn \\
&& \mu[i,f]=  \delta[i,f]  \left( \iden - \lambda \  \gamma[i] \right)   + \lambda \  \ \Psi[i,f] + \lambda \  \xi^*\cdot \V_{i, \bb{a}}\cdot    \GH[\bb{a},\bb{b}] \cdot \U_*(\bb{b},f; i ). 
 \label{fundapair}
\earray 
We can usefully invert \disp{ginvdef} and write
\beq
\GHI[i,m]    = \{ \GHI_0[i,m] +     \lambda \  \gamma_i \cdot \V_{i, m} - \lambda \ \xi^*\cdot \V_{i, \bb{a}}\cdot   \ \GH[\bb{a},\bb{b}] \cdot \Lambda_*(\bb{b},m;i)   - \lambda \ \ Y_1[i,m] - \lambda \   \Phi[i,m] \}. \label{ginvdef2} 
\eeq
We see that $\GH$ satisfies a canonical equation, with a delta function of weight unity on the right, and $\mu$ soaks up the remaining factors on the right hand side of \disp{EOM-1}.
This is  decomposition is not unique, one has  the obvious freedom of respectively  post-multiplying $\GH$ and  pre-multiplying $\mu$ by a common function and its inverse. However, requiring $\GH$ to be canonical fixes  the function to be unity.
The motivation of introducing $\lambda$ in the above equations, is  to establish  adiabatic, or more properly, parametric continuity with  the Fermi gas\cite{isothermal}.
At this stage some remarks are necessary
\begin{itemize}
\item At $\lambda=1$  \disp{ginvdef} and \disp{fundapair}  becomes the exact equations for the EC phase, while it has the virtue that as   $\lambda=0$ it gives a canonical  equation for   $\GH$, with $\mu[i,j] = \iden \delta[i,j]$. Procedurally, we can  calculate objects to a given order in $\lambda$ iteratively, and set $\lambda=1$ at the end of the calculation. We thus establish and maintain  continuity with the Fermi gas in the equations of motion. 
\item
The process of introducing $\lambda$ into the EOM is  not unique. For example the terms of \disp{EOM-1}  in the underbraces cancel at $i =\bb{j}$ 
from  the  vanishing of \disp{four-three-operators}. However this cancellation is  exact
 only at $\lambda=1$, so we will find below that an expansion in $\lambda$ has the annoying feature of a slight violation of the contraction of indices result \disp{four-three}. We will show below that this is inconsequential to the orders in $\lambda$ considered here.  With  hindsight, a better strategy would  be to  impose the constraint \disp{four-three} to the order of the calculation. This can be achieved if we 
 multiply the terms in  underbraces by a sufficiently  high power of $\lambda^r$,  say  with $r \geq r_0$ , and thereby avoid dealing with this problem at low orders $r < r_0$.  Below we will analyze the minimal choice $r=1$,   record the  issues that crop up and make suitable approximations later.  The impatient may simply ignore the terms with underbraces.
 
 \item Another type of freedom is available at this stage:   if necessary, we could add an arbitrary term that  varies smoothly with $\lambda$ and vanishes at both end points e.g.  $\propto \lambda (1- \lambda) , $ to  either  side of \disp{ginvdef} and \disp{fundapair}. It will turn out that the first order term $[ \ \GHI \ ]_1$ calculated below, does need a simple term of this type to fulfill the Fermi surface sum rule.  In general, however,   the natural and minimal choice made in \disp{EOM-1}, without such a term, seems adequate for higher terms.
 \item We note that {\em the Shift theorems (I-II)} are preserved by $X,Y$ above in \disp{xydef},  and  this invariance survives  the introduction of $\lambda$ in \disp{EOM-1}.   As a result the various objects $\Phi, \Psi, \GHI, \mu$ satisfy these theorems individually.  This property  leads  to a powerful consistency check on the approximations to each order in $\lambda$.
 
 \item  Note that a $\lambda$  expansion of $\gamma[i]$ implies that the high frequency fall off of the $\G\sim \frac{c_0}{i \omega}$,  now occurs with a coefficient $c_0= 1- \lambda \ \gamma  $ that is different from $1- \frac{n}{2}$ at finite orders of $\lambda$. While it is tempting to freeze this coefficient at the exact value, it would be  inconsistent since we  take its derivatives to find $\Psi$ etc. The departure of this coefficient from the exact value  becomes increasingly significant near $n\sim 1$, and provides a criterion for the validity of a given order of approximation.

\end{itemize}

\section{Explicit equations and the Zero source limit in  Fourier space}
When we turn off the sources, the various matrix function $\G, \GH, \mu$ become spin diagonal. We will also take Fourier transforms (only) in this limit, since translation invariance in space and time is regained when the sources vanish. 

We next express $\Phi$ and $\Psi$ explicitly in terms of the vertex functions.
We need to take the Fourier transform of \disp{phidef} and \disp{psidef}.
In the ECFL  theory, a rotationally invariant liquid    phase is  obtained by turning off the sources.  We can use the standard spin  rotational symmetry analysis illustrated here with $\Lambda$ as in \refdisp{ecql}.  We define the three non vanishing matrix elements as 
$\Lambda^{(1)}= \Lambda^{\si \si}_{\si \si}$,  $\Lambda^{(2)}= \Lambda^{\si \si}_{\sib \sib}$ and
$\Lambda^{(3)}= \Lambda^{\si \sib}_{\si \sib}$. We also   record  the  Nozi\`eres identity  for the two expressions of a particle hole singlet: $\Lambda^{(1)}- \Lambda^{(2)}= \Lambda^{(3)}$, which provides an important check on the theory. We further   use a notation for the frequently occurring  antisymmetric combination $\Lambda^{(a)}=\Lambda^{(2)}-\Lambda^{(3)} $.  Armed with these, 
we next drop the matrix structure by utilizing an identity arising with a fixed $\si$ (such as in the expression for $\Phi_{\si \si}$ above):  
\barray
\langle \si| \ \xi^* \cdot \GH \cdot \Lambda_* \ | \si \rangle &=& \sum_{\si_a \si_b } \si \si_a \ \GH_{\si_a \si_b} \ \Lambda^{\si_b \si}_{\sib \sib_a} \nn \\
&=& \sum_{\si_b} (  \GH_{\si \si_b} \ \Lambda^{\si_b \si}_{\sib \sib} -  \GH_{\sib \si_b} \ \Lambda^{\si_b \si}_{\sib \si} )\nn \\
&=&  (  \GH_{\si \si} \ \Lambda^{\si \si}_{\sib \sib} -  \GH_{\sib \sib} \ \Lambda^{\sib \si}_{\sib \si} )\nn \\
&=& \GH \ (\Lambda^{(2)}-\Lambda^{(3)}) \equiv \  \GH \ \Lambda^{(a)}.
\earray
Note that we dropped the spin index on $\GH$ due to the isotropy of the state.

We use the FT convention for the two, three and four site objects illustrated with the examples:
\barray
\G[a,b]&=& \sum_k e^{i k (a-b)} \G(k) \nn \\
\Lambda^{\si_1 \si_2}_{\si_3 \si_4}[a,b;c]&=&\sum_{p_1,p_2} e^{i p_1(a-c)+ i p_2 (c-b)} \Lambda^{\si_1 \si_2}_{\si_3 \si_4}(p_1,p_2) \nn \\
\Lambda^{\si_1 \si_2}_{\si_3 \si_4}[a,b;c,d]&=&\sum_{p_1+p_4=p_2+p_3} e^{i (p_1a-  p_2 b -   p_3 c +  p_4 d)} \Lambda^{\si_1 \si_2}_{\si_3 \si_4}(p_1,p_2;p_3,p_4).
 \earray
The identity \disp{four-three} in momentum space  implies: 
\barray
\Lambda(p_1,p_2) & =& \sum_{p_3, p_4} \Lambda(p_1,p_2;p_3,p_4),\nn \\
\U(p_1,p_2) & =& \sum_{p_3, p_4} \U(p_1,p_2;p_3,p_4). \label{four-three-momentum}
\earray
 
 At zero source  we get the exact relations between self energies and vertices by Fourier transforming  \disp{phidef} and \disp{psidef}  
 \barray
 \Phi(k)&=& \sum_p  \left( \varepsilon_p  + \frac{1}{2} \varepsilon_k  + \frac{1}{2} J_{k-p}\right) \ \GH[p] \ \Lambda^{(a)}(p,k)      + \sum_{p q}  \frac{1}{2}\varepsilon_{q+p-k} \ \GH[p] \ \ \Lambda^{(a)}(p,k;q+p-k,q)  \nn \\
 &=&  \sum_{p q}   \left( \varepsilon_p  + \frac{1}{2} \varepsilon_k  + \frac{1}{2}\varepsilon_{q+p-k} + \frac{1}{2} J_{k-p}\right)  \ \GH[p] \ \ \Lambda^{(a)}(p,k;q+p-k,q)   \nn \\
  \Psi(k)&=& \sum_p  \left( \varepsilon_p  + \frac{1}{2} \varepsilon_k  + \frac{1}{2} J_{k-p}\right) \ \GH[p] \ \U^{(a)}(p,k)      + \sum_{p q}  \frac{1}{2}\varepsilon_{q+p-k}  \  \ \GH[p] \ \U^{(a)}(p,k;q+p-k,q), \nn \\
 &=&  \sum_{p q}   \left( \varepsilon_p  + \frac{1}{2} \varepsilon_k  + \frac{1}{2}\varepsilon_{q+p-k} + \frac{1}{2} J_{k-p}\right)  \ \GH[p] \ \ \U^{(a)}(p,k;q+p-k,q)   \label{phi-psi-kspace}
 \earray
 A convergence factor $e^{ i \omega_p 0^+}$ arises from the time ordering and is implied wherever necessary and the last line in both equations is valid provided the identity \disp{four-three-momentum}
 is satisfied.  Here  $\Lambda^{(a)}= \Lambda^{(2)} -\Lambda^{(3)}  $ and $\U^{(a)}= \U^{(2)} -\U^{(3)}    $. 

 With $k=(\vec{k} , \ i \omega_k)$ and $\omega_n =  \pi ( 2 n + 1) k_B T$, the Greens functions at a fixed $\lambda$ read:
\barray
\G(k)&=& \GH[k] \times \mu(k) \nn \\
\GHI(k)&=& i \omega_n + \chem - \varepsilon_k  - \frac{1}{4} J_0  - \lambda \ Y_1(k) - \lambda \Phi(k) \nn \\
\mu(k)&=& 1 - \lambda \ \gamma + \lambda \ \Psi(k).   
\earray
 The sum rule for the  number  of physical  particles and the auxiliary Fermions is given by
\barray
\sum_p \mu[p] \  \GH[p] & = & \frac{n}{2} \label{sum-rule-1} \\
\sum_p   \GH[p] & = & \frac{n}{2} \label{sum-rule-2}
\earray
While the sum rule \disp{sum-rule-1} clearly counts the number of physical electrons, the origin the  sum rule \disp{sum-rule-2}  for $\GH$ requires some discussion taken from  \refdisp{ecfl}.  We recall that it is meant to enforce the Luttinger Ward theorem of a conserved Fermi volume for the auxiliary Fermions. By so doing  and through the composition $\G= \GH \times \mu$,  it also preserves it  for the physical Fermions.   While $\chem$ provides us with one obvious Lagrange multiplier to enforce { one of the sum rules}, the  more subtle parameter $u_0$, introduced in \disp{heff-1},  is required  to enforce the second sum rule \disp{sumrule-2}. 
Explicit expressions for $\gamma, Y_1, \Phi, \Psi$  can be calculated  order by  in $\lambda$ as demonstrated below. 

\section{Summarizing}
Before proceeding to  the iterative scheme, we collect all the relevant equations for convenience in Table ~ I.
%\vspace{.15in}
\begin{table}[h]
\begin{center}
\begin{tabular}{||p{.65in}|p{5in}|p{.5in}||} \hline
{\bf  Object} & \hspace{2in} {\bf Defining Equation} & {\bf Eq. No.} \\ \hline \hline
$\GHI[i,m]$ & $ \{ \GHI_0[i,m] +     \lambda \  \gamma_i \cdot \V_{i, m} - \lambda \ \xi^*\cdot \V_{i, \bb{a}}\cdot   \ \GH[\bb{a},\bb{b}] \cdot \Lambda_*(\bb{b},m;i)   - \lambda \ \ Y_1[i,m] - \lambda \   \Phi[i,m] \} $ & \disp{ginvdef2}
 \\ \hline
 $\mu[i,m] $ & $ \delta[i,m]  \left( \iden - \lambda \  \gamma[i] \right)   + \lambda \  \ \Psi[i,m] + \lambda \  \xi^*\cdot \V_{i, \bb{a}}\cdot    \GH[\bb{a},\bb{b}] \cdot \U_*(\bb{b},m; i )$& \disp{fundapair}
  \\ \hline
  $Y_1[i,m]$&$ t[i,m]  \ (   \gamma[i] + \frac{1}{2} \gamma[m] ) - \delta[i,m] \ \frac{1}{2} \left( J[i,\bb{k}] \ \gamma[\bb{k}] - t[i,\bb{k}] \gamma[\bb{k},i] \right)$ &\disp{ydef} \\ \hline
  $\gamma[i]$ & $ \mu^{(k)}[\bb{a}, i] \cdot  \GH^{(k)}[i, \bb{a}]$& \disp{gamma} \\ \hline
  $\gamma[i,m]$ & $\mu^{(k)}[\bb{a}, i] \cdot  \GH^{(k)}[m, \bb{a}] $  & \disp{gamma}  \\ \hline
  $ \Phi[i,m]$&  $ - t[i,\bb{j}] \ \xi^* \cdot  \GH[\bb{j},\bb{c}]
 \cdot \left( \Lambda_{*}[\bb{c}, m;i]+ \frac{1}{2}  \Lambda_{*}[\bb{c}, m;\bb{j}] \right) -  \  \frac{1}{2} t[i,\bb{k}] \  \   \xi^* \cdot  \GH[i,\bb{c}] \cdot  \Lambda_{*}[\bb{c}, m;\bb{k},i ]  +   \  \frac{1}{2} J[i,\bb{k}] \  \   \xi^* \cdot  \GH[i,\bb{c}] \cdot  \Lambda_{*}[\bb{c}, m;\bb{k}] $ & \disp{phidef} \\ \hline
  $ \Psi[i,m]$&  $ - t[i,\bb{j}] \ \xi^* \cdot  \GH[\bb{j},\bb{c}]
 \cdot \left( \U_{*}[\bb{c}, m;i]+ \frac{1}{2}  \U_{*}[\bb{c}, m;\bb{j}] \right) -  \  \frac{1}{2} t[i,\bb{k}] \  \   \xi^* \cdot  \GH[i,\bb{c}] \cdot  \U_{*}[\bb{c}, m;\bb{k},i ]  +   \  \frac{1}{2} J[i,\bb{k}] \  \   \xi^* \cdot  \GH[i,\bb{c}] \cdot  \U_{*}[\bb{c}, m;\bb{k}] $ & \disp{psidef} \\ \hline \hline
\end{tabular}
\caption{\label{tab:1} {\bf  Summary of defining equations: }   The  computation of  the Greens function $\G = \GH . \mu$ in \disp{product}  requires  several intermediate variables. The complete set of   variables in this theory (first column),  and their mutual  and  $\lambda$ dependence  (second column)   are collected here for convenience. The corresponding equation number in the paper is given in the last column.}
\end{center}
\vspace{-0.6cm}
\end{table}
%%%%%%%%%%%%%%%%
The  various vertex functions are found from relationships summarized in Table~ II.
\begin{table}[h]
\begin{center}
\begin{tabular}{||p{1in}|p{1.5in}|p{.5in}||} \hline
{\bf  Vertex} & \hspace{.1in} {\bf Defining Equation} & {\bf Eq. No.} \\ \hline \hline
$\Lambda^{\si_a \si_b}_{\si_c \si_d}[i,m; j]$& $ -(\frac{\delta}{\delta \V_{j}^{\si_c \si_d}}) \ \GHI_{\si_a \si_b}[i,m]$& \disp{vertexdef-3}  \\ \hline
 $\Lambda^{\si_a \si_b}_{\si_c \si_d}[i,m; j,k]$& $ -(\frac{\delta}{\delta \V_{j,k}^{\si_c \si_d}}) \ \GHI_{\si_a \si_b}[i,m]$& \disp{vertexdef-4} \\ \hline
 $\U^{\si_a \si_b}_{\si_c \si_d}[i,m; j]$& $ (\frac{\delta}{\delta \V_{j}^{\si_c \si_d}}) \ \mu_{\si_a \si_b}[i,m]$& \disp{vertexdef-3} \\ \hline
 $\U^{\si_a \si_b}_{\si_c \si_d}[i,m; j,k]$& $ (\frac{\delta}{\delta \V_{j,k}^{\si_c \si_d}}) \ \mu_{\si_a \si_b}[i,m]$& \disp{vertexdef-4} \\  \hline
 \hline 
\end{tabular}
\caption{\label{tab:2} {\bf  Vertex Functions: }   The   theory  requires   three point and four point vertices. Their nomenclature (first column) and definition (second column) are given, along with  the corresponding equation number in the paper.}
\end{center}
\vspace{-0.6cm}
\end{table}

 It is worthwhile providing one non trivial example  of the matrix notation.
In component form  note that $\Phi[i,m]$ can be written out as:
\barray
&&\Phi_{\si_i \si_m}[i,m] = - t[i,\bb{j}] \ \si_i \si_1 \  \GH_{\si_1 \si_2}[\bb{j},\bb{c}]
  \left( \Lambda_{\sib_i \sib_1}^{\si_2 \si_m}[\bb{c}, m;i]+ \frac{1}{2}  \Lambda_{\sib_i \sib_1}^{\si_2 \si_m}[\bb{c}, m;\bb{j}] \right)  \nn \\
&& -  \  \frac{1}{2} t[i,\bb{k}] \  \   \ \si_i \si_1 \    \GH_{\si_1 \si_2}[i,\bb{c}] \  \Lambda_{\sib_i \sib_1}^{\si_2 \si_m}[\bb{c}, m;\bb{k},i ]  +   \  \frac{1}{2} J[i,\bb{k}] \  \   \ \si_i \si_1 \    \GH_{\si_1 \si_2}[i,\bb{c}] \  \Lambda_{\sib_i \sib_1}^{\si_2 \si_m}[\bb{c}, m;\bb{k}], \nn
\earray
%\newpage

\section{   $\lambda$ expansion and the iterative scheme}

Taking functional derivatives w.r.t. $\V$, we generate a self energy - vertex hierarchy of Fermionic theory, paralleling the standard (i.e. canonical) theory, but with greater complexity due to the two kinds of vertex functions and self energies.  We describe the  $\lambda$ expansion and the  iterative process next.  The iterations are analogous to the skeleton diagram expansion in  standard many body theory, where $\lambda$ plays the role of the interaction constant. Various objects are expanded in terms of $\lambda$ and $\GH$, while $\GH$ itself is left intact. Potentially confusing is the treatment of $\GHI$, which {\em is expanded in $\lambda$ and $\GH$}, ignoring its obvious  relationship as the inverse of $\GH$. This becomes understandable when we recall that $\GHI$ is, apart from $\GHI_0$,  the Dyson self energy of the auxiliary system, and is to be  regarded as a functional of $\GH$, as in the Luttinger Ward functional \refdisp{luttinger-ward}. One  example of this expansion may be useful. Consider $\gamma[i,m]$, we will  expand it as:
\barray
\gamma^{(k)}[i,m] = \GH[m,\bb{a}] \cdot \mu[\bb{a},i] = \GH[m,\bb{a}] \cdot \left(  \left[\mu[\bb{a},i]\right] _0 + \ \lambda \  \left( \mu[\bb{a},i]\right] _1 + \lambda^2 \ \left[ \mu[\bb{a},i]\right] _2 + O(\lambda^3) \right) 
\earray
keeping $\GH$ intact, i.e. unexpanded in $\lambda$. A similar expansion is carried out also for $\gamma[i]$, leading to a correction of the high frequency fall of coefficient $c_0$ as noted above.

{\bf Iterative process:}
We now describe the various steps of the iteration process. First note that
 all variables (except $\GH$) are  expanded as
\beq
{\cal A}= [{\cal A}]_0 + \lambda \ [{\cal A}]_1 + \lambda^2 \ [{\cal A}]_2 + \cdots+ \lambda^p \ [{\cal A}]_p + \cdots 
\eeq
The iteration scheme can be summarized in the two following tables. 
Table (III) lists the seed objects needed at any order and gives the derived objects. 
\vspace{.5in}
\begin{table}
\begin{center}
\begin{tabular}{| c |    c |}\hline
{\bf Seed object} & {\bf Derived objects}  \\ \hline \hline
$[ \ \mu[i,m] \ ]_p$ & $ \left[ \gamma[i], \ \gamma[i,m], \ Y_1[i,m],  \ \U[a,b;c], \ \ \U[a,b;c,d] \right]_p $
 \\ \hline
$\left[ \ \U[a,b;c], \ \U[a,b;c,d]  \ \right]_p $ & $\left[ \Psi[i,m] \right]_p$
\\ \hline \hline
$\left[ \GHI[i,m] \right]_p $ & $  \left[ \ \Lambda[a,b;c], \ \Lambda[a,b;c,d] \ \right]_p  $
\\ \hline
$\left[ \ \Lambda[a,b;c], \ \Lambda[a,b;c,d] \ \right]_p $ & $\left[ \Phi[i,m] \right]_p$
\\ \hline
\end{tabular}
\end{center}
\caption{ {\bf Iteration level $p$  calculations} The auxiliary inverse  Greens function $\GHI$ and  the adaptive spectral weight $\mu$ play the  role of  seed objects at the p$^{th}$ order.
By computing  them   to p$^{th}$ order in the parameter $\lambda$, we  obtain 
and the vertex functions and the  other variables listed in the second column to the same order as described in \disp{eq78} - \disp{eq811}. }
\end{table}
\vspace{.15in}
Table~(IV) lists  the higher order objects   and the needed lower level objects for stepping up.
\vspace{.15in}
\begin{table}
\begin{center}
\begin{tabular}{| c |    c |} \hline
{\bf Level $(p+1)$ object} & {\bf Required level $p$ objects } \\ \hline \hline
$[ \ \mu[i,m] \ ]_{p+1}$ & $ \left[ \gamma[i], \  \ \Psi [i,m],  \ \U[a,b;c], \ \ \U[a,b;c,d]  \right]_p $
 \\ \hline \hline
$\left[ \ \GHI[i,m] \right]_{p+1} $ & $\left[ Y_1[i,m], \ \Phi[i,m], \ \Lambda[a,b;c] \  \right]_p$
\\ \hline
\end{tabular}
\caption{{\bf Iteration  level  step-up  calculations:} In proceeding upwards in the iterative process in \disp{eq82}   the computed (p+1)$^{th}$ order objects are listed in the first column, and the (p)$^{th}$ order objects needed  are in the second column. Since $\GHI$ and $\mu$ at a given level suffice to  determine all other  objects  at that level through Table~(III), the iterative nature of the scheme becomes transparent. }
\end{center}
\end{table}

\begin{itemize}
\item{\bf I. Initialization  at p=0:}
The iterations require the following starting relations.
\barray
\GHI_0[i,m]&=&
 \{ [( \chem - \partial_{\tau_i} - \frac{1}{4} J_0 )\iden - \V_i] \delta[i,m] + t[i,m] - \V_{i,m} \}    \nn \\
 \left[ \mu[i,f]  \right]_0 &=& \iden \  \delta[i,f] 
\earray

\item{\bf  II.  Computation of  derived objects  at level p  from Table (I) :}

{\em The set of equations requiring $\left[ \mu[i,m] \right]_p$} 
\barray
\left[ \gamma[i] \right]_p &=&  \left[ \mu^{(k)}[\bb{a}, i] \ \right]_p \ \cdot  \GH^{(k)}[i, \bb{a}]   \nn \\
\left[ \gamma[i,m]\right]_p & =&  \left[ \mu^{(k)}[\bb{a}, i] \ \right]_p \ \cdot  \GH^{(k)}[m, \bb{a}] \nn \\
\left[ \ \U^{\si_a \si_b}_{\si_c \si_d}[i,m; j] \ \right]_p &=& (\frac{\delta}{\delta \V_{j}^{\si_c \si_d}}) \ \left[ \mu_{\si_a \si_b}[ i,m]  \right]_p \nn \\
\left[ \ \U^{\si_a \si_b}_{\si_c \si_d}[i,m; j,k] \ \right]_p &=& (\frac{\delta}{\delta \V_{j,k}^{\si_c \si_d}}) \ \left[ \mu_{\si_a \si_b}[ i,m]  \right]_p  \label{eq78}
\earray
\barray
\left[ \ Y_1[i,m] \ \right]_p &=&  t[i,m]  \ \left[    \gamma[i] + \frac{1}{2} \gamma[m] \right]_p  - \delta[i,m] \ \frac{1}{2} \left[ J[i,\bb{k}] \ \gamma[\bb{k}] - t[i,\bb{k}] \gamma[\bb{k},i] \right]_p \label{eq791}
\earray

\barray
&&\left[ \ \Psi[i,m] \ \right]_p  
 = - t[i,\bb{j}] \ \xi^* \cdot  \GH[\bb{j},\bb{c}]
 \cdot \left( \U_{*}[\bb{c}, m;i]+ \frac{1}{2}  \U_{*}[\bb{c}, m;\bb{j}] \right)_p   \nn \\
&& -  \  \frac{1}{2} t[i,\bb{k}] \  \   \xi^* \cdot  \GH[i,\bb{c}] \cdot \ \left(  \U_{*}[\bb{c}, m;\bb{k},i ] \ \right)_p  +   \  \frac{1}{2} J[i,\bb{k}] \  \   \xi^* \cdot  \GH[i,\bb{c}] \cdot  \ \left( \U_{*}[\bb{c}, m;\bb{k}] \ \right)_p, \label{eq80}
\earray

{\em The set of equations requiring $\left[ \GHI[i,m] \right]_p$} 
\barray
  && \left[ \  \Lambda^{\si_a \si_b}_{\si_c \si_d}[i,m; j] \ \right]_p = -(\frac{\delta}{\delta \V_{j}^{\si_c \si_d}}) \ \left[  \GHI_{\si_a \si_b}[i,m] \ \right]_p \nn \\
&&\left[ \ \Lambda^{\si_a \si_b}_{\si_c \si_d}[i,m; j,k] \ \right]_p = -(\frac{\delta}{\delta \V_{j,k}^{\si_c \si_d}}) \ \left[ \GHI_{\si_a \si_b}[i,m]  \right]_p\nn \\
&&\left[ \ \Phi[i,m] \ \right]_p  
 = - t[i,\bb{j}] \ \xi^* \cdot  \GH[\bb{j},\bb{c}]
 \cdot \left( \Lambda_{*}[\bb{c}, m;i]+ \frac{1}{2}  \Lambda_{*}[\bb{c}, m;\bb{j}] \right)_p   \nn \\
&& -  \  \frac{1}{2} t[i,\bb{k}] \  \   \xi^* \cdot  \GH[i,\bb{c}] \cdot \ \left(  \Lambda_{*}[\bb{c}, m;\bb{k},i ] \ \right)_p  +   \  \frac{1}{2} J[i,\bb{k}] \  \   \xi^* \cdot  \GH[i,\bb{c}] \cdot  \ \left( \Lambda_{*}[\bb{c}, m;\bb{k}] \ \right)_p, \label{eq811}
\earray
\item{\bf III. Level $p$  to  Level $(p+1)$:  step up equations:}
\barray
  \left[ \mu[i,m] \right]_{p+1}  & = & - \delta[i,m]   \left[  \gamma[i] \right]_p     +   \  \left[ \Psi[i,m] \right]_p +   \left[\  \xi^*\cdot \V_{i, \bb{a}}\cdot    \GH[\bb{a},\bb{b}] \cdot \U_*(\bb{b},m; i)\right]_p, \nn \\
  \left[  \GHI[i,m] \right]_{p+1}    & =&     \left[ \gamma_i \cdot \V_{i, m} -  \xi^*\cdot \V_{i, \bb{a}}\cdot   \ \GH[\bb{a},\bb{b}] \cdot \Lambda_*(\bb{b},m;i) \  \right]_{p}   - \left[ \ Y_1[i,m] +    \Phi[i,m] \ \right]_p  \label{eq82}
\earray
 \item{\bf IV.  If required  level is reached exit,  else return to Step II.}
\end{itemize}
{\color{black} This iterative  procedure can thus  be applied to obtain equations for the Greens functions to any desired order. In practice the higher order terms grow very rapidly, as in the Feynman diagram series. However, as explained in the introduction, a low order expansion is expected to capture already the significant features of extreme correlations, an important  reason being that the range of is finite and small, i.e.  $\lambda \in [0,1]$. In this work we will be content to work to $O(\lambda^2)$ where all the relevant objects can be calculated explicitly.}

{\color{black} {\bf Second order Greens function:}}
Having formulated the iterative process,
we next apply this  to obtain the second order Greens functions.  The calculations are detailed in the Appendix \ref{calculations}, and we directly present the first and second order results here.  Displaying the so far hidden $u_0$ coefficient, we write the complete set of equations to $O(\lambda^2)$ from \disp{eq136} and \disp{eq176}. 

\begin{figure}[h]
\includegraphics[width=4in]{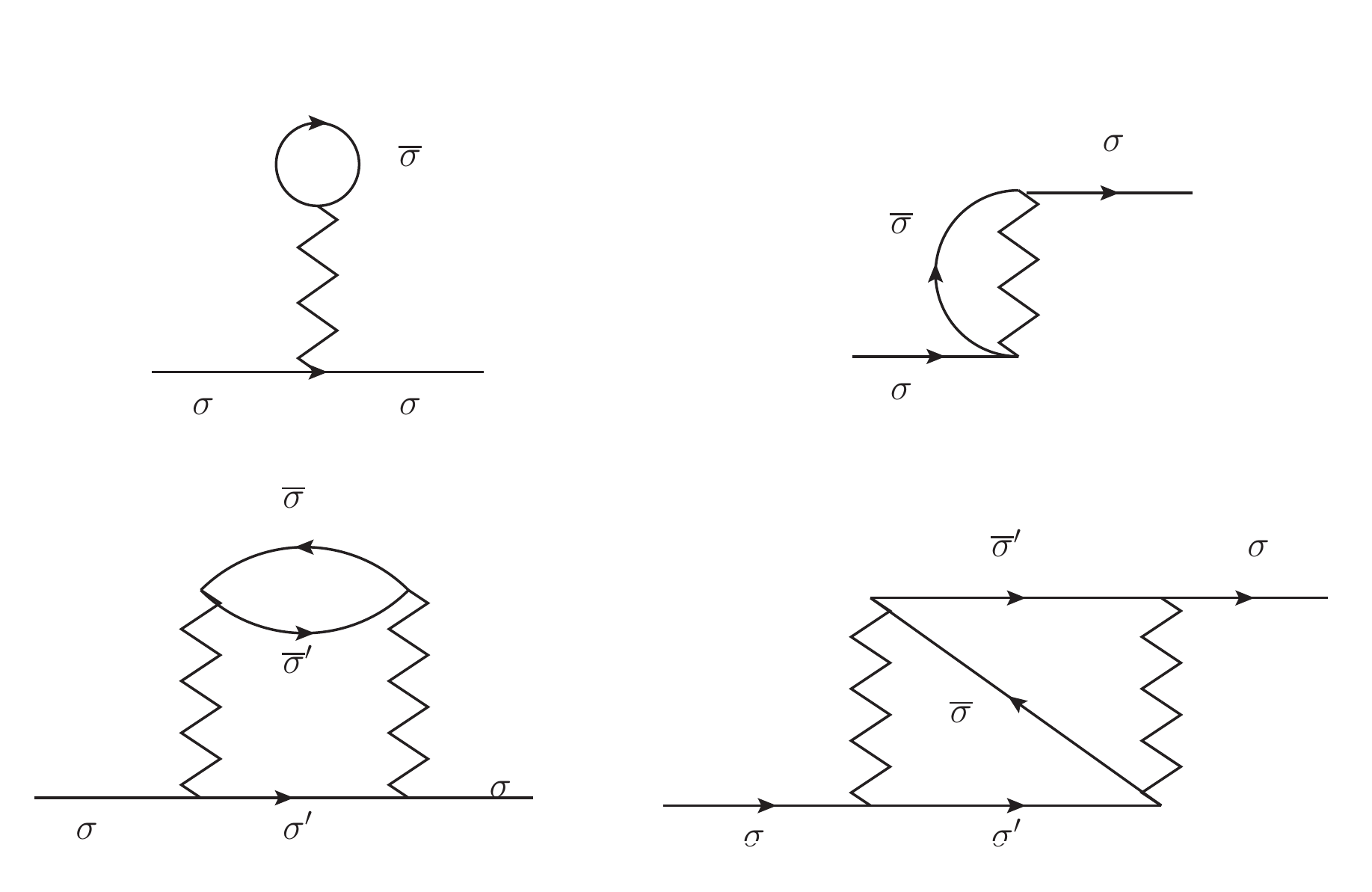}
\caption{The self energy graphs to second order from $W_{eff}$ and the effective Hamiltonian $H_{eff}$. These  determine the $\Phi$ self energy. }
\label{selfenergy}
\end{figure}

\barray
\G[k]&=& \GH[k] \times \mu[k] \nn \\
\mu[k] &=& 1- \lambda  \frac{n}{2}  + \lambda^2 \ \frac{n^2}{4} - \lambda^2   \sum_{ p,q}   \left( \varepsilon_p +\varepsilon_{k+q-p}  +  \varepsilon_k   +\varepsilon_q +  J_{k-p}-u_0\right) \ \GH[p] \ \GH[q] \ \GH[ q+ k -p ] +O(\lambda^3) \label{final-mu} \\
 \GHI(k) & = & i \omega_n + \chem' - \left( 1- \lambda \ n + \lambda^2 \ \frac{3n^2}{8}  \right)\varepsilon_k + \lambda \  \sum_q \frac{1}{2} J_{k-q} \ \GH[q] - \lambda^2  \left[  \Phi(k) \right]_1  + O(\lambda^3) \label{final-g} \\
\left[  \Phi(k) \right]_1 & = & - \sum_{q,p} \GH[q] \ \GH[p] \ \GH[k+q-p] \nn  \\
&& \times \left(  \varepsilon_{k} +\varepsilon_{p}+  \varepsilon_{q} +  \varepsilon_{k+q-p} + J_{k-p} -u_0 \ \right)  \  \{ \varepsilon_{k} +\varepsilon_{p}+ \varepsilon_{q} + \varepsilon_{k+q-p} + \frac{1}{2} \left(J_{k-p} + J_{p - q} \right) - u_0 \   \}. \label{final-phi}
  \earray 
The shifted chemical potential $\chem'$ is related to the physical (i.e. thermodynamical)  chemical potential $\chem$ and $u_0$ through 
\barray
 \chem'&=&\chem - u_0 \ \frac{\lambda n}{2}   (  1 - \frac{\lambda n}{4}) + \left[ J_0 \frac{ \lambda n}{4} ( 1- \frac{ \lambda n}{2}) + 2 \lambda ( 1 - \frac{ \lambda n}{8}) \  \sum_q \varepsilon_q \GH[q] \right]. \label{chemdef2}
\earray
In using this expansion, one must  first  set $\lambda \to 1$.
These expressions   satisfy the {\em Shift theorem (I.1)} and {\em Shift theorem (II)},   as one can verify by shifting $\varepsilon_k$ and $J_k$ by $k$ independent constants, and using $\sum_q \GH[q] = \frac{n}{2}$. The self energy from a Feynman diagram theory to second order from $H_{eff}$ in \disp{heff-1} matches the above expression for $\GHI$. The required diagrams are shown in 
  Fig.~(\ref{selfenergy}) up to second order where the zigzag line $W_{eff}$ is defined in Fig. (\ref{Fig_2}). 
Apart from a single term (the expansion of  $Y_1$ in $\lambda$), the expansion of the auxiliary Fermi liquid is  largely ``autonomous'', i.e. proceeds without requiring the  knowledge of $\mu$, and is represented in Feynman diagrammatic terms. 
The  caparison  term $\mu$ has no obvious interpretation in terms of  $H_{eff}$, but is easy to compute along  lines similar to the ones shown here, and 
 the full theory splices the two factors to yield  $\G$,  as described here. 
  
   A consistent first order, i.e. $O(\lambda)$ theory for $\GHI$ and $\mu$   can be  found  after dropping all $O(\lambda^2)$ terms. As it stands, we would get $\mu= 1- \lambda \frac{n}{2}$ to this order, and this would violate the Fermi surface volume theorem ( \refdisp{simple-remark}).  To recover from this, we   may  however  set   $\mu[k] $ to unity instead. Formally this is achieved   by adding $\lambda(1- \lambda) \frac{n}{2}$ to $\mu[k]$ as discussed below \disp{EOM-1}, since this added term vanishes at both endpoints $\lambda=0$ and $\lambda=1$. This procedure is  within the permissible adjustments of the  continuity argument, and at second order cancels out so that the quoted second order result is unchanged. Further all vertices are unchanged since this is a static term. In this way the first order theory can also be arranged  to satisfy the  Luttinger Ward  Fermi volume theorem. This theory   has a band dispersion $(1-n) \varepsilon_k$ that  that shrinks in width by a factor $(1-n)$ as in the Gutzwiller-Brinkman-Rice theory \cite{gutzwiller,brinkman}, with an enhanced effective mass $m/m* = (1-n)$. The second order result presented  here provides a more interesting  and  frequency dependent correction to the Fermi gas.

  In summary, the physical Greens function is obtained from:
  \beq
  \G[k] = \GH[k] \ \mu[k].
  \eeq
   The  number of the  physical electrons  is fixed by the first  sum rule: 
  \barray
  \frac{n}{2} &=& \sum_k \G[k] \ e^{i \omega_n 0^+},  \label{sumrule-1}
  \earray
while  the auxiliary Fermion satisfy an identical   sum rule:
  \barray
  \frac{n}{2}&=& \sum_k \GH[k] \ e^{i \omega_n 0^+}   \label{sumrule-2}
  \earray
  We can determine the two independent real parameters $\chem$ and $u_0$ in order to satisfy both these equations simultaneously, and thus the role of $u_0$ as a Lagrange multiplier, similar to that of $\chem$  is now evident.
 It is also clear that the  shifts of $t$ or $J$ can be absorbed in the two Lagrange multipliers $\chem$ and $u_0$.  It is worth noting that the Simplified ECFL model used in \refdisp{ecfl} and \refdisp{gweon}  can be obtained from \disp{final-mu} and \disp{final-phi} by throwing out 
the band energies and exchange energies in the coefficients of $\GH[q] \GH[p] \GH[k+q-p]$ while retaining $u_0$, so that the Lagrange multiplier of that approximation $\Delta_0$ is related to  $u_0$.              

The role of the two sum rules in fixing the number of Fermions and also the Luttinger Ward Fermi surface is 
 already  discussed in \refdisp{ecfl} and above. We can add to that discussion with the help of the explicit functional forms found  above.  It should  be noted from \disp{final-mu} and \disp{final-phi}  that the functional derivatives
 \barray
 I[k,p] \equiv \frac{\delta \left[ \Phi[k]\right]_1}{\delta \GH[p]},\;\;\;\;
 J[k,p] \equiv \frac{\delta \left[ \Psi[k]\right]_1}{\delta \GH[p]},
  \earray
 are symmetric functions under $k \leftrightarrow p$. This symmetry  therefore guarantees the existence of  two  Luttinger Ward type  functionals of the auxiliary Greens function $\GH$,    
 \barray
 \Omega_\Phi[ \GH]& =& - \frac{1}{4} \sum_{k,p,q,r} W(k,q;r,p) \left[ W(k,q; r,p)+W(k,q; p,r) \right] \ \GH[k]  \ \GH[p] \ \GH[q] \ \GH[r]  \nn \\
 \Omega_\Psi[ \GH]& =&  \frac{1}{4} \sum_{k,p,q,r} W(k,q; r,p) \ \GH[k]  \ \GH[p] \ \GH[q] \ \GH[ r],
 \earray
  such that the two self energies can be found from these functionals:
 \barray
 \left[ \Phi[k] \right]_1 = \frac{\delta \Omega_\Phi}{\delta \GH[k]}, \;\;\;
 \left[ \Psi[k] \right]_1 = \frac{\delta \Omega_\Psi}{\delta \GH[k]}.
 \earray
 The form of these two functionals follows  to this order from \disp{final-phi}, and it is natural to conjecture that such functionals exist to all orders in $\lambda$. 
The existence of the $\Omega_\Phi$ functional guarantees a (FS) volume conserving   Luttinger Ward Fermi surface for the $\GH$ electrons, and the smooth behavior of $\Psi(k)$ near this surface guarantees likewise for the physical electrons.

\section{Ward Identities \label{ward}}
This theory admits Ward identities involving the vertices $\Lambda$ and $\U$ that guarantee current conservation in a similar fashion as \refdisp{ecql}. This is displayed with the help of  sources, the charge potential $u[m]=\sum_\si \V_m^{\si \si}$ and an added  source $v[m]$ coupling to the kinetic energy as
\beq
t[i,j]\to t[i,j] (1+ v[j]-v[i]),
\eeq
so that $v[j]-v[i]$ acts as a discrete version of the Peierls phase factor of electromagnetic coupling in tight binding systems.
 We define
\beq
D_m \equiv \partial_{\tau_m} \frac{\delta}{\delta u[m]}- \frac{\delta}{\delta v[m]},
\eeq
so that the Ward identity expressing the conservation of current, from \refdisp{ecql}  reads
\beq
D_m  \ \G[i,f] = (\delta[i,m]-\delta[f,m]) \ \G[i,f]. \label{WI-1}
\eeq
This is a discrete (Takahashi type) version of the usual Ward identity appropriate to the lattice Fermi system at hand, and electromagnetic coupling only requires the long wavelength limit of this identity.
 We will define the $\ttau$ vertices (summing over $\si$)
\barray
\Lambda^{\ttau}(i,j;m) &=& -  \frac{\delta}{\delta v[m]} \GHI_{\si \si}[i,j]/_{u,v\to 0} \nn \\
\U^{\ttau}(i,j;m) &=&  \frac{\delta}{\delta v[m]}  \mu_{\si \si}[i,j]/_{u,v\to 0}. 
\earray
It is easy to see that the bare $\tau$ vertices are given by differentiating  $\GHI_0$ in \disp{g0} as
\beq
\lambda^{\ttau}(i,j;m)= t[i,j] \left( \delta[ i,m]-\delta[j,m]\right), \;\;\;\;\; \lambda^{\ttau}[p_1,p_2] = \varepsilon_{p_1}-\varepsilon_{p_2},
\eeq 
while the singlet (i.e. density) vertices are already known from $\Lambda^{(s)}= \sum_{\si \si'}\Lambda^{\si \si}_{\si' \si'}$.
Note that the $\ttau$ type vertices are antisymmetric in $i \leftrightarrow j$ or $p_1 \leftrightarrow p_2$.
 
Taking Fourier transforms  in \disp{WI-1} and writing $\G= \GH \times \mu$, we get the conservation law:
\barray
(i \omega_{p_1}- i \omega_{p_2}) \ \left( \GH[p_1]  \Lambda^{(s)} (p_1,p_2) \GH[p_2] \mu(p_2)+ \GH[p_1] \U^{(s)}(p_1,p_2) \right) - &&
 \left( \GH[p_1]   \Lambda^{\ttau} (p_1,p_2) \GH[p_2] \mu(p_2)+ \GH[p_1] \U^{\ttau}(p_1,p_2) \right) = \nn \\ &&  \GH[p_2]\mu(p_2)- \GH[p_1] \mu(p_1).
\earray
 Canceling out $\GH[p_1] \GH[p_2]$ we get the Ward identity:
\barray
(i \omega_{p_1}- i \omega_{p_2}) \ \left(  \Lambda^{(s)} (p_1,p_2)\mu(p_2)+  \U^{(s)}(p_1,p_2) \GHI(p_2) \right) - &&
 \left(  \Lambda^{\ttau} (p_1,p_2)\mu(p_2)+  \U^{\ttau}(p_1,p_2) \GHI(p_2) \right) = \nn \\ &&  \GHI(p_1)\mu(p_2)- \GHI(p_2) \mu(p_1).
 \earray
   With $i\omega_n \to z_n$, we rewrite this as
\barray
&& {\cal W}_g(p_1,p_2) \ \mu(p_2) + \GHI(p_2) \  {\cal W}_{\mu}(p_1,p_2) =0, 
\earray
where we have defined the two Ward functions:
\barray
&&{\cal W}_g(p_1,p_2)= ( z_1-z_2) \   \Lambda^{(s)} (p_1,p_2) - \Lambda^\ttau(p_1,p_2)+ \GHI(p_2)-\GHI(p_1)
\nn \\
&&{\cal W}_{\mu}(p_1,p_2) = ( z_1-z_2) \   \U^{(s)} (p_1,p_2) - \U^\ttau(p_1,p_2) + \mu(p_1) - \mu(p_2)
\earray
Since $p_1$ and $p_2$ are arbitrary, the two terms must vanish separately giving us the pair of Ward identities:
\barray
{\cal W}_g(p_1,p_2)&=&0, \label{ward-1} \\
{\cal W}_\mu(p_1,p_2)&=&0. \label{ward-2}
\earray

\section{Random Phase Approximation \label{RPA}}

Since the Greens functions are known to $O(\lambda^2)$, we can  take the derivatives of \disp{ginverse-first-order}
 and \disp{mu-first-order}, to get vertices to this order.  Here we calculate  by taking the equations to $O(\lambda)$ only, but  assuming
$\frac{\delta}{\delta \V} \GH = \GH \Lambda \GH$ rather than  $\frac{\delta}{\delta \V} \GH = \GH  \GH$, thereby obtaining  the analog of the RPA. Since the spin susceptibility is also of considerable interest, we will calculate the required vertices in the that channel as well. Summarizing the results we write linear integral equations for the $\U$ vertices:
\barray
\U^{\ttau}[p_1,p_2]&=& - \lambda \sum_q \GH[q] \Lambda^{\ttau}(q,q+p_2-p_1) \GH[q+p_2-p_1] + O(\lambda^2) \nn \\ 
\U^{(s)}[p_1,p_2]&=& - \lambda \sum_q \GH[q] \Lambda^{(s)}(q,q+p_2-p_1) \GH[q+p_2-p_1] + O(\lambda^2)\nn \\ 
\U^{(t)}[p_1,p_2]&=&  \lambda \sum_q \GH[q] \Lambda^{(t)}(q,q+p_2-p_1) \GH[q+p_2-p_1] + O(\lambda^2), \label{rpa-u}
\earray
and similarly for the $\Lambda$ vertices:
\barray
\Lambda^{\ttau}[p_1,p_2]&=& \left( \varepsilon_{p_1} - \varepsilon_{p_2}\right) \left( 1- \underbrace{ \lambda \ n } \right)  - \lambda \sum_{q} \GH[q] \Lambda^{\ttau}(q, q+p_2-p_1) \GH[q+p_2-p_1] \ {\cal F}(q,p_1,p_2)  +O(\lambda^2), \nn \\
\Lambda^{(s)}[p_1,p_2]&=& 1 - \lambda \sum_{q}  \GH[q] \Lambda^{(s)}(q, q+p_2-p_1) \GH[q+p_2-p_1] \ {\cal F}(q,p_1,p_2)    +O(\lambda^2), \nn \\
\Lambda^{(t)}[p_1,p_2]&=& 1 + \lambda \sum_{q}  \GH[q] \Lambda^{(t)}(q, q+p_2-p_1) \GH[q+p_2-p_1] \ {\cal F}(q,p_1,p_2)  +O(\lambda^2). \label{rpa-lambda}
\earray
where we use the shorthand ${\cal F}(q,p_1,p_2) \equiv  \{ \varepsilon_{p_1} +\varepsilon_{p_2}+\varepsilon_{q}+\varepsilon_{q+p_2-p_1} -u_0 + \frac{1}{2} \left(J_{p_1-p_2} + J_{q-p_1} \right)\}$. The  term in underbrace receives an  $O(\lambda)$ contribution from differentiating the explicit $v$ dependence of the transformed  $t[i,m]\to t[i,m] (1+v[m]-v[i]) $ term in  \disp{ginverse-first-order}.  It is readily shown by examining the kernel of the integral equations that the solution for  $\Lambda^{\ttau}(p_1,p_2)$ is antisymmetric under exchanging  $p_1 \leftrightarrow p_2$, while  $\Lambda^{(s)}(p_1,p_2)$ and $\Lambda^{(t)}(p_1,p_2)$ are symmetric.

These vertices are shown to be compatible with Ward identities to $O(\lambda)$ if used with  the first order versions of the Greens functions \disp{final-mu} and \disp{final-g}:
\beq
\GH[p]= i \omega_n + \chem'- (1- \lambda n) \varepsilon_k  + \frac{\lambda}{2} \sum_q J_{k-q} \  \GH[q] +O(\lambda^2), \;\;\;\mbox{and}\;\;\; \mu(p)= 1, \label{g-firstorder}
\eeq
by substituting in the expressions \disp{ward-1} and \disp{ward-2}, and showing the self consistency of this result. The details of this verification parallel the standard proof in QED and are omitted here. Note that $\mu$ must be chosen to be unity rather than $1-\lambda \frac{n}{2}$ as discussed in the second para below \disp{final-g}, although  this choice is irrelevant to the verification of the  Ward identity.

\section{Two particle response}

We are interested in the pair correlations of the density $n_a = \sum_\si \X{a}{\si \si}$ and the spin density
${S}^z_a = \frac{1}{2} \sum_{\si_1, \si_2} {\tau}^z_{\si_1 \si_2} \X{a}{\si_1 \si_2}$, where ${\tau}^z$ is the  usual Pauli matrix. These can be obtained from taking the functional derivatives of the Greens function
\beq
\up^{\si_1 \si_2}_{\si_3 \si_4}[i,j] = \frac{\delta}{\delta \V_j^{\si_3 \si_4}} \G_{\si_1 \si_2}[i^-,i], \label{eq92}
\eeq
and  can be conveniently found from taking a limit  of the three site object $  \up^{\si_1 \si_2}_{\si_3 \si_4}(p,q;r)=\frac{\delta}{\delta \V_r^{\si_3 \si_4}} \G_{\si_1 \si_2}[p,q]$. With the singlet and triplet objects denoted with a superscript $\alpha = s,t$, we note the following relationships with the standard charge and spin susceptibilities of interest:
\barray
   \lll  \ n_a(\tau_a) \ n_b(\tau_b) \ \rrr      &=& n^2 - 2 \ \up^{(s)}(a,b) \nn \\
   \lll  \  S^z_a(\tau_a) \ S^z_b(\tau_b) \ \rrr     &=& -  \frac{1}{2} \ \up^{(t)}(a,b) \label{susceptibilities}
\earray
 Owing to the Bosonic nature of the densities, we have the symmetry $\up^{(\alpha)}(b,a) = \up^{(\alpha)}(a,b)$ from which the Fourier transform at $Q \equiv (\vec{Q}, i \Omega_q)$ satisfies the relation:
\beq
\up^{(\alpha)}(Q) =\up^{(\alpha)}(-Q).  \label{parity}
\eeq
This symmetry can be used as another test of the consistency of any approximation.

The Greens function in \disp{eq92} can be decomposed in to $\GH$ and $\mu$ as before and  we find
\barray
 \up^{\si_1 \si_2}_{\si_3 \si_4}(a,b;r)& = & \frac{\delta }{\delta \V_r^{\si_3 \si_4}} \ \left\{  \GH_{\si_1 \si_a}(a,\bb{a}) \mu_{\si_a \si_2}(\bb{a},b)\right\}, \nn \\
 &=&\left\{  \GH[a,\bb{b}] \Lambda_{\si_3 \si_4}(\bb{b},\bb{s};r)\GH[\bb{s},\bb{a}] \mu(\bb{a}, q) \right\}_{\si_1 \si_2} + \left\{  \GH[a,\bb{b}] \  \U_{\si_3 \si_4}(\bb{b}, b;r) \right\}_{\si_1 \si_2},
\earray
where the vertex and $\up$ carry upper spin indices  that are part of the matrix product. Turning off the sources, we find the expressions for singlet and triplet response
\barray
 \up^{(\alpha)}(a,b;r) &=&  \GH[a,\bb{b}] \Lambda^{(\alpha)}(\bb{b},\bb{s})\GH[\bb{s},\bb{a}] \mu(\bb{a}, b)   +    \GH[a,\bb{b}] \  \U^\alpha(\bb{b}, b;r) , \nn \\
 \up^{(\alpha)}(p_1,p_2) &=&  \GH[p_1] \Lambda^{(\alpha)}(p_1,p_2)\GH[p_2] \mu(p_2)   +    \GH[p_1] \  \U^{(\alpha)}(p_1,p_2) ,
 \earray
where $\alpha = s,t$.  The definitional distinction between left and right derivatives leads to the asymmetry in the above equations making it necessary to test the consistency \disp{parity}  term by term.

Using the  zero source limit notation from \refdisp{ecql}: 
\barray
Q^{(1)} &=& Q^{\si \si}_{\si \si}; \;\; Q^{(2)} = Q^{\si \si}_{\sib \sib}; \; \; Q^{(3)} = Q^{\si \sib}_{\si \sib} \nn \\
Q^{(a)} &=&Q^{(2)}-Q^{(3)}; \;\; Q^{(s)}= Q^{(1)}+Q^{(2)}; \;\; Q^{(t)}= Q^{(1)}-Q^{(2)} = Q^{(3)}. 
\earray

 The charge $\alpha= s$ and spin $\alpha= t$ susceptibilities at finite $Q \equiv (\vec{Q}, i \Omega_q)$   are given by setting $p_2\to p$ and $p_1 \to p+Q$ and summing over $p$.
  \barray
\up^{(\alpha)}(Q) &\equiv&   \sum_p \up^{(\alpha)}(p,p+Q) = \sum_p  \ \left(\GH[p] \ \Lambda^{(\alpha)}(p,p+Q) \ \GH[p+Q] \ \mu(p+Q) + \GH[p] \ \U^{(\alpha)}(p,p+Q)\right),
 \label{twoparticle}
\earray
These are exact expression for the susceptibilities, but as usual  require a knowledge of the vertices and Greens functions to give practical results. We can now use the RPA vertices calculated in Section.~({\ref{RPA}) to give the corresponding expressions.

We denote the susceptibility of the auxiliary Fermions as
\beq
\chi^{(\alpha)}_{\Lambda}(Q) \equiv - \sum_q \GH[q] \Lambda^{(\alpha)}(q, q+Q) \GH[q+Q], \label{chilambda} 
\eeq
and within RPA we note that $\mu(p)$ is independent of $p$, and  from \disp{rpa-u} we denote that the $\U$ vertices are functions of the momentum difference only:
\barray
\U^{(\alpha)}[p_1,p_2] = \lambda \ \xi_\alpha  \ \chi^{(\alpha)}_{\Lambda}(p_2-p_1)
\earray
where $\xi_\alpha$  is $  1$ for $\alpha=$ singlet and $-1$ for $\alpha=$ triplet.  Therefore we can
 sum over the $p$ dependence of the second term and  rewrite \disp{twoparticle} as
\beq
\left( \up^{(\alpha)}(Q)\right)_{RPA} =   -   C_\alpha  \ \chi_{\Lambda}^{(\alpha)}(Q) 
\eeq
where $C_\alpha = \left( \mu - \xi_\alpha \ \lambda \  \frac{n}{2} \right)$. It seems more appropriate to reset $\mu=(1- \lambda \frac{n}{2}) $ from unity at this level, in order to recover  the expected high frequency behavior in the charge as well as spin channel, so that $C_{singlet} = (1- \lambda n)  \to 1-n $ and $C_{triplet}=1$. The vertices $\Lambda$ are to be computed from \disp{rpa-lambda} and form a consistent set of equations for two particle response in the sense of the usual RPA.

The integral equations must be solved numerically. However
in order to display some flavor of the results, we pursue this to the lowest order in $\lambda$ by iteration, where explicit results can be obtained.
Let us define a few  frequently occurring  generalized polarizability functions for convenience. We will now reinstate $J_k \to J_k - u_0$
\barray
\chi_0(Q)&=& - \sum_q \GH[q] \ \GH[q+Q] \nn \\
\chi_1(Q)&=& - \sum_q \  \GH[q] \ \GH[q+Q]  \  \left\{ \varepsilon_q+ \varepsilon_{q+Q} \right\}\nn \\
\chi_2(Q)&=& \frac{1}{2} \sum_{r,p} \GH[r] \GH[r+Q] \GH[p] \GH[p+Q]  \    J_{p-r}  \nn \\
F(p+Q,p)&=& \sum_{r} \GH[r] \GH[r+Q]  \ \left\{ \varepsilon_p+ \varepsilon_{p+Q} + \varepsilon_r+ \varepsilon_{r+Q} + \frac{1}{2} ( J_Q + J_{p-r} )\right\}. \label{needed}
\earray
Here $\chi_0(Q)$ is the standard Lindhard function and is positive in the static limit as $\vec{Q} \to 0$, while the other functions are generalizations thereof. 

%We now list the vertices to lowest two orders:
%\barray
%~[ \mu(p) ]_0 &=& 1, \;\;\;\;
%~[ \mu(p) ]_1 = - \frac{n}{2} \nn \\ 
%\left[ \U^{(\alpha)}(p+Q,p) \right]_0 &=& 0, \;\;\;\;
%\left[ \Lambda^{(\alpha)}(p+Q,p) \right]_0 = 1. \nn \\
%\left[ \U^{(s)}(p+Q,p) \right]_1 &=& \chi_0(Q), \;\;\;\;
%\left[ \U^{(t)}(p+Q,p) \right]_1 = -\chi_0(Q), \nn \\
%\left[ \Lambda^{(s)}(p+Q,p) \right]_1 &=& F(p+Q,p) + u_0 \ \chi_0(Q), \;\;\;
%\left[ \Lambda^{(t)}(p+Q,p) \right]_1 = -F(p+Q,p) - u_0 \ \chi_0(Q).
%\earray
%Using the vertices derived above, we  can expand the susceptibilities to various orders in $\lambda$.
%\barray
%\left[ \up^{(s)}(Q) \right]&=& \left[ \up^{(s)}(Q) \right]_0+ \lambda \ \left[ \up^{(s)}(Q) \right]_1 + O(\lambda^2) \nn \\
%\left[ \up^{(s)}(Q) \right]_0 & = & - \chi_0(Q) \ ,\nn \\
%\left[ \up^{(s)}(Q) \right]_1 & = &  n \ \chi_0(Q) -2 \chi_0(Q) \ \chi_1(Q) + \left( u_0 - \frac{1}{2} J_Q \right) \chi^2_0(Q) - \chi_2(Q)
%\earray

%\barray
%\left[ \up^{(t)}(Q) \right]&=& \left[ \up^{(t)}(Q) \right]_0+ \lambda \ \left[ \up^{(t)}(Q) \right]_1 + O(\lambda^2) \nn \\
%\left[ \up^{(t)}(Q) \right]_0&=& - \chi_0(Q)\nn \\
%\left[ \up^{(t)}(Q) \right]_1&=&  2 \chi_0(Q) \ \chi_1(Q) + \left( \frac{1}{2} J_Q -  u_0  \ \right) \chi^2_0(Q) + \chi_2(Q).
%\earray
The answers are
\barray
 \up^{(s)}(Q)&=& - (1- \lambda n) \chi_0(Q)  - \lambda [ 2  \ \chi_0(Q) \ \chi_1(Q) - \left( u_0 - \frac{1}{2} J_Q \right) \chi^2_0(Q) + \chi_2(Q) ]
 \nn \\
  \up^{(t)}(Q)&=&  - \chi_0(Q) + \lambda [ 2 \chi_0(Q) \ \chi_1(Q) + \left( \frac{1}{2} J_Q -  u_0  \ \right) \chi^2_0(Q) + \chi_2(Q)
] .
\earray
It is clear that the role of $u_0$ enhances the spin susceptibility while decreasing the charge susceptibility. To this order we see that the parity test \disp{parity} is satisfied to this order by using the symmetries of the objects in  \disp{needed}.

SInce the Greens function remain infinitely sharp within the RPA, its usefulness is limited- especially in view of the large frequency dependent   corrections with characteristic asymmetry seen in second order results \refdisp{ecfl}, \refdisp{gweon} and \refdisp{asymmetry}. A second order version of RPA seems most desirable, although even without vertex corrections to second order, the single particle spectral results are very interesting already.   It also seems interesting to also study phenomenologically, the analog of the ``bubble'' diagram for purposes of extracting the optical  conductivity; a scheme  that reflects the width of the physical Greens function and  satisfies  the parity requirement \disp{parity} is given by:
\beq
\left[ \U(Q) \right]_{phen} = - \frac{1}{1-n/2}  \sum_q \G(q) \G(q+Q),
\eeq
although this expression is not the result  a systematic expansion of \disp{twoparticle} .

\section{Discussion and Conclusions}

We have described above a controlled technique of dealing with the \tJ model. 
This  extremely correlated Fermi liquid theory  is  a strong coupling approach, specifically designed to deal with a hard many body problem. The considerations begin with the strong coupling limit of the Hubbard model, leading to the \tJ model with a hard constraint of eliminated double occupancy. The Schwinger method gives us a crucial initial platform to deal with this problem. The ensuing   exact functional differential equations  are made tractable by the introduction of the exact  {\em product ansatz}:  $\G = \GH \times \mu$, with $\GH$    a canonical Greens function of auxiliary electrons and $\mu$ the caparison factor. The latter,    in turn, is  understood as  an adaptive spectral weight balancing the requirements at the high and low frequency ends of the spectrum. Both  objects are expanded in powers of a parameter $\lambda$, that plays the role of  fractional  double occupancy. Thus $\lambda=1$ corresponds to complete elimination of double occupancy whereas $\lambda < 1$ has some residual double occupancy.  We thus  replace the hard constraint: of complete elimination of  double occupancy  by a softer one or partial removal. In order to provide a natural  description of  the canonical electrons, we introduce the effective Hamiltonian $H_{eff}$,  depending  parametrically on $\lambda$.  In order to obey  the {\em Shift theorems (I-II)}, we find it obligatory to (re)introduce a Hubbard type $u_0$ parameter in this model. It also  plays the role of a second chemical potential as explained above. The   set of steps followed, in our starting as well as ending up with a Hubbard type interaction  has a slightly  circular feel to it.   This recipe  is perhaps best understood as  a renormalization group type  procedure, where the constraint of single occupancy is enforced incrementally and  the density of  doubly occupied sites is   thinned out smoothly.  The infinite starting value of $U$ in the \tJ model is pushed downward  to $u_0$, typically  a fraction of the bandwidth from our numerical studies, albeit in a more general model $H_{eff}$, and    is therefore amenable to a perturbative expansion. The form of the $H_{eff}$ and the important role of the shift symmetries in validating the approximations is noteworthy.  The hopping $t_{ij}$ is elevated to an interaction constant of the model, this unfamiliar  step is kept under  check by requiring  the two  important shift invariances. The  Schwinger equation  \disp{eom_2} for $\G$, being an exact statement of the problem,   provides us with  a rigorous backdrop to the entire  procedure. Further  our procedure has the advantage of being systematically improvable through the iterative scheme developed here.

%%%%%%%%%%%%%%%%%%%%%%%%%%%%%%%%%%%%%%%%%
%%%%%%%%%%%%%%%%%%%%%%%%%%%%%%%%%%%%%%%%%%%%%

We can explore superconductivity at a qualitative level,  by studying the pairing instabilities of the auxiliary Fermions given by $H_{eff}$ via  its BCS  gap function $\Delta(k)$.
  In this first approximation,  the physical electron order parameter $\langle \X{k}{\uparrow 0} \X{-k}{\downarrow 0} \rangle$  is proportional to that of the auxiliary electrons  $\langle f^\dagger_{\uparrow }({k}) f^\dagger_{\downarrow }({-k}) \rangle$, together  with the  single  occupancy constraint of vanishing upon summing over the wavevector $k$.   Within a generalized Hartree Fock theory, retaining the self energy correction to first order (as in \disp{g-firstorder})  as well as the pairing field average, we obtain  an   equation for the gap function $\Delta(k)$:
\beq
\Delta(k) = \frac{1}{N_s} \sum_p \left\{ \varepsilon_k+ \varepsilon_p - u_0 + \frac{1}{2} J_{k-p} \right\}\ \Delta(p) \ \frac{\tanh{\beta E(p)/2}}{E(p)} \label{gap-equation}
\eeq
where $E(p)= \sqrt{\Delta^2(p)+ \xi^2_p}$, and $\xi_p= \varepsilon_p (1-n) - \frac{1}{2} \sum J_{q-p} n_q - \chem$.    In the computation below, we will  neglect the numerically small  $J$ term in the single particle energy. Other than $u_0$ and  the two single particle energies in \disp{gap-equation} required for satisfying the {\em Shift theorems (I-II)}, this is the same equation as the one found within the resonating valence bond theory   in \refdisp{pwa}, \refdisp{bza} and \refdisp{kotliar}. The transition temperature for a d-wave state with  a gap function $\Delta(k)=  \Delta_d \ [\cos(k_x)- \cos(k_y)]$ is  obtained by solving \disp{gap-equation} for the case of the nearest neighbour square lattice \tJ model, with parameters indicated in the caption. 
  \begin{figure}[h]
\includegraphics[width=3.5in]{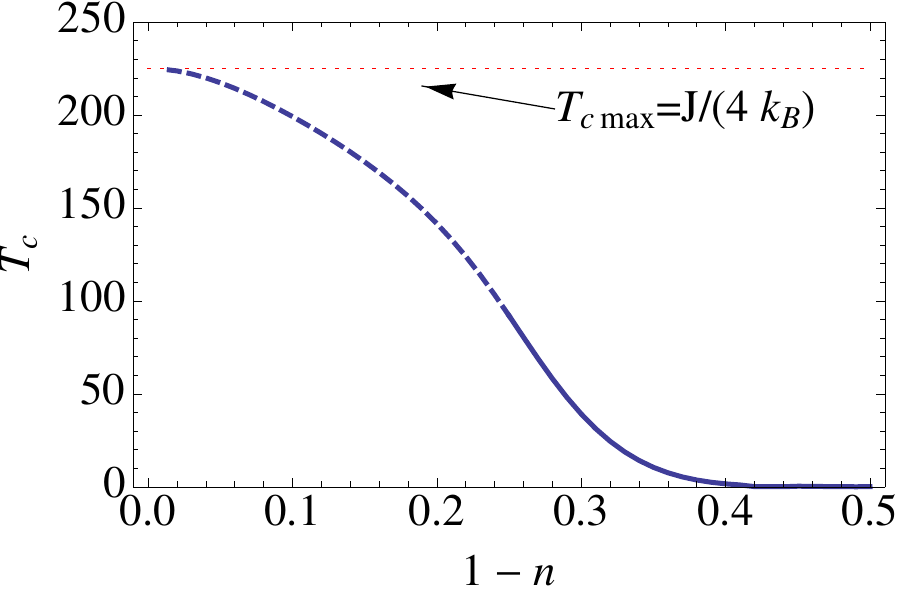}
\caption{The transition temperature in Kelvin,  from solving \disp{gap-equation} assuming $t=3000$K and $J=900$K. The solid line indicates the likely regime of validity of the $O(\lambda)$ theory. Its  dotted  extension to lower hole density is   speculative and is most likely to change with higher order corrections reflecting the nearby Mott insulating state. The dotted red line indicates the maximum $T_c$ obtainable from this scheme, and is seen to depend solely upon the magnitude of $J$.  }
\label{Fig-tc} 
\end{figure}
It is straightforward to see that the $T_c$ equation has a maximum scale of order $J/(4 k_B)$ as already noted in \refdisp{bza} and \refdisp{kotliar}. This value   is attained  in this solution at a   higher particle density,  or equivalently,   a lower hole density,  than is warranted by the first approximation. The solid line represents a plausible regime of validity of this scheme. 

 The extended s-wave order is usually described  by a gap function  $\Delta(k)= \Delta_{s, 0} + \Delta_{s, 1}\ [\cos(k_x)+\cos(k_y)]$.  The constant term $\Delta_{s,0}$ leads to a finite probability of double occupancy, since it survives a wavevector sum.  After it is  dropped as per the above discussion,   the assumed (purely extended) s-wave order is  supported by the $J$ term in the kernel of \disp{gap-equation} , but not by   the  $u_0$ dependent and single particle energy terms. The latter thus do not play a  role in determining $T_c$ for either d-wave or s-wave orders despite  their large magnitude relative to $J$.

A detailed    calculation of the gap equation  is  planned  for the pairing of  physical particles, 
parallel to  the  $O(\lambda^2)$ theory of the normal state. The finite   lifetime  effects are then expected to  become relevant. Such an  improvement of the pairing scheme should yield a greater  understanding of the balance between the different orders and    a greater range of validity in density than the  schematic theory treated here.

\section{Acknowledgements}
  I  have  benefitted from discussions  with       T. Banks,   G.-H. Gweon,   D. Hansen and E.  Perepelitsky  at Santa Cruz,  and with       P.W. Anderson,  A. Georges, A. C. Hewson,    G. Kotliar and Y. Kuramoto elsewhere. This work was supported by DOE under Grant No. FG02-06ER46319.

\appendices

\section{Atomic limit $t=J=0$ \label{app-1}}
\subsection{Double occupancy interpretation of $\lambda$ from the atomic limit}
In order to understand the role of $\lambda$ we study the atomic limit $t, J \ \to 0$ where this parameter can be introduced into the  physical Greens function in the form:
\beq
\G[\lambda, \ i \omega_n]= \frac{1 - \lambda \ \frac{n}{2}}{i \omega_n + \chem }, \label{atomic-3}
\eeq
and study its dependence on  $\lambda $ in the interval $[0,1]$.
 The  chemical potential $\chem$ can be calculated from the sum rule on the  density $n$  of the number of particles $N$ 
with
$n= N/N_s$ and temperature T as
\beq 
\chem= k_B T \ \ln (\frac{n}{2- ( 1 + \lambda) n}).  \label{atomic-chem}
\eeq

Thermodynamics tells us that  the entropy S can be expressed as
\beq
S(n) = -  N_s \ \int_0^n \ dn' \ \  \frac{\partial   \mu(n') }{\partial  T} 
\eeq
and since we know $\mu$ from \disp{atomic-chem} we obtain with $y=(1+ \lambda) n$
\beq
\frac{S(n,\lambda)}{k_B N_s} = \frac{1}{1+ \lambda} \left\{  \ln{4} -  y \ln{n}
- (2 - y ) \ \ln{(2 - y)}
\right\}. \label{entropy-1}
\eeq
  we see that its $\lambda$ derivative:   $\frac{1}{k_B N_s} \frac{\partial S}{  \partial \lambda} =  \frac{2}{(1+ \lambda)^2}
  \left[ \frac{y}{2}+ \ln{(1 - \frac{y}{2}) }\right]$ is negative definite. Thus we see that the entropy at a fixed density interpolates monotonically, between the free Fermi limit and the infinite $U$ limits as  $\lambda$ ranges over its domain $0 \leq \lambda \leq 1$. The maximum allowed density is reduced from 2 to $\frac{2}{1+ \lambda}$ and thus at $\lambda=1$ we have a maximum of one electron per site- as expected physically. Thus increasing  $\lambda$ from zero  effectively removes the available states contributing to  entropy, its role   may be viewed as that of  (continuous) removal  of states. Thus  for the equations of motion it is somewhat  analogous to the role of  Gutzwiller's parameter $g$ in his projection operator $\prod_i \left[ 1- (1-g) n_{i \uparrow} n_{j \downarrow} \right]$  at the wave function level.

In the atomic limit we can also calculate the entropy at a fixed density  of doubly occupied sites $d = \frac{1}{N_s} \sum_i n_{i \uparrow} n_{i \downarrow}$ as
\beq
\frac{S(n, d)}{k_B N_s} = - d  \ \ln  d - (n - 2 d) \ \ln(\frac{n}{2} - d) - (1+d-n) \ \ln( 1+d-n). \label{entropy-2}
\eeq 
   \begin{figure}[h]
\includegraphics[width=3.5in]{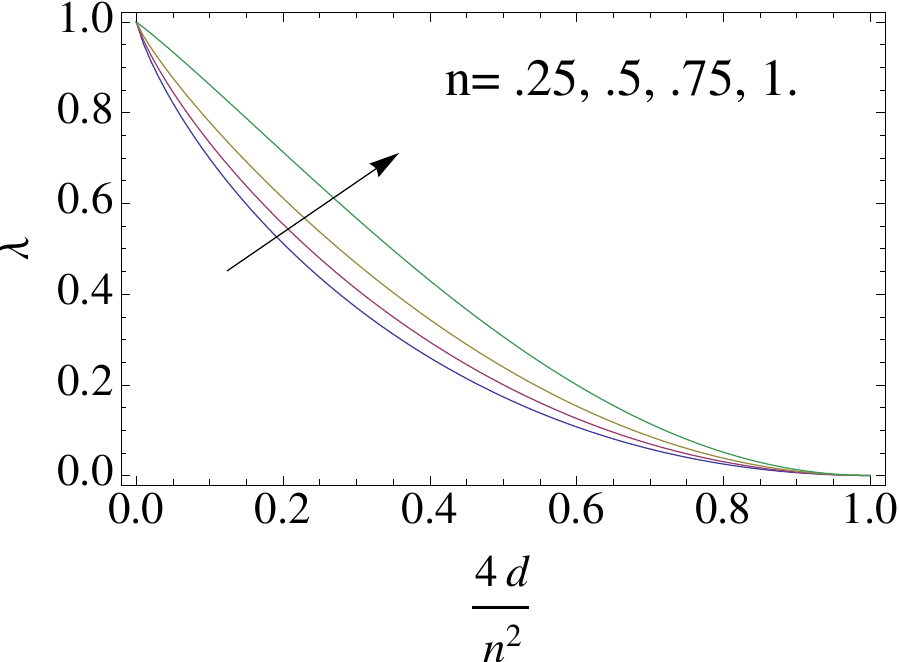}
\caption{ The parameter $\lambda$ is determined in terms of the double occupancy $d$ at various densities  in the atomic limit as described in the text. The arrow indicates increasing density $n$. Note that the  parameter $d$ is scaled into the unit interval. }
\label{Fig_1atomic}
\end{figure}
   An uncorrelated system  corresponds to    $d= \frac{n^2}{4}$, where the entropy \disp{entropy-2} is a maximum,  while $d=0$  for the fully projected  \tJ model. Comparing the two expressions for entropy \disp{entropy-1} and \disp{entropy-2}, we can express $\lambda$ in terms of $d$ at any density.  We have thus demonstrated that $\lambda$ is a conjugate variable to the double occupation density in this limit. Their explicit relationship is illustrated in \figdisp{Fig_1atomic}.

%%%%%%%%%%%%%%%%%%%%%%%%%

\subsection{Expansion in  $\lambda$ in the Atomic Limit:}

In the atomic limit we set $t\to 0$ and $J \to 0$ so that \disp{ginvdef} and \disp{fundapair} becomes
\barray
\GH[i,f]  & = &   \GH_0[i,f;\chem_a]    \nn \\
\mu[i,f]  & = &  \delta[i,f]   \left( \iden - \lambda \ \gamma[i] \right)    - \lambda \ \chem_b \ \GH[{i},\bb{f}]\cdot \mu[\bb{f},f]. \label{atomic-1}
\earray 
Here we split the chemical potential into two pieces $\chem=\chem_a+ \lambda \chem_b$. 
Thus in this limit $\GH$ is the free Fermi Greens function independent of $\lambda$, and $\chem_a$ is the free value $\chem_a \to \chem_0$,  the latter  determined from the non interacting theory  in terms of the number of particles. If we turn off the source $\V$ the Fourier transforms can be taken as
\barray
\GH[i \omega_n]  & = &   \GH_0[ i \omega_n ;\chem_a] = \frac{1}{i \omega_n + \chem_0}    \nn \\
\mu[i \omega_n]  & = &     \left( 1 - \lambda \ \frac{n}{2}\right)    - \lambda \ \chem_b \ \GH[i \omega_n ] \mu[i \omega_n], \nn \\
&=& \frac{1 - \lambda \frac{n}{2}}{1+ \lambda \chem_b \ \GH[ i \omega_n]}   . \label{atomic-2}
\earray 
Thus the physical Greens function
\beq
\G[i \omega_n]= \frac{1 - \lambda \ \frac{n}{2}}{i \omega_n + \chem_0 + \lambda \chem_b } . \label{atomic-4}
\eeq
We fix the chemical potentials from the number sum rule as usual and thus
\barray
\frac{n}{2}&=& \frac{1}{1+e^{- \beta \chem_0 }}, \nn \\
\frac{n}{2}&=& (1 - \lambda \ \frac{n}{2}) \ \frac{1}{1+e^{- \beta (\chem_0 + \lambda \chem_b)}}. 
\earray
We may then solve for $\chem$'s in terms of the density and obtain
\barray
\chem_0&=&  k_B T \ln(\frac{n}{2-n}) \nn \\
\lambda \ \chem_b&=&  k_B T \ln(\frac{2-n}{2- ( 1 + \lambda) n}).
\earray
Thus the chemical potential $\chem_1$ has a power series representation
\beq
 \chem_b = \sum_{m=0}^{\infty} \ \lambda^m \  \chem_b^{(m)} =  k_B T \ \ \sum_{m=0}^{\infty}  \frac{\lambda^m}{m+1} \left[  \frac{n}{2-n} \right]^{m+1}. \label{chem-exp}
\eeq
We see explicitly from \disp{chem-exp} that the $\lambda$  expansion   of the the atomic limit is  an  expansion in $\lambda n/(2-n)$ i.e. a density expansion as well.

\section{ The low order calculations of Greens functions \label{calculations}}

\subsection{Greens function to  $O(\lambda$)}
We evaluate the complete starting point of the hierarchy. We start with terms of $O(\lambda^0)$ and end with  $[\mu]_1$ and $[\GHI]_1$, which are the seeds for the $O(\lambda)$ terms.

\subsubsection{\bf Seed terms and Initialization}
\barray
\GHI_0[i,m]&=&
 \{ [( \chem - \partial_{\tau_i} - \frac{1}{4} J_0)\iden - \V_i] \delta[i,m] + t[i,m] - \V_{i,m} \}    \nn \\
 \left[ \mu[i,f]  \right]_0 &=& \iden \  \delta[i,f] 
 \earray
 
 {\bf Derived objects}
 \barray
 \left[ \gamma[i] \right]_0 &=& \GH^{(k)}[i,i] \nn \\
 \left[ \gamma[i,m] \right]_0 &=& \GH^{(k)}[m,i] \nn \\
\left[ \U^{\si_a \si_b}_{\si_c \si_d}[i,m; j] \right]_0 &=& 0 \nn \\
\left[ \U^{\si_a \si_b}_{\si_c \si_d}[i,m; j,k] \right]_0 &=& 0. \nn \\
 \earray

\barray
\left[ Y_1[i,m] \right]_0 
&=& t[i,m]  \ (   \GH^{(k)}[i,i] + \frac{1}{2} \GH^{(k)}[m,m] ) - \delta[i,m] \ \frac{1}{2} \left( J[i,\bb{j}] \ \GH^{(k)}[\bb{j},\bb{j}] - t[i,\bb{j}] \ \GH^{(k)}[i,\bb{j}] \right) 
\nn \\
\left[ \Psi[i,m] \right]_0 &=& 0. \
\earray

\barray
&&\left[  \ \Lambda^{\si_a \si_b}_{\si_c \si_d}[i,m; j] \right]_0 = \delta_{\si_a \si_c} \delta_{\si_b,\si_d} \ \delta[i,j] \ \delta[j,m], \nn \\
&&\left[ \Lambda^{\si_a \si_b}_{\si_c \si_d}[i,m; j, k] \right]_0 = \delta_{\si_a \si_c} \delta_{\si_b,\si_d} \ \delta[i,j] \ \delta[m,k] {\color{black} \ \delta(\tau_j-\tau_k)}. \label{fourpoint}
\earray
{\color{black} In the four point vertex above, we have introduced the delta function $ \delta(\tau_j-\tau_k)$, so that the labels $i,m,j,k$ can be viewed as four independent space time variables.
}
Thus
\barray
&& \left[ \Phi[i,m] \right]_0 = \delta[i,m] \ t[i,\bb{j}]  \ \GH^{(k)}[\bb{j},i ] +  \frac{1}{2}  \ t[i, m] \ \GH^{(k)}[m,m] + \frac{1}{2} \delta[i,m] \ t[i,\bb{j}] \  \ \GH^{(k)}[i,\bb{j}] - \frac{1}{2} J[i,m] \ \GH^{(k)}[i,m]
\earray
Combining the two we get
\barray
\left[ \ Y_1[i,m] \right]_0 +   \left[  \Phi[i,m] \right]_0 && = \delta[i,m] \ t[i,\bb{j}] \left( \GH^{(k)}[i, \bb{j}] + \GH^{(k)}[ \bb{j},i ] \right) + t[i,m]  \ (   \GH^{(k)}[i,i] +  \GH^{(k)}[m,m] )\nn \\
&& -\frac{1}{2} J[i,m] \ \GH^{(k)}[i,m] - \delta[i,m] \ \frac{1}{2} J[i,\bf{j}] \ \GH^{(k)}[\bf{j},\bf{j}]
\earray

\subsubsection{\bf Stepping  up and final Greens function to  $O(\lambda)$.}

To first order in $\lambda$ we collect the above results to obtain the Greens function
\barray
\left[ \GHI[i,m] \right]_1 &=&  \GH^{(k)}[i,i] \cdot \V_{i, m} +  \delta[i,m] \ \GH^{(k)}[\bb{a},i] \cdot \V^{(k)}_{i, \bb{a}}   + \delta[i,m] \ \frac{1}{2}  J[i,\bb{j}] \ \GH^{(k)} [\bb{j},\bb{j}]  + \frac{1}{2} J[i,m] \ \GH^{(k)}[i,m] \nn \\
 &&- t[i,m]  \ (   \GH^{(k)}[i,i]  +  \GH^{(k)}[m,m]   )   - \delta[i,m] \ t[i,\bb{j}] \left( \GH^{(k)}[i, \bb{j}] + \GH^{(k)}[ \bb{j},i ] \right)   \label{ginverse-first-order}
 \earray
 and the caparison factor:
\barray
  \left[ \mu[i,m] \right]_{1} 
  &=& - \GH^{(k)}[i^-,i] \ \delta[i,m]   \label{mu-first-order}
\earray

The FT's of these on turning off the sources is found using $\GH[i^-,i]\to \frac{n}{2} $ as
\barray
\left[ \GHI[k]   \right]_1 & = &   {n} \  \varepsilon_k - \frac{n}{2} u_0 + \frac{1}{2} \sum_q J_{k-q} \ \GH[q] +
\left( \frac{1}{4} J_0 \ n + 2 \sum_q \varepsilon_q \ \GH[q]  \right)  \label{eq139} \\
\left[ \mu[k]   \right]_1 & = & - \frac{n}{2}
\earray 
The term $- \frac{n}{2} u_0$ in \disp{eq139}  arises when we reinstate $J[i,j]\to J[i,j]- u_0 \delta[i,j]$ in \disp{ginverse-first-order}. Let us note that under the shift \disp{shift-1},  the first order correction  $ \left[ \GHI[k]   \right]_1$ shifts by $ 2 n u_t + \frac{n}{2} u_J $. Therefore this term is invariant under the {\em Shift theorem-II} and also  the  {\em Shift theorem-(I.1)}, provided $u_0$ is simultaneously transformed  as  specified in  \disp{u-shift}. 

\subsection{ Greens function to  $O(\lambda^2)$}
\subsubsection{$\mu$ derived objects}
 We next start with  seed terms of $O(\lambda)$ calculated above and end with  $[\mu]_2$ and $[\GHI]_2$.
\barray
  \left[ \mu[i,m] \right]_{1}  &=& - \GH^{(k)}[i^-,i] \ \delta[i,m]  \label{eq79}
  \earray
 
Let us calculate the derived quantities from the above at the same level:
\barray
\left[ \ \gamma[i] \ \right]_1&=& \left[ \mu^{(k)}[ \bb{a}, i] \right]_1 \cdot \GH^{(k)} [i,\bb{a}] = - \GH[i,i] \cdot  \GH^{(k)} [i,i] \nn \\
\left[ \ \gamma[i, m] \ \right]_1&=& \left[ \mu^{(k)}[ \bb{a}, i] \right]_1 \cdot \GH^{(k)} [m ,\bb{a}] = - \GH[i,i] \cdot  \GH^{(k)}[m,i] \nn \\
\left[ Y[i,m] \right]_1 &=& - t[i,m] \left(\GH[i,i] \cdot \GH^{(k)}[i,i] + \frac{1}{2} \GH[m,m] \cdot \GH^{(k)}[m,m] \right) \nn \\
&& + \frac{1}{2} \delta[i,m] \left( J[i,\bb{j}] \  \GH[ \bb{j}, \bb{j}] \cdot \GH^{(k)}[ \bb{j}, \bb{j}] - t[i,\bb{j}] \  \GH[ \bb{j}, \bb{j}] \cdot \GH^{(k)}[ i, \bb{j}]\right) \label{eq81}
\earray

Zero source Fourier transforms:
\barray
\left( \left[ \gamma[0] \right]_1 \right)_{\V \to 0} & = & - \frac{n^2}{4}  \nn \\
\left( \left[ \gamma[k] \right]_1 \right)_{\V \to 0} & = & - \frac{n}{2} \GH[-k]  \nn \\
\left( \left[ Y_1[k] \right]_1 \right)_{\V \to 0} & = &  \frac{3n^2}{8} \varepsilon_k - \frac{n^2}{8} u_0  + \left( \frac{n^2}{8} J_0 + \frac{n}{4} \sum_q \varepsilon_q \ \GH[q] \ \right) \label{eq97}
\earray
Here we reinstated $J[i,j]\to J[i,j]- u_0 \delta[i,j]$  in \disp{eq81} to obtain the $- \frac{n^2}{8} u_0 $ term in \disp{eq97}.

\subsubsection{$\mu$ derived vertices}
Next we calculate (using lowest order functional derivatives) 
\barray
\left[ \ \U^{\si_1 \si_2}_{\si_3 \si_4}[i,m; a ] \ \right]_1 
&=& - \delta[i,m] \ \si_1 \si_2  \ \GH_{\sib_2 \si_3}[i,a] \ \GH_{\si_4 \sib_1}[a,i] \nn \\
 \left[ \ \U^{\si_1 \si_2}_{\si_3 \si_4}[i,m; a ] \ \right]_1&_{\V\to 0}=  &- \delta[i,m] \ \si_1 \si_2 \ \delta_{\sib_2,\si_3} \delta_{\sib_4, \si_1} \ \GH[i,a] \ \GH[a,i] \nn \\
\earray
At zero sources so with ${\V\to 0}$
\barray
\left[ \ \U{(a)}[i,m; a ] \ \right]_1  &=  &-  2 \  \delta[i,m] \ \ \GH[i,a] \ \GH[a,i], \nn \\
\left[ \ \U^{(s)}[i,m; a ] \ \right]_1 & =&  -   \  \delta[i,m] \ \ \GH[i,a] \ \GH[a,i].
\earray
The zero source Fourier transforms  are as follows:
 \barray
 \left[ \ \U^{\si_1 \si_2}_{\si_3 \si_4}[p_1,p_2] \ \right]_1 &=& -  \ \si_1 \si_2 \ \delta_{\sib_2,\si_3} \delta_{\sib_4, \si_1} \   \ \sum_q \GH[q] \ \GH[q+p_2-p_1] \nn \\
 \left[ \U^{(a)}[p_1,p_2] \right]_1 &=& - 2  \sum_q \GH[q] \ \GH[q+p_2-p_1] 
  \earray

  Similarly we find for the four index vertices:
\barray
\left[ \ \U^{\si_1 \si_2}_{\si_3 \si_4}[i,m; a ,b ] \ \right]_1&=& -  \delta[i,m] \  \ \si_1 \si_2 \ \delta_{\sib_2,\si_3} \delta_{\sib_4, \si_1} \  \ \GH[i,a] \ \GH[b,i] {\ \color{black} \delta(\tau_a-\tau_b) }, \nn \\
\left[ \ \U^{(a)}[i,m; a, b ] \ \right]_1  &_{\V\to 0}=  &-  2 \  \delta[i,m] \   \ \GH[i,a] \ \GH[b,i]{\ \color{black} \delta(\tau_a-\tau_b) },
\earray

 The zero source Fourier transforms  are as follows:
 \barray
 \left[ \ \U^{\si_1 \si_2}_{\si_3 \si_4} [p_1,p_2;p_3,p_4] \ \right]_1 &=& -  \ \ \si_1 \si_2 \ \delta_{\sib_2,\si_3} \delta_{\sib_4, \si_1} \  \delta_{p_1+p_4,p_2+p_3} \  \GH[p_3] \ \GH[p_4] \nn \\
 \left[ \ \U^{(a)}[p_1,p_2;p_3,p_4] \ \right]_1 &=& - 2 \ \delta_{p_1+p_4,p_2+p_3} \  \GH[p_3] \ \GH[p_4]
 \earray
 \subsubsection{$\Psi$ to $O(\lambda)$}
 We compute $[\Psi]_1$ from these.
 \barray
  \left[ \Psi(k) \right]_1 &=& \sum_p  \left( \varepsilon_p  + \frac{1}{2} \varepsilon_k  + \frac{1}{2} J_{k-p}\right) \ \GH[p] \  \left[ \U^{(a)}(p,k) \right]_1     + \sum_{p q}  \frac{1}{2}\varepsilon_{q+p-k}  \  \ \GH[p] \ \left[  \U^{(a)}(p,k;q+p-k,q) \right]_1  \nn \\
    &=& - \sum_{ p,q}   \left( \varepsilon_p +\varepsilon_{k+q-p}  +  \varepsilon_k   +\varepsilon_q +  J_{k-p}\right) \ \GH[p] \ \GH[q] \ \GH[ q+ k -p ] \nn \\
  &=&\sum_{p,q} W(k,q; q + k -p,p)   \ \GH[p] \ \GH[q] \ \GH[ q+ k -p ]  \label{eq104}
 \earray
 \subsubsection{Stepping up: $\mu$ to $O(\lambda^2)$}
Stepping up, we calculate  
 \barray
  \left[ \mu[i,m] \right]_{2}
  &=& - \delta[i,m]  \  \GH[i^-,i] \GH^{(k)}[i^-,i]  +    \  \left[ \Psi[i,m] \right]_1
   \earray
Hence at zero sources, the Fourier transform reads
\barray
\left[ \mu[k]\right]_2&=&  \frac{n^2}{4} 
 - \sum_{ p,q}   \left( \varepsilon_p +\varepsilon_{k+q-p}  +  \varepsilon_k   +\varepsilon_q +  J_{k-p}\right) \ \GH[p] \ \GH[q] \ \GH[ q+ k -p ]
\earray
 
Note that $[\mu]_2, \;  [\Psi]_1$ are invariant under all   three shift theorems. It is clear that this is a more non trivial application of the theorems than those in the lowest order.

\subsubsection{$\GHI$ derived objects}
 Let us now start with: $\left[ \GHI[i,m] \right]_1$ given in \disp{ginverse-first-order}:
  \barray  
  \left[ \GHI[i,m] \right]_1\
 &=& \left[  \GH^{(k)}[i,i] \cdot \V_{i, m}  +  \delta[i,m] \ \GH^{(k)}[\bb{a},i] \cdot \V^{(k)}_{i, \bb{a}} \right]  + \left[ \delta[i,m] \ \frac{1}{2}  J[i,\bb{j}] \ \GH^{(k)} [\bb{j},\bb{j}]  + \frac{1}{2} J[i,m] \ \GH^{(k)}[i,m] \right] \nn \\
 &&\left[ - t[i,m]  \ (   \GH^{(k)}[i,i]  +  \GH^{(k)}[m,m]   )   - \delta[i,m] \ t[i,\bb{j}] \left( \GH^{(k)}[i, \bb{j}] + \GH^{(k)}[ \bb{j},i ] \right)\right].   \label{eq8011}
   \earray
 
 \subsubsection{ Vertex functions to $O(\lambda)$}
{\bf  Three point vertex:}
 \barray
 \left[ \ \Lambda^{\si_1 \si_2}_{\si_3 \si_4}[i,m; a ] \ \right]_1 &=& - (\frac{\delta}{\delta \V_{a}^{\si_3 \si_4}}) \ \left[ \GHI_{\si_1 \si_2}[ i,m]  \right]_1 \nn \\
 &=& [ (I) +(II)+(III)]^{\si_1 \si_2}_{\si_3 \si_4},
 \earray
  where the terms $(I),(II),(III)$ refer to the three square bracketed terms in \disp{eq8011}.

 For the first term we calculate
 \barray
 (I)^{\si_1 \si_2}_{\si_3 \si_4} &=& -  (\frac{\delta}{\delta \V_{a}^{\si_3 \si_4}})  \left[ (\si_1 \si_a) \ \GH_{\sib_a \sib_1}[i,i] \ \V_{i,m}^{\si_a \si_2} + \delta[i,m] \  \ (\si_1 \si_2) \ \GH_{\sib_a \sib_1}[\bb{a} ,i] \ \V_{i,\bb{a}}^{\sib_2 \sib_a }
 \right]\\
 &_{\V\to 0} =& 0.
 \earray
 
 For all other terms we can use a simple calculation:
 \barray
 {\cal X} ^{\si_1 \si_2}_{\si_3 \si_4}[p,q;r] &=&   (\frac{\delta}{\delta \V_{r}^{\si_3 \si_4}}) \GH^{(k)}_{\si_1 \si_2}[p,q] \nn \\
  &=&  (\si_1 \si_2)  \GH_{\sib_2 \si_3}[p,r] \GH_{\si_4 \sib_1}[r,q]    \nn \\
{\cal X} ^{\si_1 \si_2}_{\si_3 \si_4}[p,q;r] &_{\V\to 0} =&  (\si_1 \si_2) \delta_{\sib_2 \si_3} \ \delta_{ \sib_1 \si_4} \  \GH[p,r] \GH[r,q] \nn \\
  {\cal X} ^{(a)}[p,q;r] &_{\V\to 0}=&  2 \  \GH[p,r] \GH[r,q] \nn \\
  \earray

 Therefore
 \barray
 (II)^{(a)}[i,m;a]
 &=& - \delta[i,m] \  J[i,\bb{j}] \  \GH [\bb{j}, a] \ \GH [a,\bb{j}]   -  J[i,m] \ \GH[i,a]  \ \GH[a ,m],
 \earray
 and
 \barray
 (III)^{(a)}[i,m;a] 
 &=&  2 \left[ \  t[i,m]  \ (   \GH[i,a] \GH[a,i] +  \GH[m,a]\GH[a,m]   )   + \delta[i,m] \ t[i,\bb{j}] \left( \GH[i, a] \GH[a, \bb{j}] + \GH[ \bb{j},a ] \GH[ a,i ]  \right) \ \right]
  \earray

 Zero source Fourier transforms read as:
 \barray
 (II)^{(a)}[p_1,p_2] &=& - J_{p_2-p_1} \sum_q \GH[q] \ \GH[q+p_2-p_1]  -  \sum_q \ J_{p_1-q } \ \GH[q] \ \GH[q+p_2-p_1] \nn \\ 
 (III)^{(a)}[p_1,p_2] &=& - 2 \sum_q \GH[q] \GH[q+ p_2-p_1] \ \{ \varepsilon_{p_2} +\varepsilon_{p_1}+ \varepsilon_{q+p_2-p_1} + \varepsilon_{q}  \}
  \earray

  Hence adding up we obtain:
\barray
\left[ \ \Lambda^{\si_1 \si_2}_{\si_3 \si_4}[p_1,p_2] \ \right]_1 & = & - (\si_1 \si_2) \delta_{\sib_2 \si_3} \ \delta_{ \sib_1 \si_4} \  \sum_{p_3,p_4} \delta_{p_1+p_4,p_2+p_3} \  \GH[p_3] \GH[p_4] \ \{ \varepsilon_{p_1} +\varepsilon_{p_2}+ \varepsilon_{p_3} + \varepsilon_{p_4} + \frac{1}{2} \left(J_{p_1-p_2} + J_{p_1-p_3} \right)  \} \nn \\
&&=  \frac{1}{2}(\si_1 \si_2) \delta_{\sib_2 \si_3} \ \delta_{ \sib_1 \si_4} \  \sum_{p_3,p_4} \GH[p_3] \GH[p_4] \ \left[ W(p_2,p_3,p_4,p_1)+ W(p_2,p_3,p_1,p_4) \right]
 \nn \\
&& \ \label{three-vertex-firstorder} \\
\left[ \ \Lambda^{(a)}[p_1,p_2] \ \right]_1&=&  - \  \sum_{p_3,p_4} \GH[p_3] \GH[p_4] \ \left[ W(p_2,p_3,p_4,p_1)+ W(p_2,p_3,p_1,p_4) \right]
\earray

Note that  rotation invariance relations imply that  since $\left[ \Lambda^{(1)}[p_1,p_2] \ \right]_1=0$, we must have
\barray
\left[ \Lambda^{(s)}[p_1,p_2] \ \right]_1& =& - \left[ \Lambda^{(t)}[p_1,p_2] \ \right]_1= \frac{1}{2} \left[ \Lambda^{(a)}[p_1,p_2] \ \right]_1 
\earray

{\bf The four point vertex}. The calculation  proceeds similarly:
 \barray
 \left[ \ \Lambda^{\si_1 \si_2}_{\si_3 \si_4}[i,m; a, b ] \ \right]_1 &=& - (\frac{\delta}{\delta \V_{a,b }^{\si_3 \si_4}}) \ \left[ \GHI_{\si_1 \si_2}[ i,m]  \right]_1 \nn \\
 &=& [ (IV) +(V)+(VI)]^{\si_1 \si_2}_{\si_3 \si_4}
 \earray
 Here the terms $(IV)-(VI)$ refer to the three square bracketed terms in \disp{eq8011}. 
 For the first term we calculate {\color{black} with implicit $\tau_a=\tau_b$}:
 \barray
 (IV)^{\si_1 \si_2}_{\si_3 \si_4}   &=&-  (\frac{\delta}{\delta \V_{a,b}^{\si_3 \si_4}})  \left[ (\si_1 \si_a) \ \GH_{\sib_a \sib_1}[i,i] \ \V_{i,m}^{\si_a \si_2} + \delta[i,m] \  \ (\si_1 \si_2) \ \GH_{\sib_a \sib_1}[\bb{a} ,i] \ \V_{i,\bb{a}}^{\sib_2 \sib_a }
 \right]\\
&  =& -  \delta[i, a] \delta[ m, b] \left[  (\si_1 \si_3) \delta_{\si_2 \si_4} \GH_{\sib_3 \sib_1}[i,i] \ \right] - \delta[i, a] \delta[ i,m ] \left[ (\si_1 \si_2) \ \delta_{\sib_2, \si_3} \ \GH_{\si_4 \sib_1}[b,i] \right] \nn \\
 (IV)^{\si_1 \si_2}_{\si_3 \si_4}  &_{\V\to 0}& - \delta[i, a] \delta[ m, b]   \ \delta_{\si_1 \si_3} \ \delta_{\si_2 \si_4}\GH[i,i]   -   \delta[i, a] \delta[ i, m] \ (\si_1 \si_2)  \  \delta_{\si_1 \sib_4} \delta_{\si_2 \sib_3} \ \GH[b,i]  \nn \\
 (IV)^{(a)}&=& \underbrace{ \delta[i, a] \delta[ m, b]   \ \GH[i,i] -  2 \     \delta[i, a] \delta[ i, m] \ \GH[b,i]}.  \label{inc-3}
 \earray
 This term is seen  result in a violation of \disp{four-three} and \disp{four-three-momentum} for reasons discussed there and in the second remark below \disp{fundapair}, and therefore is dropped below. We have carried it in the calculation, and demarcated it with the underbrace, in order to see its (minor)  contribution explicitly before dropping it.
 
 For all other terms we can use a simple calculation:
 \barray
 {\cal Y} ^{\si_1 \si_2}_{\si_3 \si_4}[p,q;r, s] &=&   (\frac{\delta}{\delta \V_{r,s}^{\si_3 \si_4}}) \GH^{(k)}_{\si_1 \si_2}[p,q] \nn \\
  {\cal Y} ^{(a)}[p,q;r,s] &_{\V\to 0}=&  2 \  \GH[p,r] \ \GH[s,q] \ {\color{black} \delta(\tau_r-\tau_s) }
 \earray
 
 Therefore {\color{black} with implicit $\tau_a=\tau_b$}:
 \barray
 (V)^{(a)}[i,m;a, b] 
 &=& - \delta[i,m] \  J[i,\bb{j}] \  \GH [\bb{j}, a] \ \GH [b ,\bb{j}]   -  J[i,m] \ \GH[i,a]  \ \GH[b ,m]
 \earray
 
 \barray
 (VI)^{(a)}[i,m;a, b]  &=&  2 \left[ \  t[i,m]  \ (   \GH[i,a] \GH[b,i] +  \GH[m,a]\GH[b,m]   )   + \delta[i,m] \ t[i,\bb{j}] \left( \GH[i, a] \GH[b, \bb{j}] + \GH[ \bb{j},a ] \GH[ b ,i ]  \right) \ \right]
  \earray

\barray
(IV)^{(a)}[i,m;a,b] 
&=& 
\delta[i, a] \delta[ m, b]   \ \GH[i,i] -  2 \     \delta[i, a] \delta[ i, m] \ \GH[b,i] \nn \\
(IV)^{(a)}[p_1,p_2,p_3,p_4]&=& \underbrace{  \delta_{p_1,p_3} \delta_{p_2,p_4} \GH[0^-] - 2 \ \delta_{p_1+p_4,p_2+p_3} \  \GH[p_4] }   \nn \\ \nn \\
 (V)^{(a)}[i,m;a, b] 
 &=& - \delta[i,m] \  J[i,\bb{j}] \  \GH [\bb{j}, a] \ \GH [b ,\bb{j}]   -  J[i,m] \ \GH[i,a]  \ \GH[b ,m] \label{inc-2} \\ 
 (V)^{(a)}[p_1,p_2,p_3,p_4]&=& - \left( J_{p_2-p_1}  +  J_{p_1-p_3 } \right) \ \GH[p_3] \ \GH[p_4] \ \delta_{p_1+p_4,p_2+p_3}  \nn \\ \nn \\
 (VI)^{(a)}[i,m;a, b]  &=&  2 \left[ \  t[i,m]  \ (   \GH[i,a] \GH[b,i] +  \GH[m,a]\GH[b,m]   )   + \delta[i,m] \ t[i,\bb{j}] \left( \GH[i, a] \GH[b, \bb{j}] + \GH[ \bb{j},a ] \GH[ b ,i ]  \right) \ \right] \nn \\
 (VI)^{(a)}[p_1,p_2,p_3,p_4]&=& - 2 \{  \varepsilon_{p_2} +\varepsilon_{p_1}+ \varepsilon_{p_3} + \varepsilon_{p_4}  \}  \ \GH[p_3] \ \GH[p_4] \ \delta_{p_1+p_4,p_2+p_3} 
\earray

Hence 
\barray
\left[ \ \Lambda^{\si_1 \si_2}_{\si_3 \si_4} [p_1,p_2,p_3,p_4] \ \right]_1& = & -  \  (\si_1 \si_2) \delta_{\sib_2 \si_3} \ \delta_{ \sib_1 \si_4} \ \delta_{p_1+p_4,p_2+p_3} \  \GH[p_3] \GH[p_4] \ \{ \varepsilon_{p_1} +\varepsilon_{p_2}+ \varepsilon_{p_3} + \varepsilon_{p_4} + \frac{1}{2} \left(J_{p_1-p_2} + J_{p_1-p_3} \right)  \} \nn \\
&&  \underbrace{ - \delta_{p_1,p_3} \delta_{p_2,p_4} \delta_{\si_1 \si_3} \delta_{\si_2 \si_4}  \GH[0^-] - 2 \ (\si_1 \si_2) \ \delta_{\si_1 \sib_4} \delta_{\si_2 \sib_3} \  \delta_{p_1+p_4,p_2+p_3} \  \GH[p_4]  } \label{inc-4}
\earray
 
 Thus
\barray
\left[ \Lambda^{(a)}[p_1,p_2,p_3,p_4] \ \right]_1& = & - 2 \  \delta_{p_1+p_4,p_2+p_3} \  \GH[p_3] \GH[p_4] \ \{ \varepsilon_{p_1} +\varepsilon_{p_2}+ \varepsilon_{p_3} + \varepsilon_{p_4} + \frac{1}{2} \left(J_{p_1-p_2} + J_{p_1-p_3} \right)  \} \nn \\
&& + \underbrace{ \delta_{p_1,p_3} \   \delta_{p_2,p_4} \ \GH[0^-] - 2 \ \delta_{p_1+p_4,p_2+p_3} \  \GH[p_4]}  \label{inc-1}
\earray
Comparing \disp{three-vertex-firstorder} and \disp{inc-1}, we see that other than the term with underbraces, these vertices satisfy \disp{four-three} or \disp{four-three-momentum}.

\subsubsection{ $\Phi$ to $O(\lambda)$}
We now assemble terms:
 \barray
\left[  \Phi(k) \right]_1 &=& \sum_p  \left( \varepsilon_p  + \frac{1}{2} \varepsilon_k  + \frac{1}{2} J_{k-p}\right) \ \GH[p] \left[ \Lambda^{(a)}(p,k) \right]_1     + \sum_{p q}  \frac{1}{2}\varepsilon_{q+p-k} \ \GH[p] \ \left[  \Lambda^{(a)}(p,k;q+p-k,q) \right]_1 \nn \\
 \earray
 
Let us rewrite this as ($k \to p_2, p\to p_1, q \to p_4$)
 \barray
&& \left[  \Phi(p_2) \right]_1 = \sum_{p_1}  \left( \varepsilon_{p_1}  + \frac{1}{2} \varepsilon_{p_2}  + \frac{1}{2} J_{p_1-p_2}\right) \ \GH[p_1] \left[ \Lambda^{(a)}(p_1,p_2) \right]_1     + \sum_{p_1+p_4= p_2+p_3}  \frac{1}{2}\varepsilon_{p_3} \ \GH[p_1] \ \left[  \Lambda^{(a)}(p_1,p_2;p_3,p_4) \right]_1 \nn \\
&=& \underbrace{\frac{n}{4} \sum_{p_3} \varepsilon_{p_3} \GH[p_3] - \sum_{p_1,p_4} \varepsilon_{p_1+p_4-p_2} \GH[p_1] \GH[p_4]}  \nn \\
&& -2 \sum_{p_1+p_4= p_2+p_3} \GH[p_1] \GH[p_3] \GH[p_4] \  \left( \varepsilon_{p_1}  + \frac{1}{2} \varepsilon_{p_2}  + \frac{1}{2} J_{p_1-p_2}\right) \{ \varepsilon_{p_1} +\varepsilon_{p_2}+ \varepsilon_{p_3} + \varepsilon_{p_4} + \frac{1}{2} \left(J_{p_1-p_2} + J_{p_1-p_3} \right)  \} \nn \\
&& - \sum_{p_1+p_4= p_2+p_3} \GH[p_1] \GH[p_3] \GH[p_4] \ \varepsilon_{p_3}  \  \{ \varepsilon_{p_1} +\varepsilon_{p_2}+ \varepsilon_{p_3} + \varepsilon_{p_4} + \frac{1}{2} \left(J_{p_1-p_2} + J_{p_1-p_3} \right)  \}
\earray
The first line with underbraces arises from the term $\Lambda$ in \disp{inc-2}, or \disp{inc-3} and \disp{inc-1}
which disobey the relation \disp{four-three} or \disp{four-three-momentum}. It  gives   a static but momentum dependent contribution, and  we will  drop it as discussed below \disp{four-three} and in the second remark below \disp{fundapair}.    The rest are  combined and rearranged to give
\barray
\left[  \Phi(p_2) \right]_1 
&=& - \sum_{p_1+p_4= p_2+p_3} \GH[p_1] \GH[p_3] \GH[p_4] \ \left(  \varepsilon_{p_1} +\varepsilon_{p_2}+  \varepsilon_{p_3} +  \varepsilon_{p_4} + J_{p_1-p_2} \right)    \{ \varepsilon_{p_1} +\varepsilon_{p_2}+ \varepsilon_{p_3} + \varepsilon_{p_4} + \frac{1}{2} \left(J_{p_1-p_2} + J_{p_1-p_3} \right)  \}, \nn \\
&=& \frac{1}{2} \sum_{p_1,p_3,p_4} \GH[p_1] \GH[p_3] \GH[p_4] W(p_2,p_3;p_4,p_1) \left[W(p_2,p_3;p_4,p_1)+W(p_2,p_3;p_1,p_4) \right], 
\earray
where in the first line we symmetrized further in $p_1  \leftrightarrow p_4$.

 We can bring this into standard notation by sending $p_2 \to k, p_1 \to p, p_3 \to q, p_4 \to k+q -p$:
 \barray
\left[  \Phi(k) \right]_1 & = & - \sum_{q,p} \GH[q] \ \GH[p] \ \GH[k+q-p] \nn  \\
&& \times \left(  \varepsilon_{k} +\varepsilon_{p}+  \varepsilon_{q} +  \varepsilon_{k+q-p} + J_{k-p} \right)  \  \{ \varepsilon_{k} +\varepsilon_{p}+ \varepsilon_{q} + \varepsilon_{k+q-p} + \frac{1}{2} \left(J_{k-p} + J_{p - q} \right)  \}  \nn \\
\left[  \Phi(k) \right]_1 &=& \frac{1}{2}  \sum_{q,p} \GH[q] \ \GH[p] \ \GH[k+q-p] W(k,q; q + k -p,p) \left[W(k,q; q + k -p,p) +W(k,q; p,q + k -p)  \right] \label{eq132}
\earray

\subsubsection{\bf Stepping  up and final Greens function to  $O(\lambda^2)$.}

We are now in a position to put together the second order result for $\GHI$ and also $\mu$.  Recall that
$
\left[ \GHI[k]   \right]_2 =  -\left[ Y_1[k] + \Phi[k] \right]_1 $, where  these variables are calculated in \disp{eq97} and \disp{eq132}. Hence we can now compile the equations of the second order theory with  sources turned off:
\barray
 \GHI(k) & = &  \GHI_0(k) + \lambda \left[ \GHI(k)\right]_1 + \lambda^2 \left[ \GHI(k)\right]_2 + O(\lambda^3), \nn \\
 \left[ \GHI[k]   \right]_0 & = & i \omega_n + \chem - \varepsilon_k  - \frac{1}{4} J_0 \nn \\
\left[ \GHI[k]   \right]_1 & = &   {n} \  \varepsilon_k  - \frac{n}{2} u_0 + \frac{1}{2} \sum_q J_{k-q} \ \GH[q] + \left( \frac{1}{4} J_0 \ n + 2 \sum_q \varepsilon_q \ \GH[q]  \right) \nn \\
\left[ \GHI[k]   \right]_2 & = & - \frac{3n^2}{8} \varepsilon_k +  \frac{n^2}{8} u_0  - \left[  \Phi(k) \right]_1 - \left( \frac{n^2}{8} J_0 + \frac{n}{4} \sum_q \varepsilon_q \ \GH[q] \ \right) \label{detailed-g}
\earray 
 We thus see that all the computed $[\GHI]_j$  are  invariant under the two shift theorems.  Adding up terms to $O(\lambda^2)$,
\barray
 \GHI(k) & = & i \omega_n + \chem' - \left( 1- \lambda \ n + \lambda^2 \ \frac{3n^2}{8}  \right)\varepsilon_k + \lambda \  \sum_q \frac{1}{2} J_{k-q} \ \GH[q] - \lambda^2  \left[  \Phi(k) \right]_1 + O(\lambda^3), \label{eq176} \\
 \chem'&=&\chem - u_0 \ \frac{\lambda n}{2}   (  1 - \frac{\lambda n}{4}) + \left[ J_0 \frac{ \lambda n}{4} ( 1- \frac{ \lambda n}{2}) + 2 \lambda ( 1 - \frac{ \lambda n}{8}) \  \sum_q \varepsilon_q \GH[q] \right], \label{chemdef}
 \earray 
with   $\left[  \Phi(k) \right]_1 $ defined in \disp{eq132} and a shifted chemical potential $\chem'$. Note that both terms in square brackets in  \disp{chemdef} are  independent of frequency and wave vector;  the first (T independent) term may be safely ignored since it 
 vanishes when we finally set $J_0 \to 0$, while the second term  involving $\sum_q \varepsilon_q \GH[q]$ is  expected to be weakly T dependent.
 
 Similarly the caparison factor $\mu$ is found to $O(\lambda^2)$ as:
 \barray
 \mu[k]&=& 1 + \lambda \ \left[ \mu[k] \right]_1 + \lambda^2 \ \left[ \mu[k] \right]_2 + O(\lambda^3), \nn \\
 \left[ \mu[k] \right]_1&=& - \frac{n}{2}, \nn \\
\left[ \mu[k]\right]_2&=&  \frac{n^2}{4} + \left[ \Psi[k]  \right]_1. \label{eq135}
\earray 
Adding up terms to $O(\lambda^2)$ we obtain:
\barray
\mu[k] &=& 1- \lambda  \frac{n}{2}  + \lambda^2 \ \frac{n^2}{4} + \lambda^2  \  \left[ \Psi(k) \right]_1 +O(\lambda^3)  \label{eq136}. \earray
along with the definition in \disp{eq104}.


\begin{thebibliography}{36}
\bibitem{hansen-shastry} {\em Extremely Correlated Fermi Liquids: Self consistent solution of the second order theory },  arXiv:1211.0594(2012) [cond-mat.str-el]
.

\bibitem{ecfl} B. S. Shastry, Phys. Rev. Letts {\bf 107}, 056403 (2011); {\em ibid},  {\bf 108}, 029702 (2012).




\bibitem{gros} B. Edegger, V. N. Muthukumar and C. Gros, Advances in Physics, {\bf 56}, 927 (2007).
\bibitem{pwa} P W Anderson, Science {\bf 235}, 1196 (1987).

\bibitem{harris} A. B Harris and R. V. Lange, Phys. Rev. {\bf 157}, 295 (1967).

\bibitem{zhang-rice} F. C. Zhang and T. M. Rice, Phys. Rev. {\bf B 37}, 3759 (1988).

\bibitem{gutzwiller} M. Gutzwiller, Phys. Rev. Letts. {\bf 10}, 159 (1963).


\bibitem{anatomy} B. S. Shastry,   arXiv:1104.2633 (2011) [cond-mat.str-el]; Phys. Rev. {\bf B 84} 165112 (2011).



\bibitem{gweon} G.-H. Gweon, B. S. Shastry, and G. D. Gu, Phys. Rev. Letts {\bf 107}, 056404 (2011).

\bibitem{asymmetry} B. S. Shastry, {\em Dynamical Particle Hole Asymmetry in Cuprate Superconductors},    arXiv:1110.1032 (2011) [cond-mat.str-el]; Phys. Rev. Letts {\bf 109},  (2012).


\bibitem{ecql} B S Shastry, Phys Rev {\bf B 81}, 045121 (2010); arXiv:0911.4327 [cond-mat.str-el].


\bibitem{simple-remark}  
A simple example $\G\sim c_0/( i \omega_n + \mu - \varepsilon_k)$ can be worked out in detail   and  illustrates  this remark.


\bibitem{agd} A. A. Abrikosov, L.  Gorkov and I. Dzyaloshinski, {\em Methods of Quantum Field Theory in Statistical Physics },  Prentice-Hall,
Englewood Cliffs, NJ (1963).
\bibitem{mahan} G. D. Mahan, {\em Many-Particle Physics}, (3 rd edition, Kluwer-Plenum, 2000).


\bibitem{fn-shift}  The magnitude of the relative coefficient $4$  in the condition for cancellation  $u_J= - 4 \times u_t$  is also  consistent with a second procedure. In the latter, the shift of $t_{ij}$ and $J_{ij}$  is carried out {\em after taking the commutator of $\X{i}{0\si}$ with H}, e.g. in the expression \disp{a2}, and the extra term in $A_{i, \si_i}$ generated by this process, is required to be zero. In general these two procedures can produce different coefficients, as in the minimal theory where one does not symmetrize the expressions. In such cases  the coefficient is determined by the one appearing in the equation after taking the commutator, as in \disp{a2}, since it propagates down the hierarchy of equations of motion. We report elsewhere the {\em minimal theory}, where the condition for cancellation is $u_J= - 2 \times u_t$, with a  relative coefficient $2$.   



\bibitem{isothermal}  Since finite T  many body  theory formalism is  of an isothermal rather than adiabatic character,  where quantum numbers are  unconstrained and consequently the Fermi surface changes its shape,  the  conventional usage of the  term adiabatic continuity seems misplaced. It  might be  more appropriately replaced by   the term ``parametric continuity'' or just ``continuity''. 


\bibitem{luttinger-ward} J. M. Luttinger and J. C. Ward, Phys. Rev. {\bf 118}, 1417 (1960).




\bibitem{brinkman} W. Brinkman and T. M. Rice, Phys. Rev. {\bf B 2}, 4302 (1970).


\bibitem{bza} G. Baskaran, Z. Zhou and P. W. Anderson, Sol. St. Comm. {\bf 63}, 973 (1987).


\bibitem{kotliar} G. Kotliar, Phys. Rev. {\bf 37}, 3664 (1988).

\end{thebibliography}
\end{document}